\documentclass[aps,prb,twocolumn,floats,epsfig]{revtex4}
\usepackage[iso-8859-1]{inputenx}
\usepackage{amssymb}
\usepackage{amsbsy}
\usepackage{amsmath}
\usepackage{epsfig}
\usepackage{color}
\usepackage{float}
\usepackage[colorlinks,linktocpage,bookmarks=false,citecolor=blue,linkcolor=red,urlcolor=blue]{hyperref}

\newcommand{\bib}{\bibitem}
\newcommand{\beq}{\begin{equation}}
\newcommand{\eeq}{\end{equation}}
\newcommand{\bea}{\begin{eqnarray}}
\newcommand{\eea}{\end{eqnarray}}

\begin{document}

\title{Mobility edge and multifractality in a periodically driven Aubry-Andr\'{e} model}

\author{Madhumita Sarkar, Roopayan Ghosh, Arnab Sen, and K. Sengupta}

\affiliation{School of Physical Sciences, Indian
Association for the Cultivation of Science, 2A and 2B Raja S. C.
Mullick Road, Jadavpur 700032, India}

\date{\today}

\begin{abstract}

We study the localization-delocalization transition of Floquet
eigenstates in a driven fermionic chain with an incommensurate
Aubry-Andr\'{e} potential and a hopping amplitude which is varied
periodically in time. Our analysis shows the presence of a mobility
edge separating single-particle delocalized states from localized
and multifractal states in the Floquet spectrum. Such a mobility
edge does not have any counterpart in the static Aubry-Andr\'{e}
model and exists for a range of drive frequencies near the critical
frequency at which the transition occurs. The presence of the
mobility edge is shown to leave a distinct imprint on fermion
transport in the driven chain; it also influences the Shannon
entropy and the survival probability of the fermions at long times.
In addition, we find the presence of CAT states in the Floquet
spectrum with weights centered around a few nearby sites of the
chain. This is shown to be tied to the flattening of Floquet bands
over a range of quasienergies. We support our numerical studies with
a semi-analytic expression for the Floquet Hamiltonian ($H_F$)
computed within a Floquet perturbation theory. The eigenspectra of
the perturbative $H_F$ so obtained exhibit qualitatively identical
properties to the exact eigenstates of $H_F$ obtained numerically.
Our results thus constitute an analytic expression of a $H_F$ whose
spectrum supports multifractal and CAT states. We suggest
experiments which can test our theory.

\end{abstract}

\maketitle

\section{Introduction}
\label{intro}

Localization phenomenon in an one-dimensional (1D) fermion chain
with quasiperiodic potentials has been studied extensively in the
past \cite{aaref, gaaref,sinharef, otherref}. These studies have
received a new impetus in recent times due to experimental
realization of such potentials in ultracold atom chains
\cite{exp1,exp2}. In contrast to the more conventional models with
uncorrelated disorder which exhibits such localization for any
disorder strength in 1D \cite{rev1}, fermion chains with non-random
but quasiperiodic potentials harbor a localization-delocalization
transition \cite{aaref,gaaref}. The simplest of such models termed
as Aubry-Andr\'{e} (AA) model \cite{aaref} has an Hamiltonian given
by $H= H_k + H_p$ where
\begin{eqnarray}
H_k &=& - \frac{\mathcal{J}}{2}\sum_j c_j^{\dagger} (c_{j+1} + c_{j-1}) \nonumber\\
H_p &=& \sum_j V_0 \cos(2\pi \eta j + \phi) c_j^{\dagger} c_j.
\label{ham1}
\end{eqnarray}
Here $j$ is the site index of the chain, $c_j$ denotes fermionic
annihilation operator at site $j$, $\mathcal{J}$ is the hopping
amplitude, $\eta$ is an irrational number usually chosen to be the
golden ratio $(\sqrt{5}-1)/2$, $V_0$ is the amplitude of the
potential, and $\phi$ is an arbitrary global phase. The AA
Hamiltonian can be shown to be self-dual and hosts a
localization-delocalization transition at $V_{0c}=2\mathcal{J}$
\cite{aaref}. For $V_0>(<)V_{0c}$, all the single-particle states in
the spectrum for the model are localized (delocalized). Such
transitions also occur in models with a more general class of such
quasiperiodic potentials (termed as generalized Aubry-Andr\'{e}
(GAA) potentials). These GAA Hamiltonians may have several forms;
for example, they may be given by Eq.\ \ref{ham1} with a different
form of the quasiperiodic potential [$\cos(2\pi \eta j+\phi) \to
\cos(2\pi \eta j+\phi)/(1-\alpha \cos(2\pi \eta j+\phi))$
]\cite{gaaref} or with longer range hopping $ \mathcal{J}\to
\mathcal{J}_{ij} = \mathcal{J}/|i-j|^a$, where $a$ is an exponent
\cite{sinharef}. One of the key aspects of these GAA Hamiltonians
which is absent in the AA model is the presence of a mobility edge
in the spectrum. Moreover, the latter class of GAA models also host
band of multifractal eigenstates in the delocalized phase. These
states, unlike their delocalized counterpart, are non-ergodic; thus
their presence change the ergodicity properties of these models
\cite{sinharef}.

The study of non-equilibrium dynamics of closed quantum systems has
gained tremendous impetus in the last decade \cite{rev2,rev3}. More
recently, it was realized that periodically (or quasiperiodically)
driven systems host a wide range of interesting phenomena that have
no analog in their undriven counterparts. These include topological
transitions in driven systems \cite{oka1}, dynamical transitions
\cite{heyl1, sen1}, dynamical freezing \cite{das1, pekker1},
realization of time crystals \cite{ach1,ashv1}, and weak ergodicity
breaking behavior \cite{rydref1}. Moreover such driven systems are
known to lead to novel steady states which have no analog in
non-driven systems \cite{sen2,ach2}. For periodically driven
systems, most of these phenomena can be understood by analyzing its
Floquet Hamiltonian $H_F$ which is related to the evolution operator
via $U(T,0)= \exp[-i H_F T/\hbar]$ \cite{rev4}, where
$T=2\pi/\omega_D$ is the drive period, $\omega_D$ is the drive
frequency, and $\hbar$ is the Planck's constant.

In this work, we study the properties of the Floquet eigenstates of
a driven fermionic chain in the presence of an AA potential. The
Hamiltonian of the chain that we study is given by Eq.\ \ref{ham1}
with $\mathcal{J} \to \mathcal{J}(t)$, where $\mathcal{J}(t)$ is a
periodic function of time characterized by a drive frequency
$\omega_D$. In our study, we choose two distinct protocols for
$\mathcal{J}(t)$. The first is the square pulse protocol where
$\mathcal{J}(t)= \mathcal{J}_0(-\mathcal{J}_0)$ for $t \le (>) T/2$
while the second is continuous protocol for which $\mathcal{J}(t)=
\mathcal{J}_0 \cos \omega_D t$. For both these protocols, we choose
$\mathcal{J}_0 \gg V_0$. In the regime, of large drive frequency,
$H_F \simeq H_p$, so that all the Floquet states are localized. In
contrast, for quasistatic drive $\omega_D \simeq 0$, all the states
are expected to be delocalized since $\mathcal{J}_0 \gg V_0$. This
features ensures the presence of a localization-delocalization
transition in the Floquet spectrum; the aim of the present work is
to understand the nature of the Floquet eigenstates near the
transition. We carry out this analysis numerically using exact
diagonalization of the fermionic chain followed by numerical
computation of $U(T,0)$; this numerical study is complemented by a
semi-analytic, albeit perturbative, computation of $H_F$ using a
Floquet perturbation theory (FPT) \cite{fl1,fl2}.

The central results that we obtain from this analysis are as
follows. First, we find that for a range of drive frequencies around
the localization-delocalization transition (occurring at a critical
value of the drive frequency $\omega_c$), the Floquet spectrum of
the driven AA model supports a mobility edge. This mobility edge,
which has no analog in the static AA model, occurs for $\omega_D \le
\omega_c$ and separates the delocalized states from either localized
or multifractal band of states. We chart out the drive frequency
range for which these multifractal states are present for both the
square-pulse and continuous drive protocols. Second, we unravel the
presence of single-particle CAT states in the Floquet spectrum for
$\omega_D \ge \omega_c$. These states occur at two specific
quasienergies in the Floquet spectrum and have wavefunctions which
are localized around two or three next-nearest neighbor sites in the
chain. We tie the presence of these states to the presence of
near-flat band dispersion in the Floquet spectrum at these
quasienergies and provide a semi-analytic understanding of their
existence. Third, we study the transport in such driven chain by
tracking the steady state value of the fermion density as a function
of drive frequency starting from a domain-wall initial state. This
initial state constitutes a many-body state where all sites to the
left (right) of the chain center are occupied (empty). The density
of the fermions in the steady state stays close to its initial
profile in the localized phase; in contrast it evolves to an uniform
density profile for the delocalized phase. In between, near the
transition where the mobility edge exists, it shows an intermediate
behavior which arises from the presence of both delocalized and
localized (or multifractal) states in the Floquet spectrum.
Analogous features are found in the Shannon entropy, and the return
probability of a single-particle fermion wavefunction (initially
localized at the center of the chain) measured in the steady state.
Fourth, we construct a semi-analytic, albeit perturbative,
expression of the Floquet Hamiltonian $H_F$ using a FPT. We show
that this semi-analytic Hamiltonian qualitatively captures the
physics of the driven system and use it to explain the presence of
the mobility edge and the multifractal states in the Floquet
spectrum. Finally, we discuss possible experiments which can test
our theory.

The plan of the rest of the paper is as follows. In Sec.\
\ref{secnum}, we provide a detailed numerical study of the driven
chain charting out the phase diagram, demonstrating the existence of
the mobility edge, and determining the location of the multifractal
and CAT states. This is followed by Sec.\ \ref{secfpt} where we
construct a semi-analytic Floquet Hamiltonian using FPT. Finally, we
summarize our main results and suggest experiments which can test
our theory in Sec.\ \ref{secdiss}. Some details of
the calculation of $H_F$ and a discussion of the approach of the
driven chain to the steady state are presented in the Appendices.

\section{Numerical results}
\label{secnum}

In this section, we present exact numerical result on the driven
fermion chain for both square-pulse  (Sec.\ \ref{sqppr}) and
sinusoidal drive protocol (Sec.\ \ref{sinpr}). Henceforth, we set
the global phase $\phi=0$ in Eq.~\ref{ham1} without loss of
generality.

\subsection{Square pulse protocol}
\label{sqppr}

For the square-pulse protocol, we vary the hopping amplitude of the
AA model (Eq.\ \ref{ham1}) as
\begin{eqnarray}
\mathcal{J}(t) &=& -\mathcal{J}_0,  \quad  t \le T/2  \nonumber\\
&=&\mathcal{J}_0, \quad  t > T/2  \label{spulse1}
\end{eqnarray}
This protocol is chosen to ensure that at the high frequency limit
where $H_F \simeq \int_0^T H(t) dt/T = H_p$, the Floquet Hamiltonian
represents a localized phase. To numerically find out the Floquet
spectrum at arbitrary frequency, we first find the eigenspectrum of
$H_{\pm}= H[\pm \mathcal{J}_0]$ (Eq.\ \ref{ham1}) using exact diagonalization
(ED). We denote these eigenvalues and eigenvectors as
$\epsilon^{\pm}_m$ and $|\psi_m^{\pm}\rangle$ respectively. Next, we note
that for the protocol given by Eq.\ \ref{spulse1}, the evolution
operator at $t=T$ can be written as
\begin{eqnarray}
U(T,0) &=&  e^{- i H_+ T/(2 \hbar)} e^{- i H_- T/(2 \hbar)}
\label{unum1} \\
&=& \sum_{p^+,q^-} e^{i (\epsilon_p^+ - \epsilon_q^-)T/(2 \hbar)}
c_{p^+ q^-} |\psi_p^+\rangle \langle \psi_q^-| \nonumber
\end{eqnarray}
where the coefficients $c_{p^+ q^-}= \langle
\psi_p^+|\psi_q^-\rangle$ denote overlap between the two eigenbasis.
Next, we numerically diagonalize $U(T,0)$ and obtain its eigenvalues
$\lambda_{m}$ and $|\psi_m\rangle$. The corresponding eigenstectrum
of $H_F$ is then obtained using the relation $U(T,0)= \exp[- i H_F
T/\hbar]$ which identifies the eigenvectors of $U(T,0)$ and $H_F$
and yields
\begin{eqnarray}
U(T,0) &=& \sum_m \lambda_m |\psi_m\rangle \langle \psi_m|, \quad
\lambda_{m} = e^{-i \epsilon^F_{m} T/\hbar} \label{fleigen}
\end{eqnarray}
where $\epsilon_m^F$ are the quasienergies which satisfy $H_F
|\psi_m\rangle = \epsilon_m^F |\psi_m\rangle$. In this section, we
shall use the properties of these Floquet eigenvalues and
eigenvectors to study phase diagram of the driven chain along with
multifractality of Floquet eigenstates and transport of fermions.

\subsubsection{Phase diagram and CAT states}
\label{pdsqp}

Having obtained the eigenspectrum of $H_F$, we first analyze the
localization properties of normalized single-particle eigenstates
$|\psi_m\rangle$ as a function of the drive frequency. To this end,
we compute the inverse participation ratio (IPR) of these states
given by
\begin{eqnarray}
I_m &=& \sum_{j=1}^L  |\psi_m(j)|^4, \quad \psi_m(j) = \langle
j|\psi_m\rangle \label{ipr1}
\end{eqnarray}
where $j$ denotes the coordinate of the lattice sites of the chain
of length $L$. The IPR $I_m \sim L^{-1(0)}$ in $d=1$ for a
delocalized (localized) state and thus acts as a measure of
localization of a quantum state.

\begin{figure}
\rotatebox{0}{\includegraphics*[width= 0.8 \linewidth]{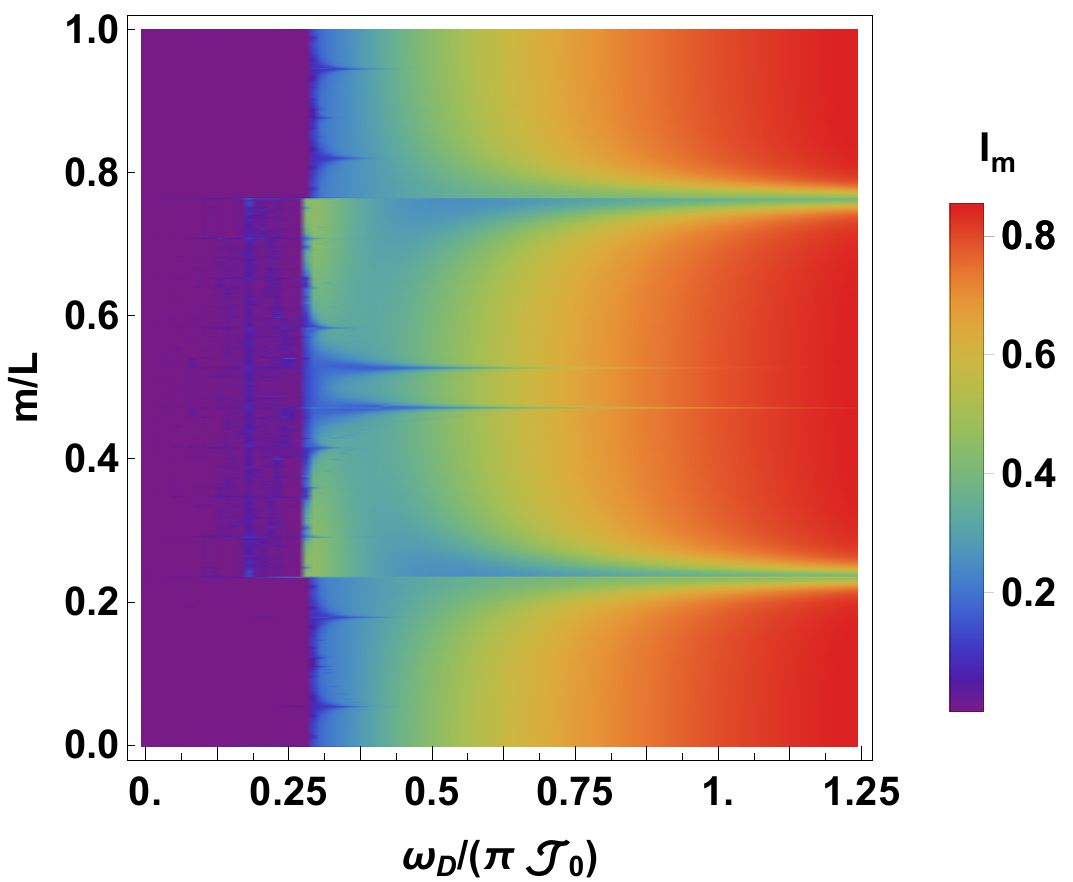}}
\caption{Plot of $I_m$ as a function of the normalized eigenfunction
index $m/L$ and $\omega_D/(\pi \mathcal{J}_0)$ showing the
localized/delocalized nature of the Floquet eigenstates
$|\psi_m\rangle$. Here $L=2048$ and we have set $\mathcal{J}_0=1$,
 $V_0/\mathcal{J}_0=0.05$, and scaled all energies(frequencies) in units of
$\mathcal{J}_0(\mathcal{J}_0/\hbar)$.} \label{fig1}
\end{figure}

This analysis leads to the phase diagram shown in Fig.\ \ref{fig1},
where we present $I_m$ as a function of the drive frequency
$\omega_D$ and for $V_0/\mathcal{J}_0=0.05$. As expected, the
Floquet eigenstates stay localized at large drive frequency; in
contrast, they are delocalized at low drive frequencies. In between,
around $\hbar \omega_D/\mathcal{J}_0 \simeq 0.3 \pi$, we find a
localization-delocalization transition. Near the transition, for
drive frequencies $0.15 \pi \le \hbar \omega_D/\mathcal{J}_0 \le 0.3
\pi$, we find the existence of a mobility edge separating a
delocalized band hosting states with $I_m \simeq 1/L \le 10^{-3}$
from those with finite $I_m > 0.1$. The nature of the states
separated by this mobility edge will be analyzed in detail in the
next subsection.

\begin{figure}
\rotatebox{0}{\includegraphics*[width= 0.49 \linewidth]{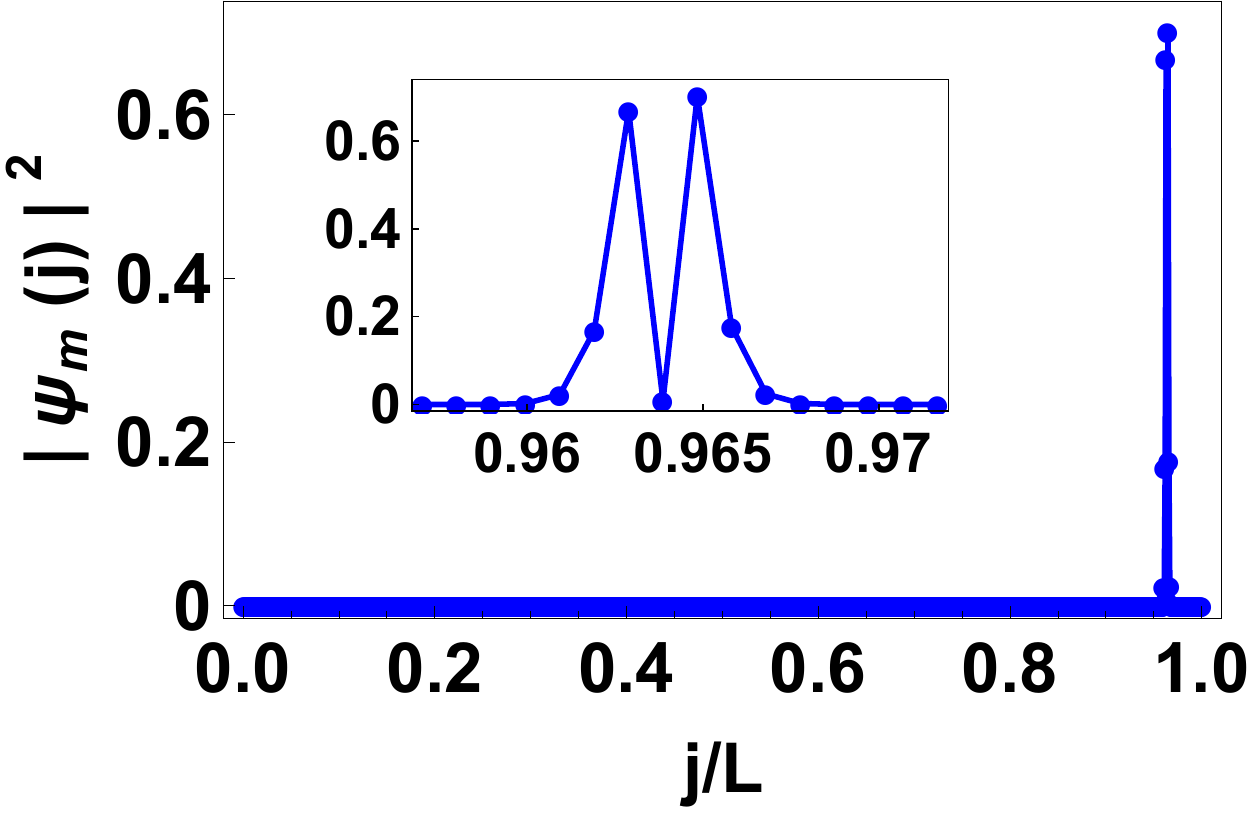}}
\rotatebox{0}{\includegraphics*[width= 0.49 \linewidth]{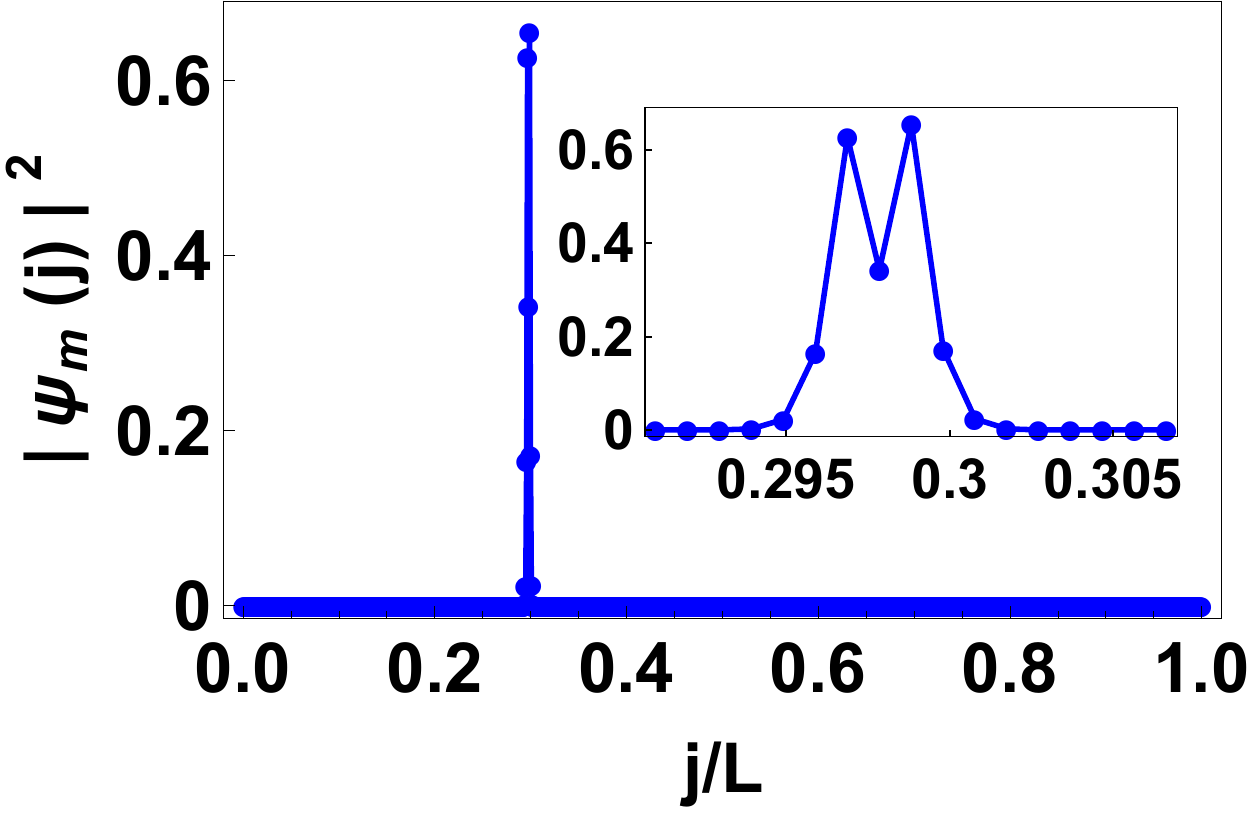}}
\rotatebox{0}{\includegraphics*[width= 0.49 \linewidth]{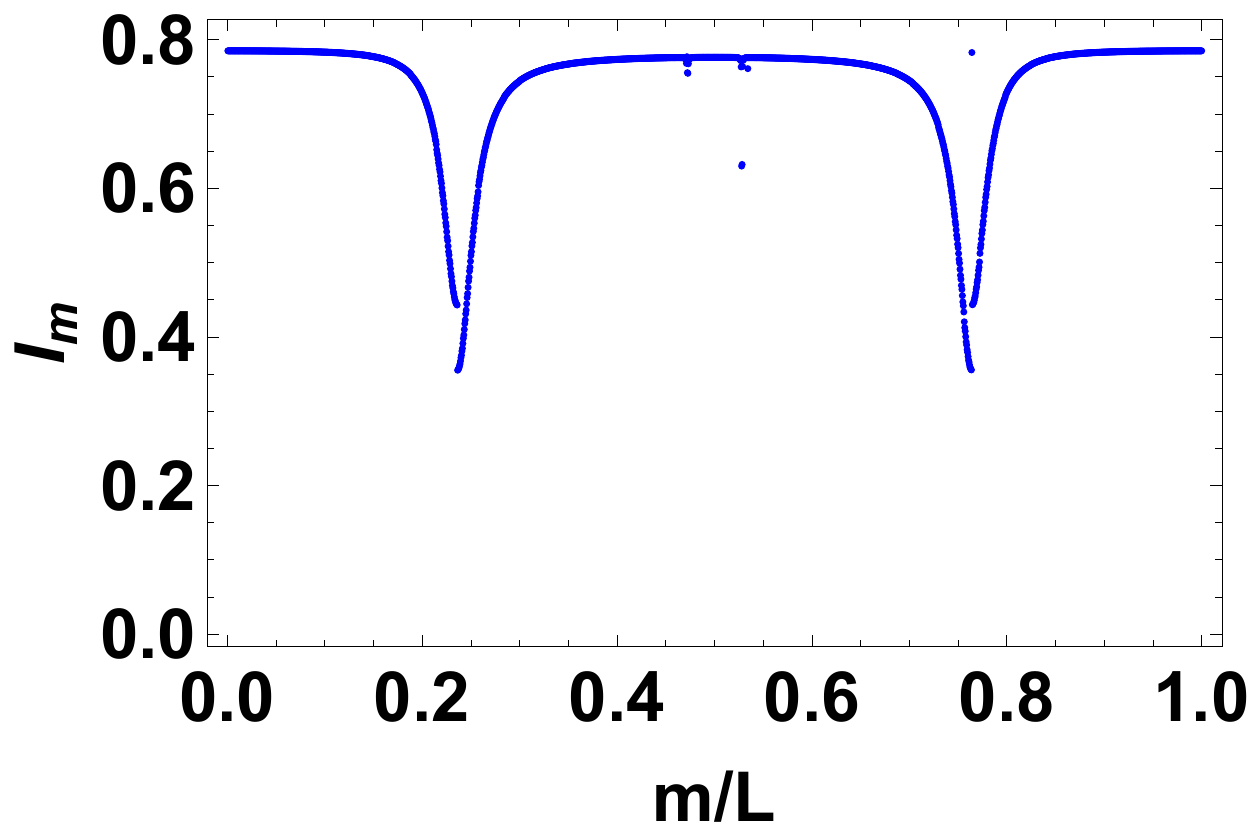}}
\rotatebox{0}{\includegraphics*[width= 0.49 \linewidth]{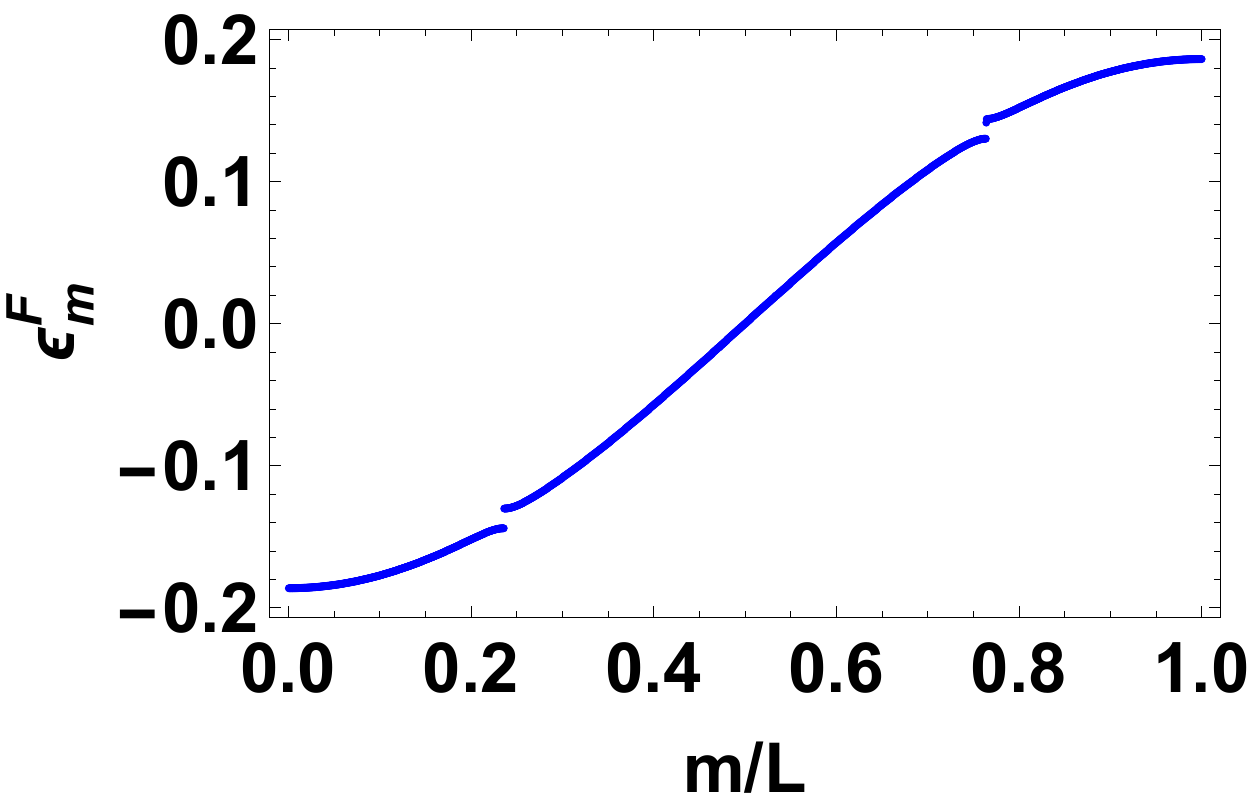}}
\caption{Top Panels: Spatial distribution of the CAT states at
$m/L=0.24$ (left panel) and $m/L= 0.76$ (right panel) and
$\omega_D/(\pi \mathcal{J}_0) =1$. Bottom left panel: Plot of $I_m$
as a function of $m/L$ showing the dip in $I_m$ for the CAT states.
Bottom right panel: Plot of $\epsilon_m^{F}$ as a function of $m/L$
at the same frequency showing flattening of Floquet bands before the
gap near $m/L\simeq 0.24, 0.76$.  All other parameters are as in
Fig.\ \ref{fig1}. See text for details.} \label{fig2}
\end{figure}

In addition to the mobility edge near the transition, we also find a
two narrow bands of states which retain a smaller value of $I_m
\simeq 0.5$ deep inside the localized phase. In what follows, we
analyze the character of these states. First, a plot of
$|\psi_m(j)|^2$ at a fixed frequency $\omega_D$, shown in the top
panels of Fig.\ \ref{fig2}, reveals that these states have their
weights spread between few lattice sites even deep inside the
localized phase. This behavior is to be contrasted with that of a
canonical localized state where $|\psi_m(j)|^2$ is finite only on a
single site. This feature makes them  perfect examples of CAT states
whose wavefunctions are localized over more than one site. These
states can also be distinguished from either localized or
delocalized states via $I_m$. This can be clearly seen in the bottom
left panel of Fig.\ \ref{fig2} where one sees a clear dip in $I_m$
for these states. The reason for the existence of such states can be
understood from the structure of the Floquet eigenenergies
$\epsilon_m^F$ shown in the bottom right panel of Fig.\ \ref{fig2}
for $\hbar \omega_D/\mathcal{J}_0= \pi$ deep inside the localized
regime. The Floquet energy dispersion becomes flat
near the gaps in the spectrum; we have checked that this
characteristic persists for all $\omega \ge \omega_c$ for large
enough $L$. We find that the CAT states reside in these flat band
regimes of the Floquet spectrum.

The existence of such CAT states and its relation to the flat
regions in the Floquet band can be understood, in the high frequency
regime, as follows. We first note that in this regime, from the
first order Magnus expansion $H_F \simeq  H_p$; thus $H_F$ is almost
diagonal in the position basis; each of its eigenstates is localized
on one of the sites and these eigenstates can be approximately
labeled by site indices of the chain. The off-diagonal terms are
generated at higher order in the Magnus expansion and are therefore
typically small in this region. Usually if the quasienergies are
well-separated from each other, these off-diagonal terms do not
change the nature of the Floquet spectrum. However, we note that
this assumption breaks down around $m/L\simeq 0.24,0.76$; the
quasienergy spacing between the states localized in this regime
approaches zero as can be seen from flattening of the Floquet band.
Consequently, the presence of off-diagonal term in $H_F$ arising
from $H_K$, however small, becomes important and leads to
hybridization of the states localized on nearby sites. This leads to
a pair of CAT states in the spectrum. We note that these states
persists only for frequencies where the Floquet spectrum has a flat
region; for $\omega_D \le \omega_c$, this feature is absent and one
does not find the CAT states in this regime.

\subsubsection{Multifractal states}
\label{mssqp}

The multifractal nature of a quantum state can not be ascertained
from the IPR alone. To this end, we now present computation of a
generalized IPR defined as \cite{mfracref1,mfracref2}
\begin{eqnarray}
I_m^{(q)} &=& \sum_{j=1}^L|\psi_m(j)|^{2 q} \label{mfrac}
\end{eqnarray}
where $I_m \equiv I_m^{(2)}$. It is well-known that $I_m^{(q)} \sim
L^{-\tau_q}$, where the fractal dimension of the state is given by
$D_q= \tau_q/(q-1)$. We note that for delocalized states $D_q=1$ for
all $q$, while for localized states $D_q=\tau_q=0$. Multifractal
states typically yield $0<D_q<1$.

\begin{figure}
\rotatebox{0}{\includegraphics*[width= 0.49 \linewidth]{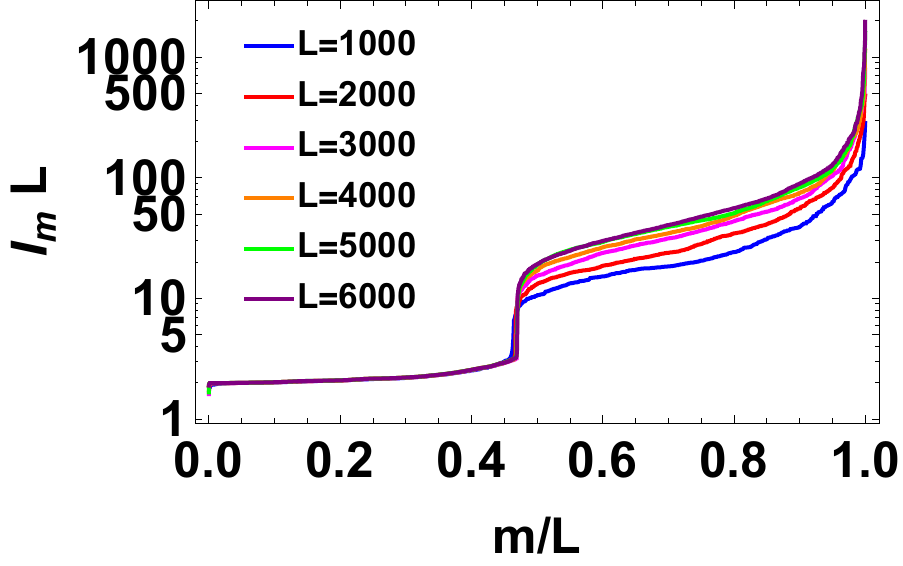}}
\rotatebox{0}{\includegraphics*[width= 0.49 \linewidth]{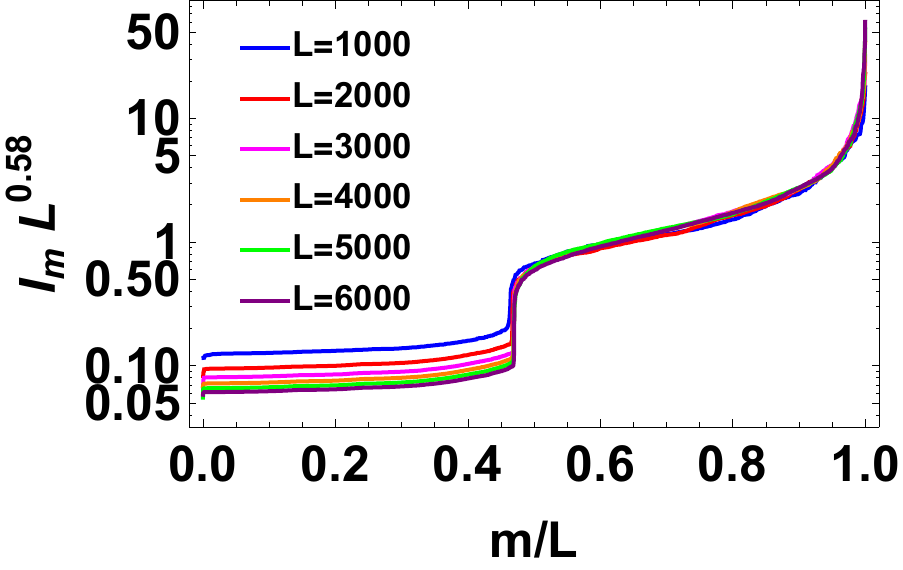}}
\caption{Left Panel: Plot of $I_m L$ as a function of $m/L$ (sorted
in increasing order of $I_m$) for $\omega_D/(\pi \mathcal{J}_0)
=0.245$ for several $L$ showing collapse of the delocalized states
below the mobility edge around the middle of the spectrum. Right
panel: Similar plot for $I_m L^{0.58}$ showing collapse of the
multifractal states above the mobility edge. All other parameters
are same as in Fig.\ \ref{fig1}.} \label{fig3}
\end{figure}

\begin{figure}
\rotatebox{0}{\includegraphics*[width= 0.48 \linewidth]{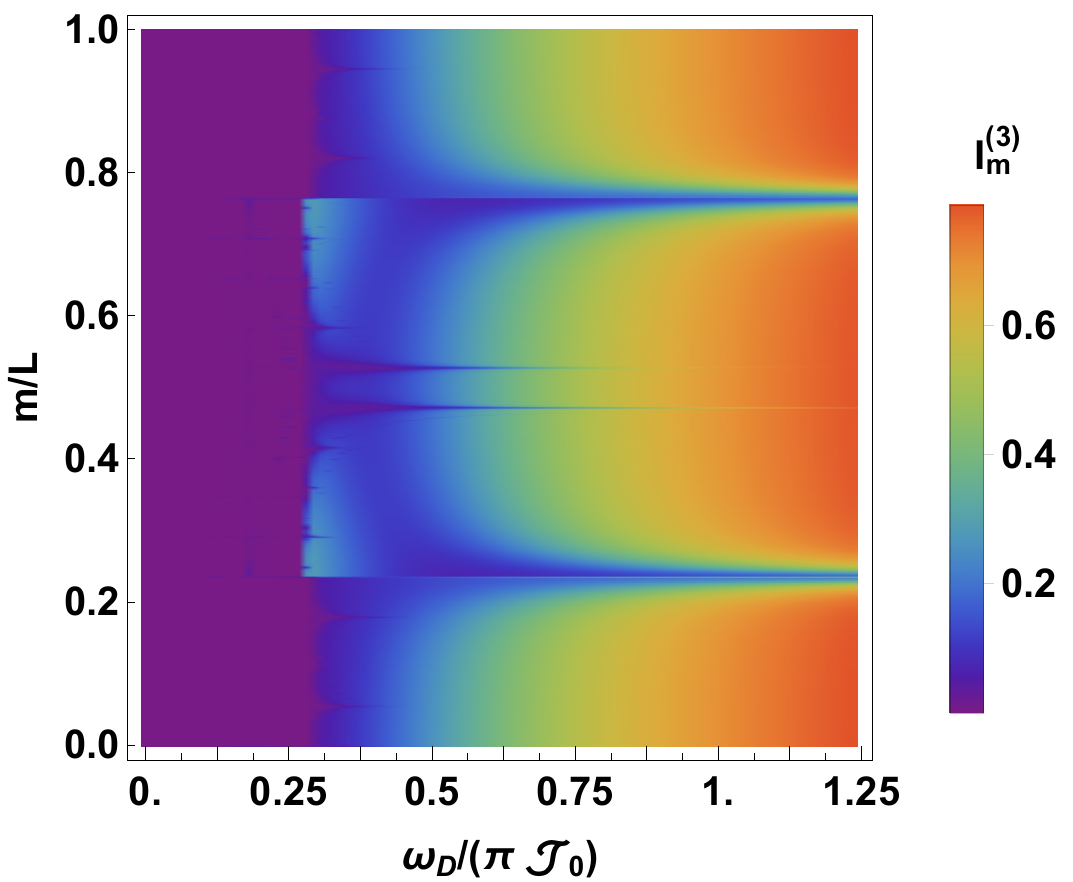}}
\rotatebox{0}{\includegraphics*[width= 0.48 \linewidth]{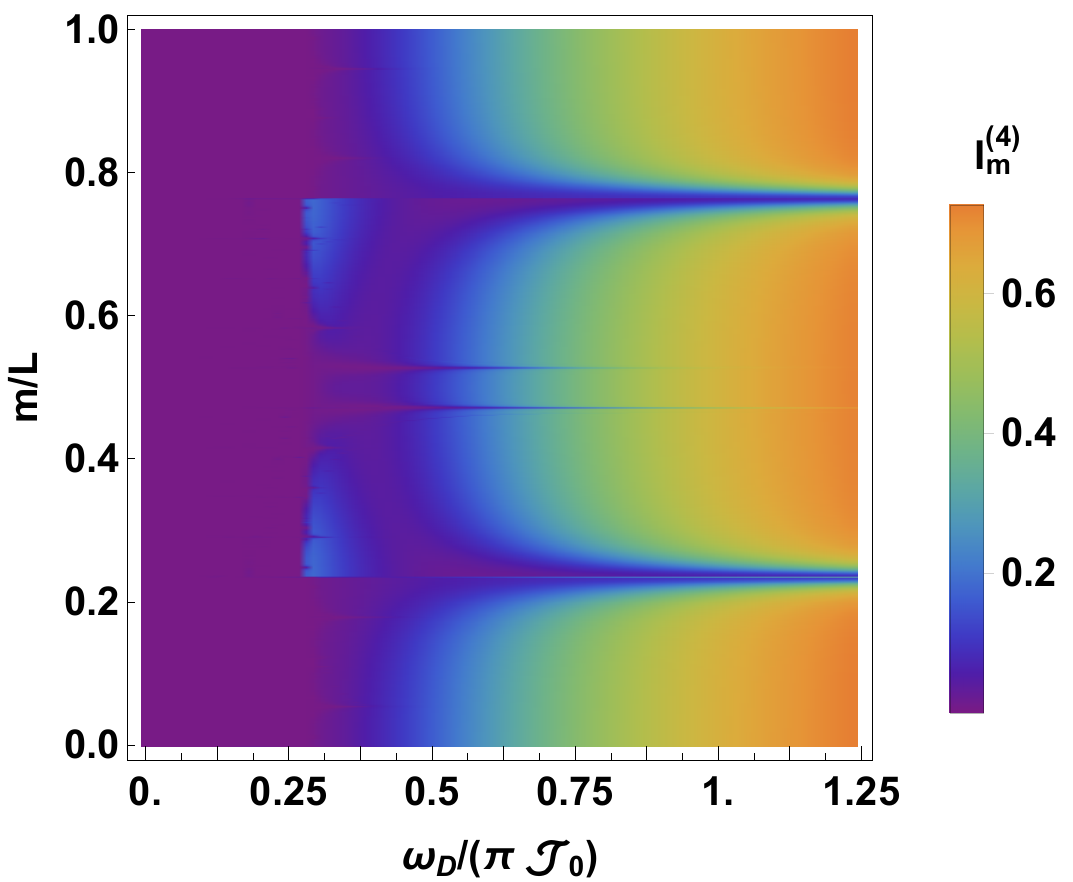}}
\rotatebox{0}{\includegraphics*[width= 0.48 \linewidth]{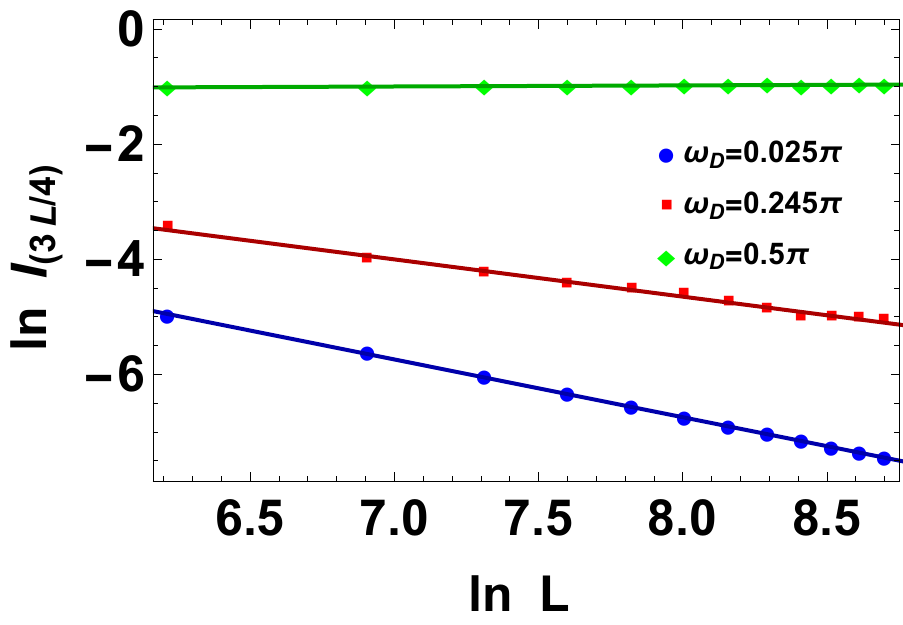}}
\rotatebox{0}{\includegraphics*[width= 0.48 \linewidth]{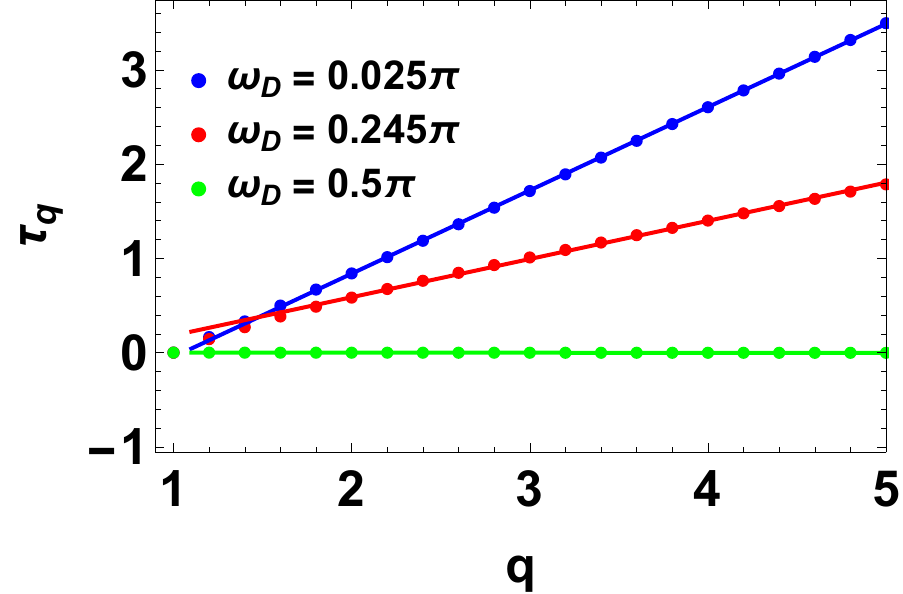}}
\caption{Top Panels: Plot of $I_m^{(3)}$ (left panel) and
$I_m^{(4)}$ (right panel) as a function of $m/L$ and $\omega_D$
indicating multifractal states at intermediate frequency. Bottom
left panel: Plot for $\ln I_m$ vs $\ln L$ used for extracting
$\tau_2$ for several representative frequencies for the state
corresponding to $m/L=0.75$. The behavior of perfectly delocalized
(blue dots at $\omega_D/(\pi \mathcal{J}_0)=0.025 $) and localized
(green dots, $\omega_D/(\pi \mathcal{J}_0) =0.5$) can be
distinguished from that of a multifractal states (red dots
$\omega_D/(\pi \mathcal{J}_0)=0.245$). Bottom right panel: Plot of
$\tau_q$ as a function of $q$ for a delocalized (blue dots at
$\omega_D/(\pi \mathcal{J}_0)=0.025 $), localized (green dots
at $\omega_D/(\pi \mathcal{J}_0) =0.5$) and multifractal (red dots
at $\omega_D/(\pi \mathcal{J}_0)=0.245$) states. All other parameters
are same as in Fig.\ \ref{fig1}.} \label{fig4}
\end{figure}

From the phase diagram shown in Fig.\ \ref{fig1}, it can be seen
that the Floquet eigenstates are mostly perfectly localized for
large $\omega_D \gg \omega_c$; in contrast, they are delocalized for
$\omega_D \simeq 0$. Thus it is evident that the presence of
multifractal states, if any, would be near the transition where the
mobility edge separates delocalized states from a bunch of states
for which $0 < \tau_2 < 1$. With this expectation, we first plot
$I_m$ for all Floquet eigenstates near the transition corresponding
to $\hbar \omega_D/\mathcal{J}_0 =0.245 \pi$ in Fig.\ \ref{fig3}
after sorting the eigenstates in terms of increasing IPR, which
clearly shows the presence of delocalized states and multifractal
states separated by a mobility edge. The left panel shows a plot of
$I_m L$ for all states; we find that for the states with $m/L <0.6$,
this quantity collapses for the different system sizes indicating
that $I_m \sim L^{-1}$ and hence, the delocalized nature of these
states. In contrast, the right panel of the Fig.\ \ref{fig3}
indicates that $I_m$ for all states with $m/L>0.6$ scale as
$L^{-0.58}$ indicating $\tau_2=D_2= 0.58$ and a multifractal nature.
To confirm this, we plot $I_m^{(q)}$ for these states for $q=3$ and
$q=4$ as shown in top panels of Fig.\ \ref{fig4}. The value of
$\tau_q$ is extracted from a plot of $\ln I_m^{(q)}$ vs $\ln L$ for
several $L$ as shown in the bottom left panel of Fig.\ \ref{fig4},
where $m/L \approx 3/4$ after sorting in increasing order of $I_m$
for each $L$. A plot of $\tau_q$ obtained using this procedure is
shown as a function $q$ for representative drive frequencies in the
right bottom panel of Fig.\ \ref{fig4}. We find that $\tau_q \sim
D_q (q-1)$ for all plots; $D_q \sim 0(1)$ for localized
(delocalized) states corresponding to $\hbar
\omega_D/\mathcal{J}_0=0.5(0.025) \pi$ while $0 < D_q <1$ for
multifractal states at $\hbar \omega_D/\mathcal{J}_0=0.245 \pi$.

\begin{figure}
\rotatebox{0}{\includegraphics*[width= 0.49 \linewidth]{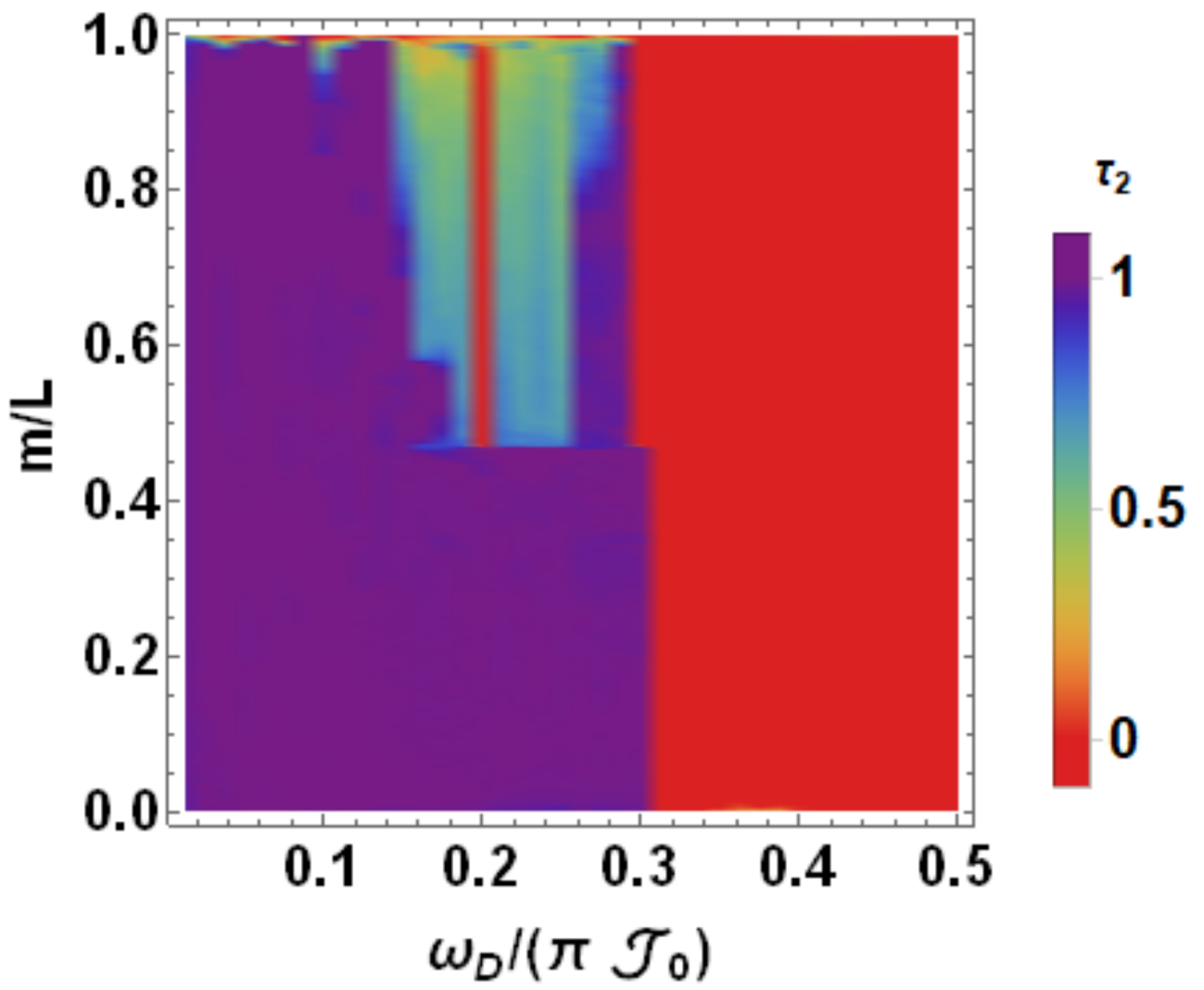}}
\rotatebox{0}{\includegraphics*[width= 0.49 \linewidth]{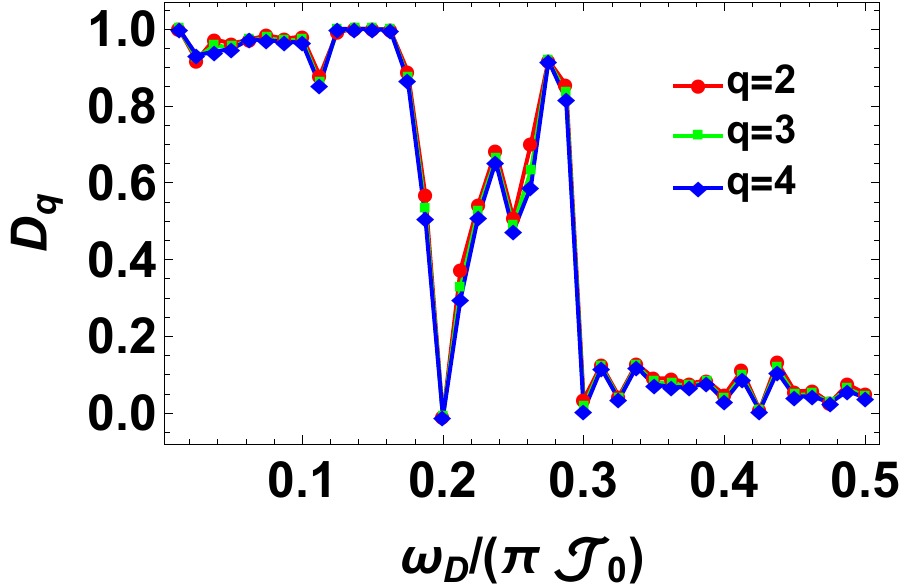}}
\caption{Left: Plot of $\tau_2$ as a function of $m/L$ (after
sorting in increasing order of $I_m$) and $\omega_D/(\pi
\mathcal{J}_0)$ showing the mobility edge separating delocalized and
multifractal states for $0.15  \le \omega_D/(\pi \mathcal{J}_0)\le
0.25$. The system sizes used for extracting $\tau_2$ are $L=500,
\cdots, 6000$ in steps of $500$. Right: Plot of $D_q$ as a function
of $\omega_D/(\pi \mathcal{J}_0)$ for $m/L=0.75$. All other
parameters are same as in Fig.\ \ref{fig1}. See text for details.}
\label{fig5}
\end{figure}

Numerically we find the presence of multifractal states for a wide
range of frequencies below $\omega_c$, till $\hbar
\omega_D/\mathcal{J}_0=0.15 \pi$. This is shown in the left panel of
Fig.\ \ref{fig5} where we plot $\tau_q$ for all states in the
Hilbert space after sorting in increasing order of $I_m$ as a
function of $\omega_D$. This clearly shows the presence of
multifractal states with quasienergies higher than the mobility edge
for $0.15 \pi \le \hbar \omega_D/\mathcal{J}_0 \le 0.25 \pi$. Our
analysis indicates that the multifractal dimension $D_q$ is a
non-monotonic function of $\omega_D$. This is shown in the right
panel of Fig.\ \ref{fig5} for a randomly chosen state corresponding
to $m/L=0.75$. The dip in the plot around $\hbar
\omega_D/\mathcal{J}_0=0.2 \pi$ corresponds to the narrow frequency
region where we find localized, rather than multifractal, states
above the mobility edge.

Another test of multifractality of a given state is the distribution
of the energy difference between the odd-even ($s^{\rm o-e}_m =
\epsilon_{2m+1}^F-\epsilon_{2m}^F$) and the even-odd ($s^{\rm e-o}_m
= \epsilon_{2m}^F-\epsilon_{2m-1}^F$) energies \cite{sinharef}. For
delocalized states these two gaps are different due to almost doubly
degenerate spectrum leading to $s^{\rm o-e}_m \simeq 0$ while for
the localized  states, this distinction is not present, hence this
gap vanishes. For multifractal states, the behavior follows neither
of the two patterns and both $s^{\rm o-e}_m$ and $s^{\rm e-o}_m$
show a scattered behavior. Thus one can distinguish between
different set of states by studying these energy gaps.

In our case, due to the drive, we study the difference of
quasi-energies. Since our spectrum has a mobility edge, the
quasi-energy spectrum is folded.\cite{chalker} There
is no general way to unfold the spectrum in such a case;
consequently, the identification of odd and even energies cannot be
done uniquely at low and intermediate drive frequencies. However,
the distribution of quasi-energy differences would still show the
same features as discussed in the last paragraph. Hence to
highlight the expected behavior, we define two new quantities,
$s_m^{min}={\rm Min}[s^{\rm o-e}_m,s^{\rm e-o}_m]$ and
$s_m^{max}={\rm Max}[s^{\rm o-e}_m,s^{\rm e-o}_m]$, which would
allow us to separate the two gaps properly in the delocalized region
of the spectrum.

\begin{figure}
\rotatebox{0}{\includegraphics*[width= 0.48 \linewidth]{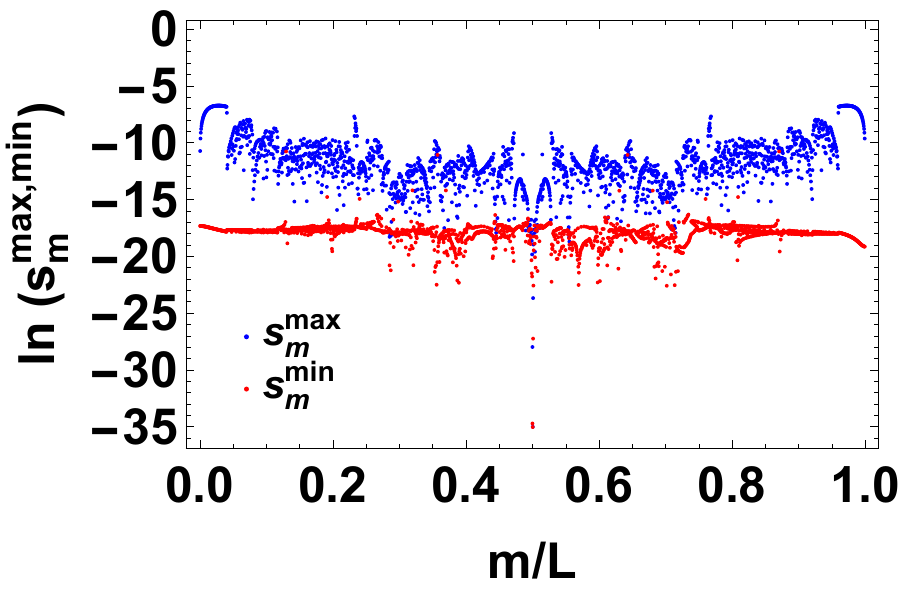}}
\rotatebox{0}{\includegraphics*[width= 0.48 \linewidth]{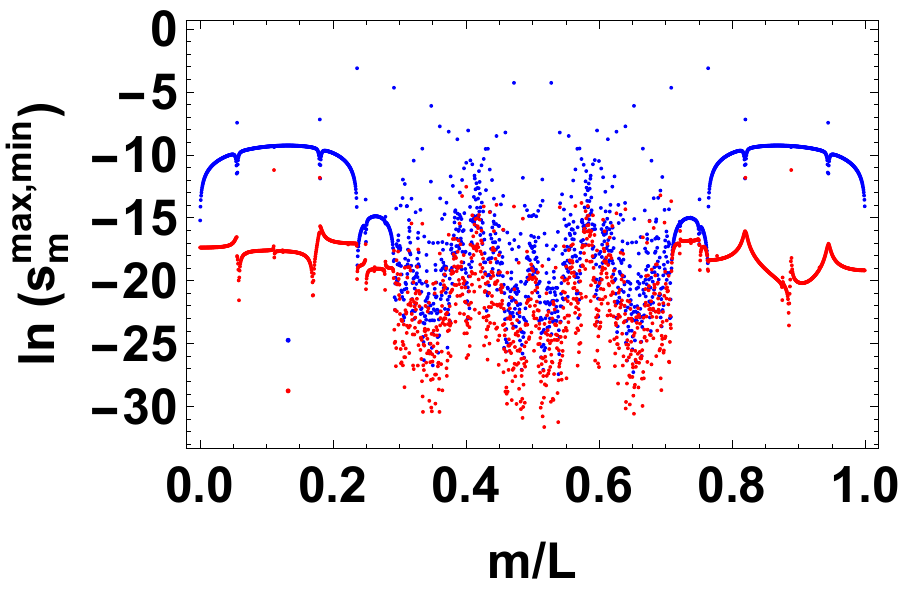}}
\rotatebox{0}{\includegraphics*[width= 0.48 \linewidth]{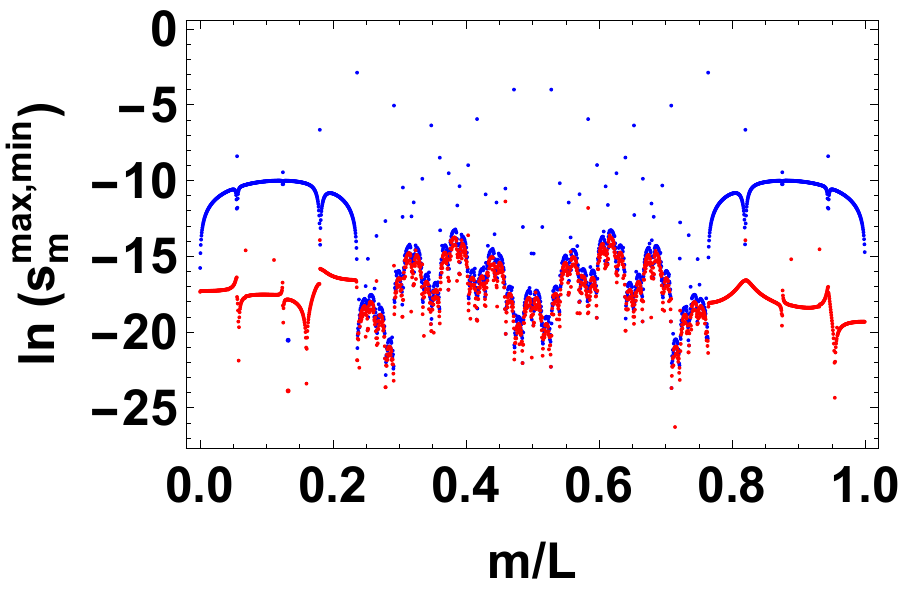}}
\rotatebox{0}{\includegraphics*[width= 0.48 \linewidth]{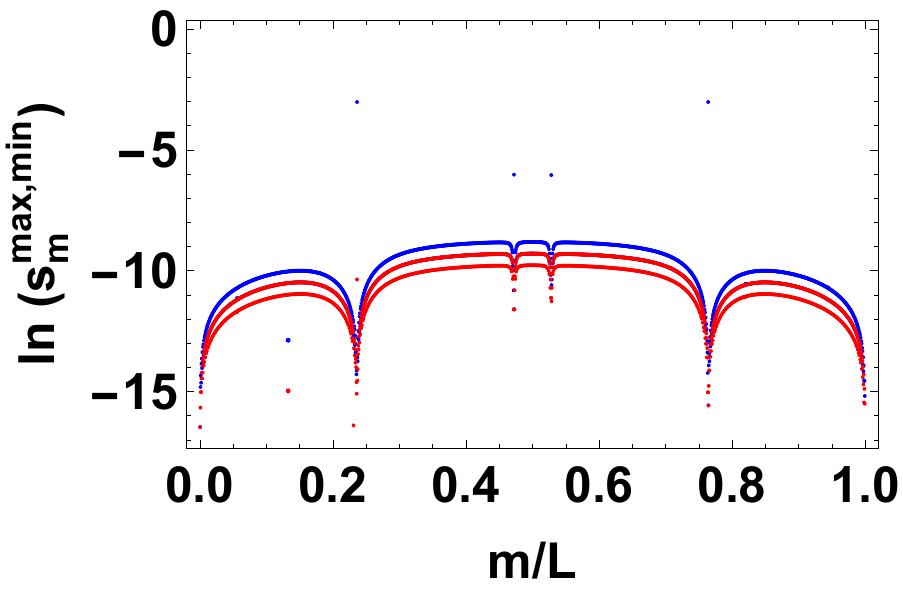}}
\caption{Top Panels: Plot of $\ln s_m^{min}$ (red dots) and $\ln
s_m^{max}$ (blue dots) as a function of $m/L$ for  $\omega_D/(\pi
\mathcal{J}_0)=0.025 $ (top left panel), $0.175 $ (top right panel),
$0.2 $ (bottom left panel), and $0.5$ (bottom right panel). We have
set $L=4181$, since $L$ needs to be a Fibonacci number. All other
parameters are same as in Fig.\ \ref{fig1}.} \label{fig6}
\end{figure}

A plot of $\ln s_m^{min}$ and $\ln s_m^{max}$ is shown in Fig.\
\ref{fig6} as a function of $m/L$ for several representative
frequencies. For $\hbar \omega_D/\mathcal{J}_0=0.025 \pi$, where all
states are delocalized, the plot shows clear separation of these two
quantities for all $m/L$; we find, in accordance to standard
expectation, that $s_m^{min} \simeq 0$ for all $m$. In contrast for
$\hbar \omega_D/\mathcal{J}_0 = 0.5 \pi$, where all states are
localized we find regular distribution of both energy gaps as shown
in the bottom right panel of Fig.\ \ref{fig6}. The small difference
between $s_m^{min}$ and $s_m^{max}$ in this regime is a finite size
effect and reduces with increasing $L$. In contrast, in the
intermediate frequency regime at $\hbar \omega_D/\mathcal{J}_0=0.175
\pi$ (top right panel of Fig.\ \ref{fig6}), we find clear signature
of a mobility edge separating delocalized and multifractal states;
the latter class of states can be recognized by strong scattering in
distribution of both $\ln s_m^{min}$ and $\ln s_m^{max}$
\cite{sinharef}. The presence of a mobility edge separating the
localized and delocalized at $\hbar \omega_D/\mathcal{J}_0=0.2 \pi$
is shown in the bottom left panel of Fig.\ \ref{fig6}. We find that
the presence of  localized states above the mobility edge can be
clearly distinguished from that of multifractal states because here
there is a overlap of $\ln s_m^{min}$ and $\ln s_m^{max}$ unlike the
scattered distribution found in the latter states.

Before ending this subsection, we would like to point out that our
analysis shows that the driven AA model, at intermediate
frequencies, shows mobility edge and multifractal states even when
the parent Hamiltonian (Eq.\ \ref{ham1}) does not host either of
these features. This distinguishes this phenomenon from earlier
studies of driven GAA model where the drive, in the high frequency
regime, creates a multifractal state by superposing localized and
delocalized states across the mobility edge of the static GAA
Hamiltonian \cite{gaadrive}. For completeness, we note here that
Ref.\ \onlinecite{sinha2} showed that periodic modulations of the
phase of the hopping amplitude (e.g., by applying a time dependent
gauge field) in the AA model also exhibited a mobility edge and
multifractal states. However, in our case, the time dependent
hopping amplitudes are real-valued. The mechanism leading to the
multifractal states for the driven AA model shall be discussed in
Sec.\ \ref{secfpt}.

\subsubsection{Transport, return probability and entropy}
\label{tresec}

In this subsection, we address the effect of the presence of
mobility edge on fermion transport, survival probability of the
fermion wavefunction in the steady state and their Shannon entropy .

For studying transport property of the fermions we start from a
domain-wall initial state \cite{domainref} defined, in the fermion
number basis, by
\begin{eqnarray}
|\psi_{\rm init}\rangle = |n_{1}=1, ... n_{L/2}=1, n_{L/2+1}=0, ...
n_{L}=0\rangle  \label{initstate}
\end{eqnarray}
where we have taken $L$ to be an even integer (chain with even
number of sites) and $n_j =\langle \hat n_j \rangle$ denotes fermion
occupation number on the $j^{\rm th}$ site and $\hat n_j =
c_j^{\dagger} c_j$ is the fermion number operator on that site. The
wavefunction after $n$ drive cycles is then given by
\begin{eqnarray} |\psi'\rangle &=& U(nT,0) |\psi_{\rm init}\rangle=
\sum_m c_m^{\rm init} e^{-i n\epsilon_m^F T/\hbar} |\psi_m\rangle
\label{finalstate}
\end{eqnarray}
where $|\psi_m\rangle$ denotes Floquet eigenstates with $L/2$
fermions and $c_m^{\rm init}= \langle \psi_m|\psi_{\rm
init}\rangle$. Using this state, one may compute the density profile
of fermions in the steady state. In what follows we study the
quantities
\begin{eqnarray}
N_{0j}(T) &=& \langle 2(\hat n_j-1/2)\rangle \nonumber \\
\quad N_{\rm av} (T) &=& \frac{4}{L} \sum_{j=1..L}  \langle \hat
(\hat n_j-1/2)\rangle ^2 \label{nnsq}
\end{eqnarray}
where the average is taken with respect to the steady state reached
under a Floquet drive starting from $|\psi_{\rm init}\rangle$. In
terms of the Floquet eigenfunctions $|\psi_m\rangle$ and the overlap
coefficients $c_m^{\rm init}$ (Eq.\ \ref{finalstate}) these can be
expressed as
\begin{eqnarray}
N_{0j}(T) &=& \sum_m |c_{m}^{\rm init}|^2  \langle \psi_m| 2(\hat
n_j
-1/2)|\psi_m\rangle \label{ssopexp} \\
 N_{\rm av}(T) &=& \frac{4}{L} \sum_{j=1..L} (\sum_m |c_m^{\rm init}|^2  \langle \psi_m| \hat
(\hat n_j-1/2) |\psi_m\rangle)^2 \nonumber
\end{eqnarray}
We note that for the initial state $ | \langle \psi_{\rm init}| \hat
2(\hat n_j -1/2) |\psi_{\rm init}\rangle |^2 =0$ and $| \langle
\psi_{\rm init}| \hat 2(\hat n_j -1/2) |\psi_{\rm init}\rangle |^2
=1$  while for free fermions, the ground state with $\mathcal{J}_0
\gg V_0$, $\langle 2(\hat n_j -1/2) \rangle=0$. Thus $N_{\rm av}(T)$
provides a measure of degree of delocalization of the driven chain.
A similar reasoning shows that $N_{0j} \to 0$ for all sites in the
delocalized regime and $N_{0j} = 1[-1]$ for $j<[>]L/2$ in the
localized regime; in contrast, in the presence of a mobility edge,
$N_{0j}$ takes values between $0$ and $1$ at different sites.

\begin{figure}
\rotatebox{0}{\includegraphics*[width= 0.48 \linewidth]{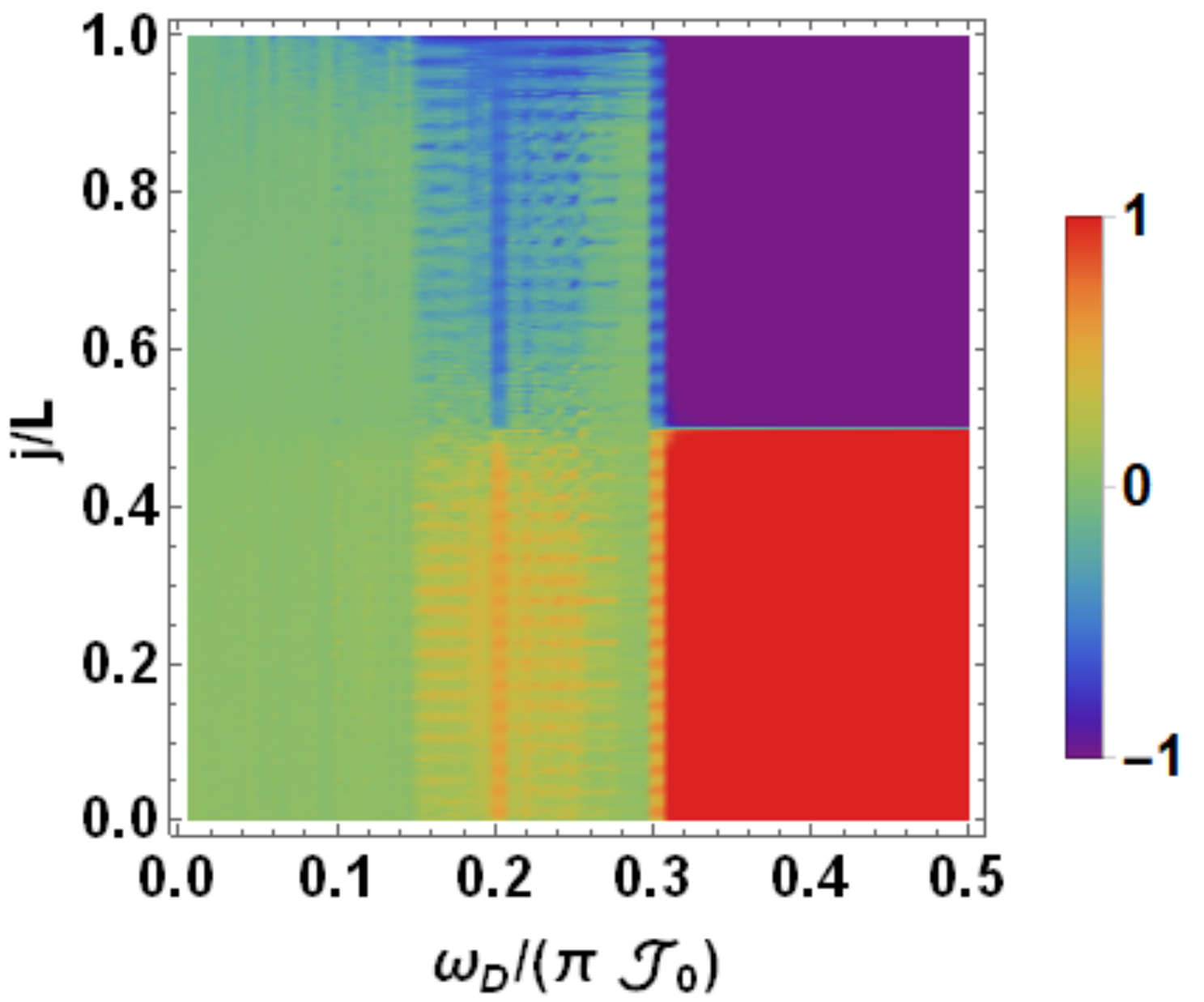}}
\rotatebox{0}{\includegraphics*[width= 0.48 \linewidth]{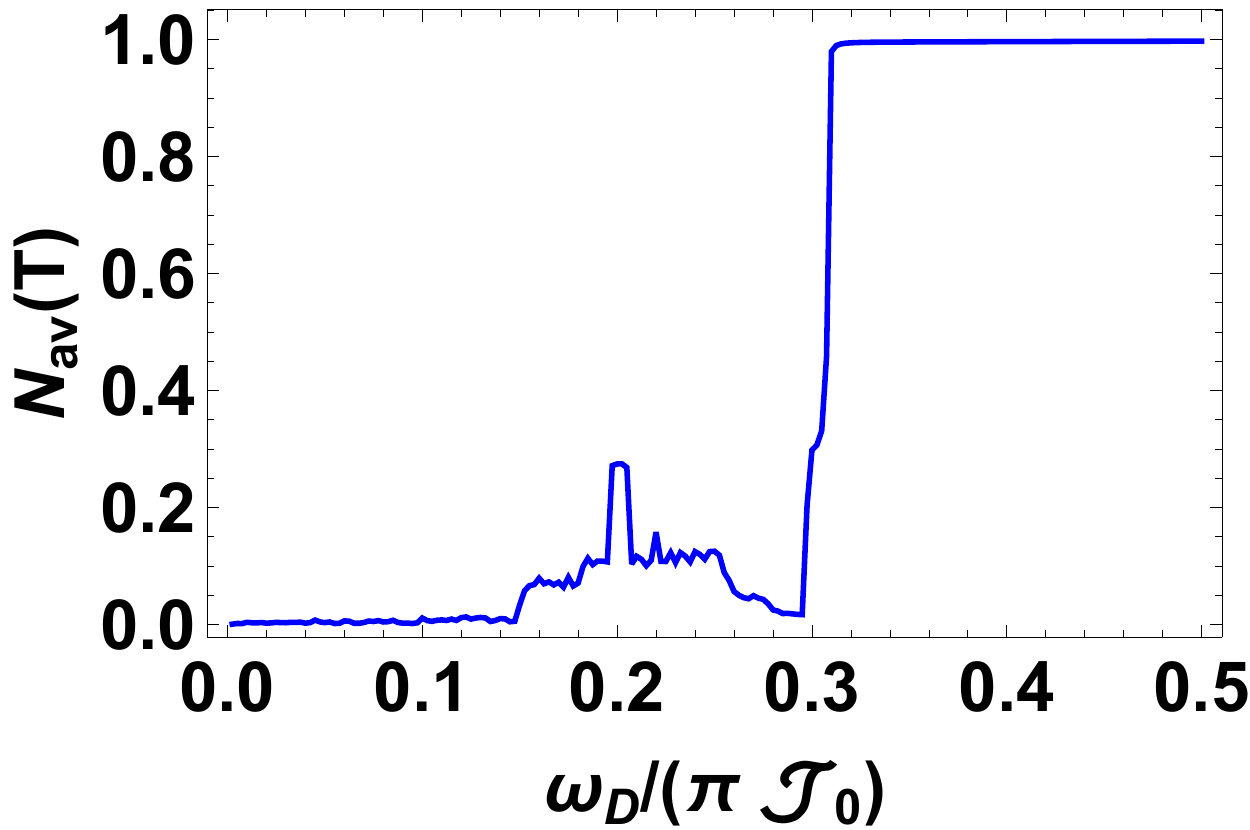}}
\rotatebox{0}{\includegraphics*[width= 0.48 \linewidth]{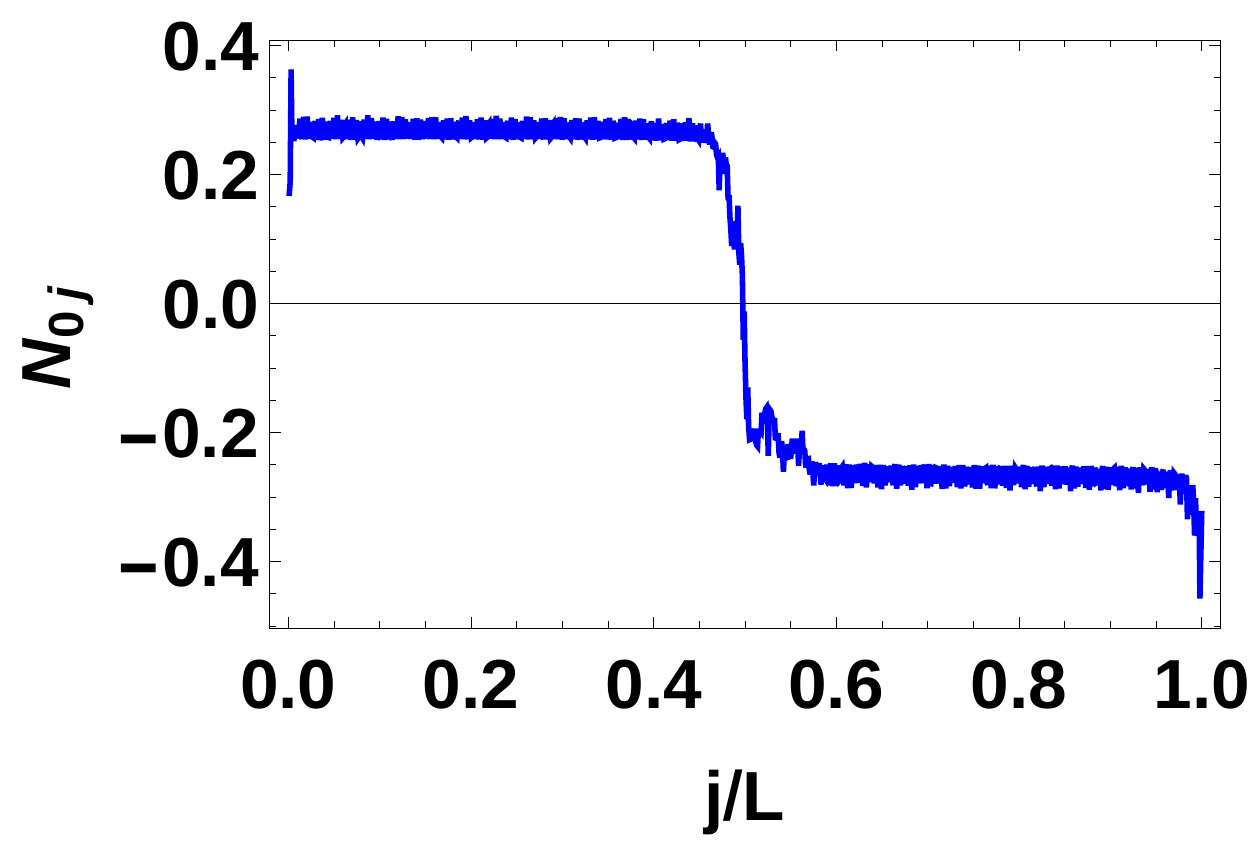}}
\rotatebox{0}{\includegraphics*[width= 0.48 \linewidth]{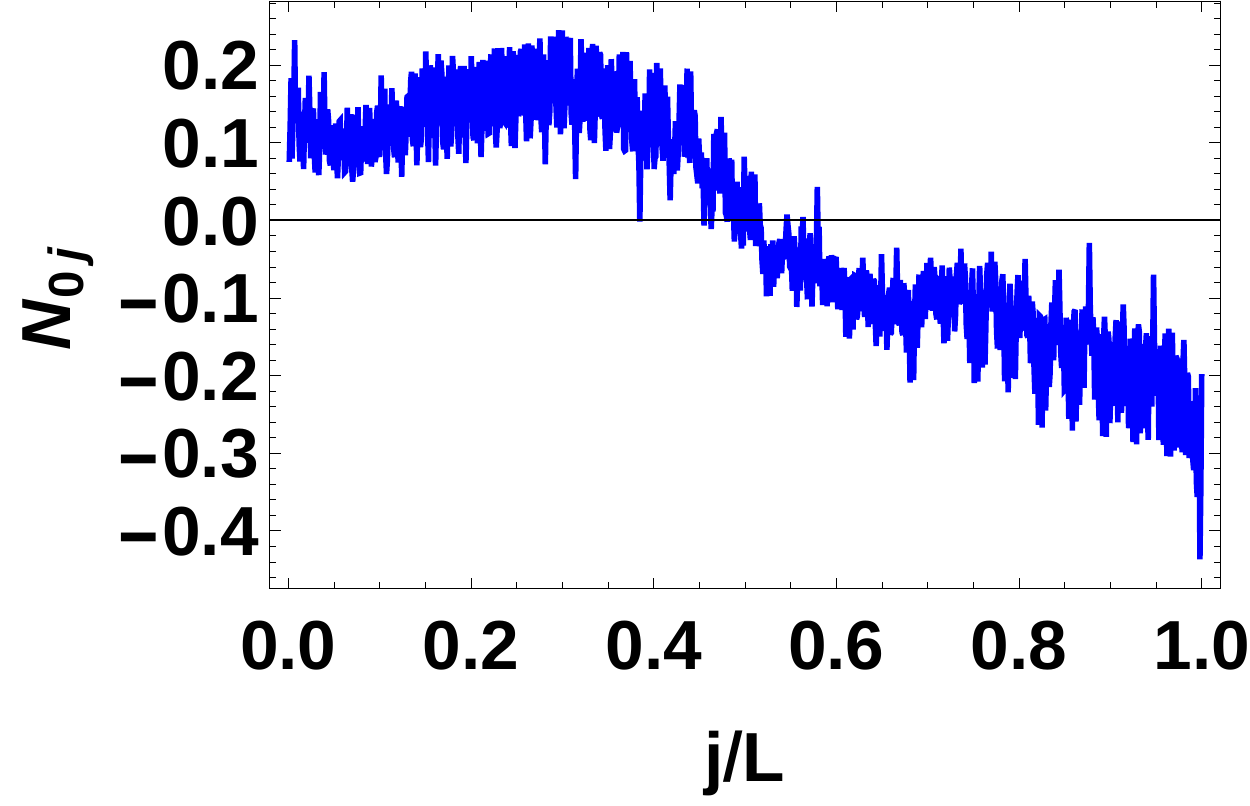}}
\caption{Top Left Panel: Plot of $N_0$ as a function of $j/L$ and
$\omega_D/(\pi\mathcal{J}_0)$ showing fermion density profile at all
sites of the chain in the steady state as a function of
$\omega_D/(\pi\mathcal{J}_0)$. Top Right Panel: Plot of $N_{\rm
av}(T)$ as a function of $\omega_D/(\pi\mathcal{J}_0)$ in the steady
state showing $0 \le N_{\rm av}(T) \le 1$ for $0.15 \le
\omega_D/(\pi \mathcal{J}_0) \le 0.25$. Bottom Panels: Plot of
$N_{0j}$ as a function of $j/L$ for $\omega_D/(\pi
\mathcal{J}_0)=0.2 $ (left) where the mobility edge separates the
delocalized and localized states and $\omega_D/(\pi
\mathcal{J}_0)=0.175$ (right) where it separates the delocalized and
multifractal states. All other parameters are same as in Fig.\
\ref{fig1}. See text for details.} \label{fig7}
\end{figure}

A plot of $N_{0j}$ as a function of site index $j/L$ and drive
frequency $\omega_D$ for the steady state is shown in the left panel
of Fig.\ \ref{fig7}. The density profile is seen to stay close to
that of the initial state confirming localization at high drive
frequency. In contrast, at low drive frequencies, it approaches zero
as expected for the delocalized regime with $\mathcal{J}_0 \gg V_0$.
In between, $N_{0j}$ indicates intermediate behavior showing
signature of partial transport such that $0<|N_{0j}|<1$.  The
distribution of $N_{0j}$ is much more spread out in the case where
the mobility edge separates delocalized and multifractal(as opposed
to localized) states as can be clearly seen from the bottom panels
of Fig.\ \ref{fig7}. Thus we find that fermion number distribution
in the steady state may provide a signature of presence of the
multifractal state in the driven AA model. A plot of $N_{\rm av}(T)$
as a function of $\omega_D$, shown in the top right panel of Fig.\
\ref{fig7}, also confirms this behavior. We note that an increased
value of $N_{\rm av}(T)$ (between $0$ for perfectly delocalized
states and $1$ perfectly localized states) for $0.15 \pi <\hbar
\omega_D/\mathcal{J}_0 < 0.25 \pi$ is a signature of presence of
both localized (or multifractal) and delocalized states in the
Floquet spectrum and hence provides an indication of the presence of
mobility edge in the spectrum. Moreover, the value of $N_{\rm
av}(T)$ seems to be larger in a narrow frequency range around $\hbar
\omega_D/\mathcal{J}_0=0.2 \pi$ where the mobility edge separates
delocalized and localized states. Thus our results show that the
local fermion density in the steady state starting from a
domain-wall initial condition in these chains may serve as a
detector of mobility edge in the Floquet spectrum.

\begin{figure}
\rotatebox{0}{\includegraphics*[width= 0.8 \linewidth]{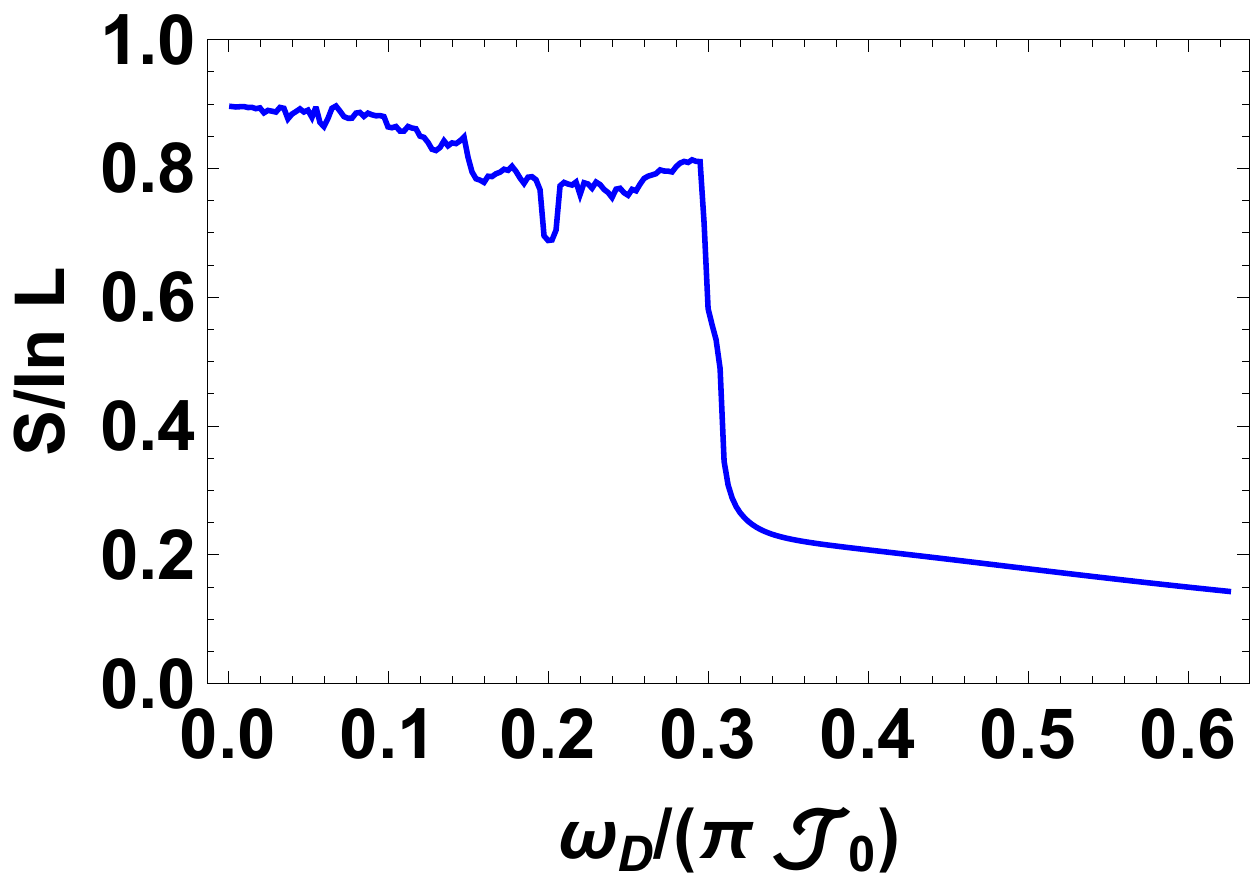}}
\caption{Plot of the mean Shannon entropy $S/\ln L$
as a function of $\omega_D/(\pi\mathcal{J}_0)$. All other parameters are same as in
Fig.\ \ref{fig1}. See text for details.} \label{fig8}
\end{figure}

Next, we compute the Shannon entropy of the driven chain. This is
defined in terms of the overlap coefficients obtained by computing
overlap of the single-particle Floquet eigenstates $|\psi_m\rangle$
with the eigenstates of $H_p= H_F(T=0)$ $|j\rangle$: $\psi_m(j) =
\langle j|\psi_m\rangle$. The Floquet eigenstates can be written as
$|\psi_m\rangle= \sum_j \psi_m(j) |j\rangle$. The Shannon entropy of
the $m^{\rm th}$ Floquet eigenstate is then given by
\cite{shannonref}
\begin{eqnarray}
S_m &=& -\sum_j |\psi_m(j)|^2 \ln |\psi_m(j)|^2, \quad S =
\frac{1}{L} \sum_m S_m  \nonumber\\ \label{shannon1}
\end{eqnarray}
where $S$ is the mean entropy. We note that for high frequency when
$H_F \simeq H_p$, $\psi_m(j) \simeq \delta_{mj}$ leading to $S_m \simeq
0$ for $\hbar \omega_D/\mathcal{J}_0 \gg 1$ by construction. Since
eigenfunctions of $H_p$ are localized this means that $S \to 0$ for
localized states. In contrast for $\hbar \omega_D/\mathcal{J}_0 \ll 1$ when all
Floquet eigenstates are delocalized, $\psi_m(j) \simeq 1/\sqrt{L}$ for
all $m$ leading to maximum entropy of $S \simeq \ln L$. A plot of
$S/\ln L$ as a function of the drive frequency, shown in Fig.\
\ref{fig8}, indicates this change. We find that the
localization-delocalization transition is marked by a sharp rise in
$S$ around $\omega_D=\omega_c=0.3 \pi \mathcal{J}_0/\hbar$. The appearance of
the mobility edge just below the transition leads to $S/\ln L \le
1$; this value would have been closer to unity if all the Floquet
eigenstates would be delocalized for $\omega_D \le \omega_c$. We
note that $S$ shows a narrow dip around $\hbar \omega_D/\mathcal{J}_0=0.2 \pi$.
This can be understood to be due to the fact that around this
frequency the mobility edge separates localized, rather than
multifractal, states from the delocalized ones; the presence of
these localized states in the spectrum leads to a lower value of
$S$.

Finally we compute the survival probability which is defined as the
probability of finding a fermion, initially localized at a given site,
within a neighborhood of length $R$ around that site after $n$ drive
cycles. This is given by
\begin{eqnarray}
F_n(R) &=& \sum_{j=j_0-R/2}^{j_0+R/2} |\psi_n(j)|^2 \nonumber\\
\psi_{n}(j) &=& U(nT,0) \psi_{\rm init}(j)
\end{eqnarray}
where $j$ denotes lattice sites, we shall consider the initial
wavefunction to be localized at the center of the chain ($j_0=L/2$)
for the rest of this section. The limiting values of $F_n(R)$ can be
easily deduced. For example, if the wavefunction remains localized
$F_n(R) \simeq 1$ for all $R$ and $n$; in contrast if the drive
leads to delocalization, $F_n(R)$ should linearly increase with $R$
for large $n$. In the presence of a mobility edge separating
delocalized and multifractal states, $F_n(R)$ should again increase
with $R$, but with a sublinear growth for large $n$ . Moreover, the
steady state value of $F_n(R)$ can be obtained in terms of Floquet
eigenfunctions as
\begin{eqnarray}
F_s (R) &=& \sum_{j=j_0-R/2}^{j_0+R/2} \sum_m |\psi_m(j)|^2
\end{eqnarray}
and is therefore controlled by the coefficients $\psi_m(j)$.

\begin{figure}
\rotatebox{0}{\includegraphics*[width= 0.48 \linewidth]{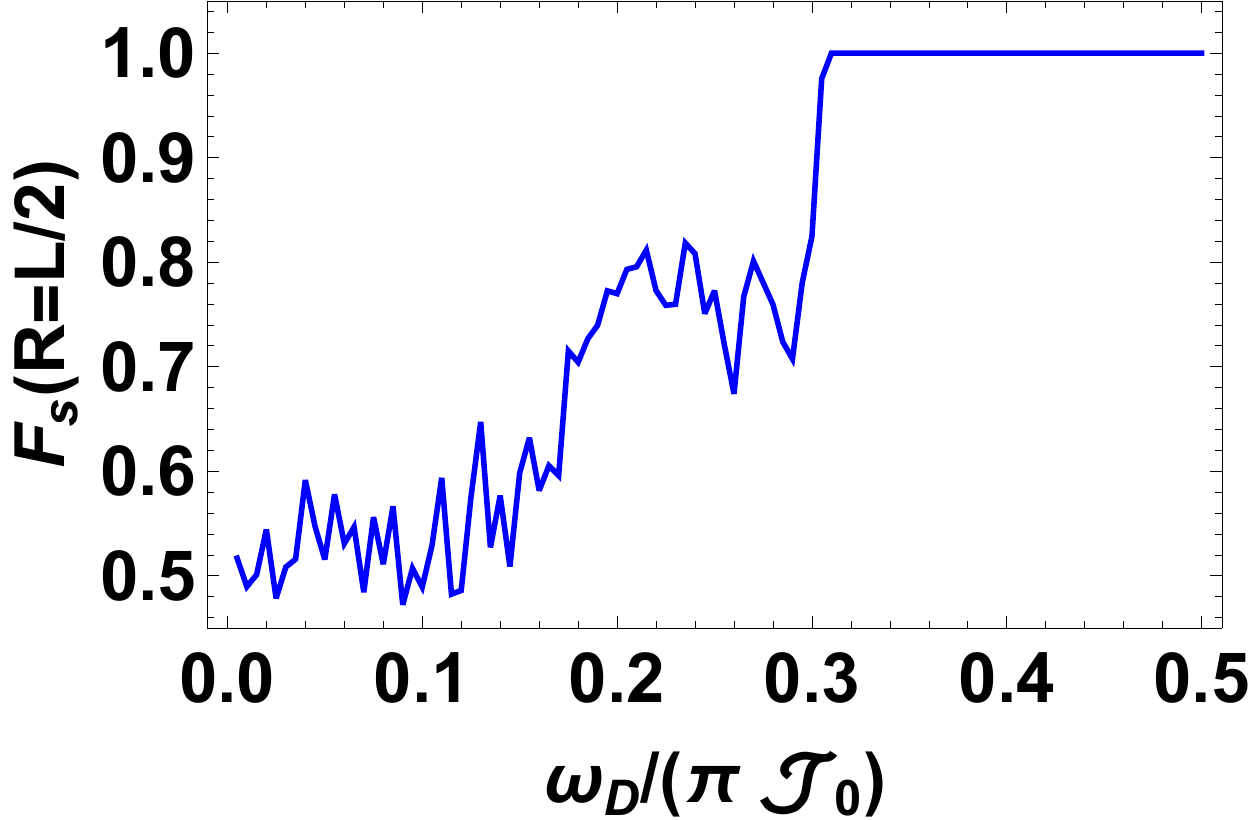}}
\rotatebox{0}{\includegraphics*[width= 0.48 \linewidth]{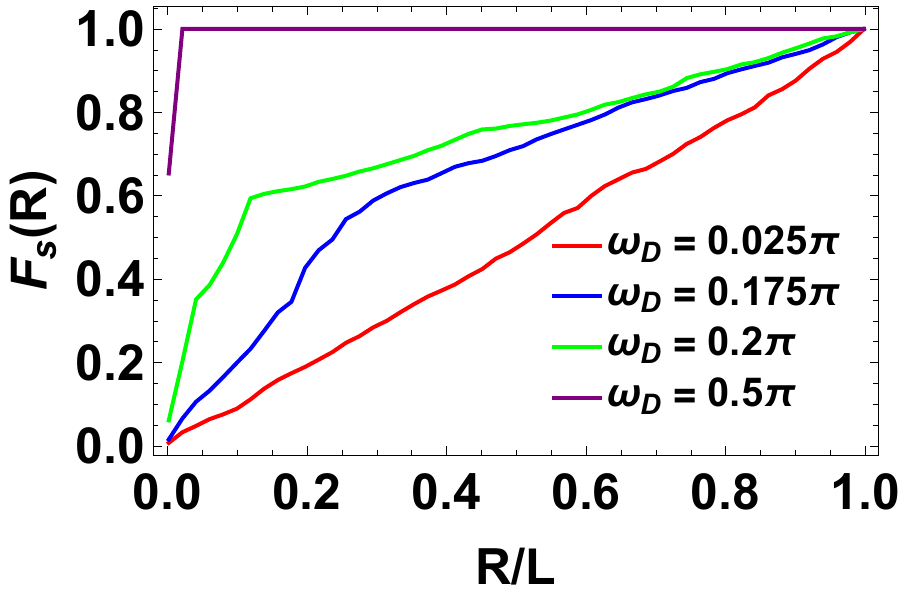}}
\caption{Left Panel: Plot of the survival probability $F_s(L/2)$ as
a function of $\omega_D/(\pi\mathcal{J}_0)$. Right panel: Plot of $F_s(R)$ as a
function of $R$ for several representative value of $\omega_D/(\pi\mathcal{J}_0)$.
All other parameters are same as in Fig.\ \ref{fig1}. See text for
details.} \label{fig9}
\end{figure}

A plot of $F_s(R=L/2)$ as a function of the drive frequency
$\omega_D$ is shown in the left panel of Fig.\ \ref{fig9}. We find
that $F_s(L/2)$ shows a sharp dip at the localization-delocalization
transition. Below the transition, the decay of $F_s(L/2)$ is gradual
and non-monotonic; this seems to be a direct consequence of the
presence of the mobility edge. The right panel of Fig.\ \ref{fig9}
shows the $R$ dependence of $F_s(R)$ for several representative
drive frequencies. We find that at high drive frequencies $\hbar
\omega_D/\mathcal{J}_0=0.5 \pi$, the system remains localized leading to $F_s
\simeq 1$ for almost all $R$; in contrast it linearly decreases to
zero as $R$ is decreased in the low frequency limit $\hbar
\omega_D/\mathcal{J}_0=0.025 \pi$. In between, in the regime where the mobility
edge separates delocalized states from multifractal or localized
states in the Floquet eigenspectrum, we find sublinear decay of
$F_s(R)$ as a function of $R$; this decay is faster if states with
quasienergies above the mobility edge are localized ($\hbar
\omega_D/\mathcal{J}_0=0.2 \pi$). Thus $F_s(R)$ distinguishes between mobility
edge separating delocalized states with multifractal or localized
states.

\begin{figure}
\rotatebox{0}
{\includegraphics*[width=0.49\linewidth]{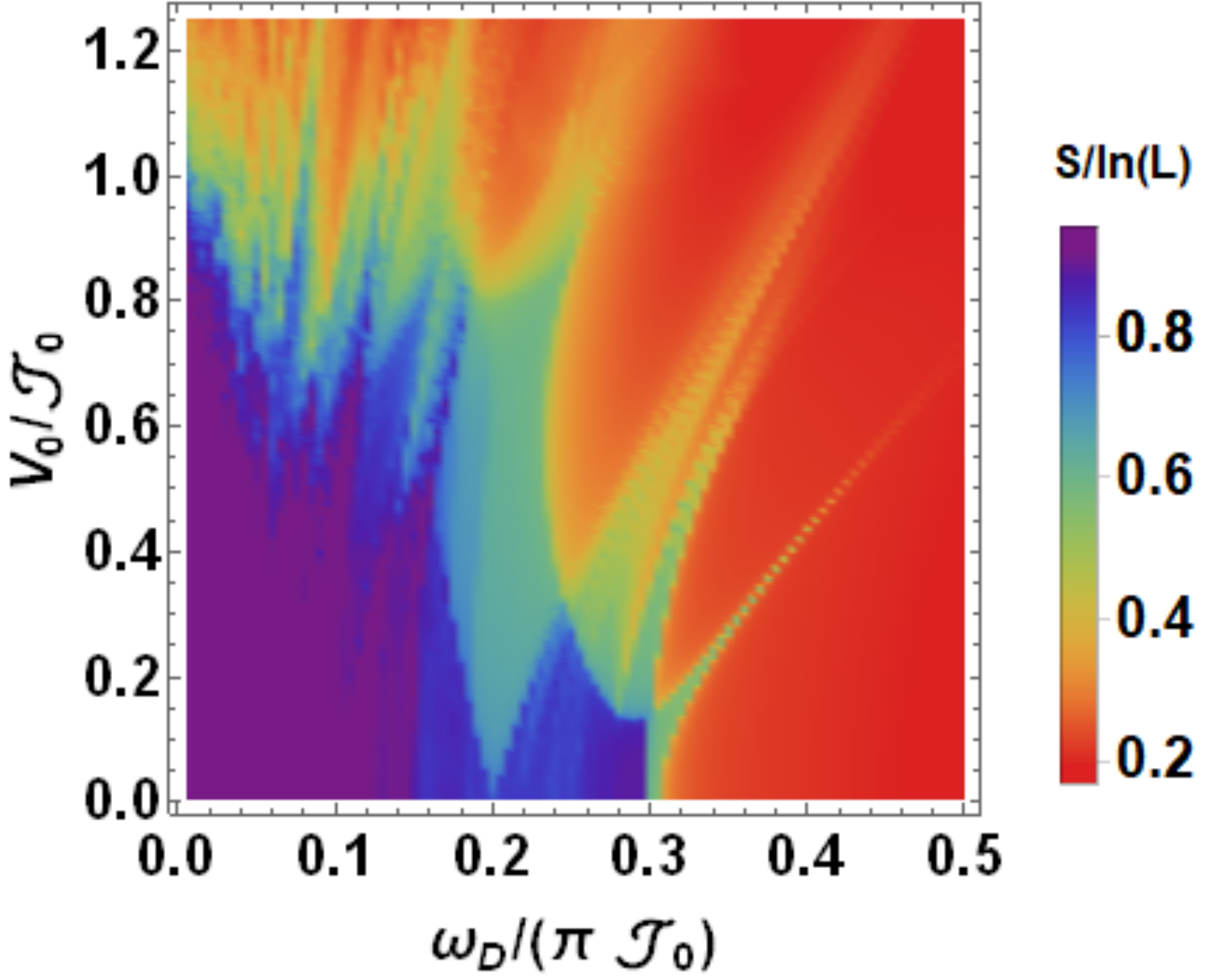}}
{\includegraphics*[width= 0.49 \linewidth]{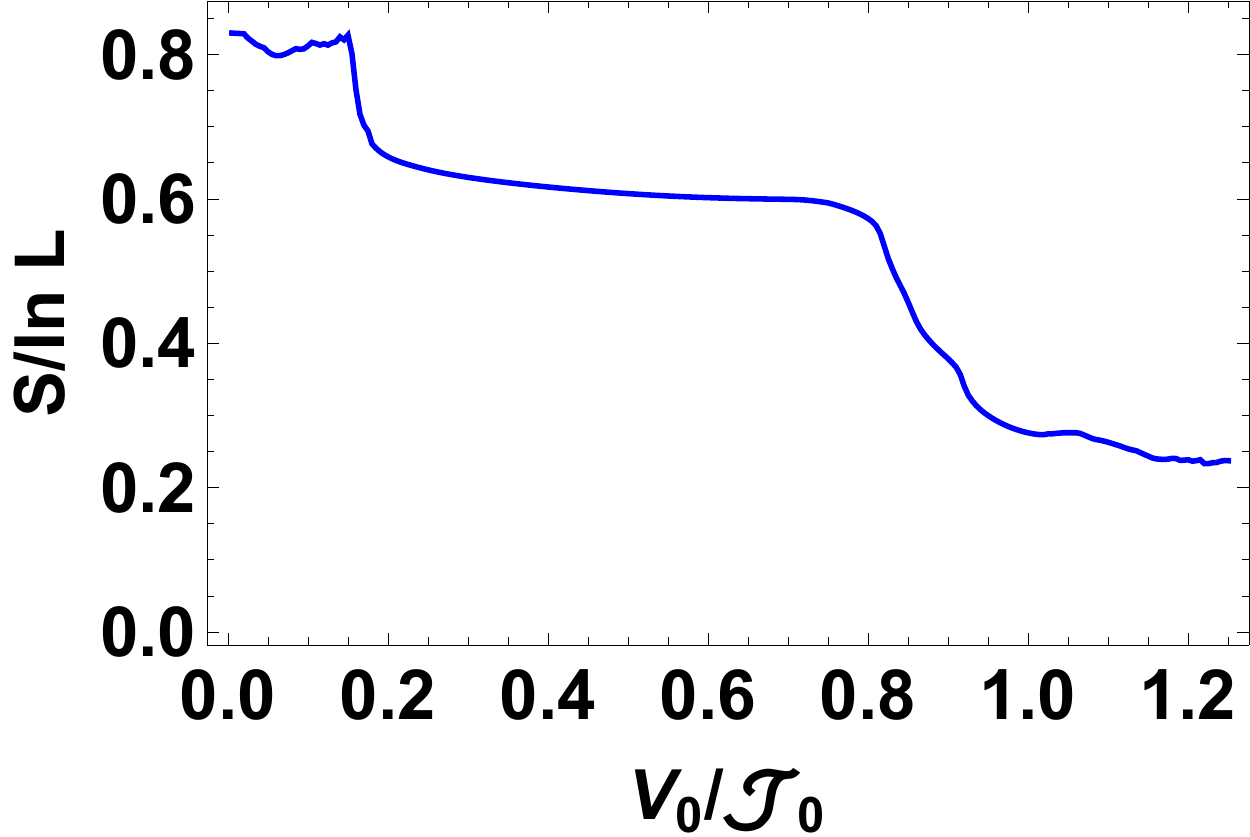}}
\caption{Left panel: Distribution of $S/\ln L$ as a function of the
Aubrey Andre Strength $V_0/\mathcal{J}_0$ and the drive frequency
$\omega_D/(\pi \mathcal{J}_0)$. The regions with all localized
states are indicated by red, delocalized states by violet. The blue
regions denote mixture of delocalized and multifractal states and
green mixture of delocalized and localized states. Right Panel: Plot
of $S/\ln L$ as a function of $V_0/{\mathcal J}_0$ for $\hbar
\omega_D/(\pi \mathcal{J}_0)=0.22$. . All other parameters are same
as in Fig.\ \ref{fig1}.} \label{fig10a}
\end{figure}

Finally, we study the $V_0$ dependence of our results. In particular
we concentrate on obtaining an estimate of the range of
$V_0/{\mathcal J_0}$ over which the multifractal states exist. To
this end, we plot the mean Shannon entropy of the Floquet
eigenstates as a function of $V_0$ and $\omega_D$. This plot, shown
in the left panel of Fig.\ \ref{fig10a}, indicates that a mobility
edge separating delocalized and multifractal states (indicated by
blue in the plot) are present of over a range of frequency whose
width tend to be maximal around $V_0 \ll {\mathcal J_0}$. For $V_0
\ge {\mathcal J}_0$, the Floquet states are either all localized
(red region) or display a mobility edge separating delocalized and
localized states (green regions). For $V_0, \hbar \omega_D \ll
{\mathcal J}_0$, the Floquet states are all delocalized (violet
regions). The right panel shows a plot of $S/\ln L$ as a function of
$V_0$ for $\hbar \omega_D/{\mathcal J}_0 = 0.22 \pi$. The plot shows
indication of a mobility edge separating delocalized and localized
states for $0.8 \le V_0/{\mathcal J}_0 \le 0.2$; in contrast, the
mobility edge separates delocalized and multifractal states for
$V_0/{\mathcal J}_0 \le 0.2$. Thus our results show that the
multifractal states are indeed present in the Floquet spectrum for a
wide region in the $(V_0/{\mathcal J}_0, \hbar \omega_D/{\mathcal
J}_0)$ plane.

\subsection{Sinusoidal Protocol}
\label{sinpr}

In this section, we study the properties of AA model in the presence of
a continuous drive. Such a drive is implemented by choosing
\begin{eqnarray}
\mathcal{J}(t) = \mathcal{J}_0 \cos \omega_D t
\end{eqnarray}
A numerical study of the AA model in the presence of such a
continuous drive involves decomposition of the evolution operator
into $N$ Trotter steps such $H(t)$ does not change significantly in
the interval $t_j$ and $t_j+ T/N$ for any time instant $t_j$. One
can define the eigenvalues and eigenfunctions of the instantaneous
Hamiltonian $H_j= H[t_j +T/(2N)]$ as $\epsilon_n^j$ and
$|\psi_n^j\rangle$; these are obtained numerically by exact
diagonalization of $H_j$on a lattice of size $L$. One can then
construct the evolution operator as
\begin{eqnarray}
U(T,0) &=& \prod_{j=1}^N  \sum_n e^{- i \epsilon_n^j T/N}
|\psi_n^j\rangle\langle \psi_n^j | \label{evolcont}
\end{eqnarray}
We note this procedure requires numerical diagonalization of $N$
instantaneous Hamiltonians; this make numerical study of continuous
protocols significantly more costly compared to their discrete
counterparts. Having constructed $U(T,0)$, we diagnosable it
numerically to obtain the eigenvalues and eigenfunctions of the
Floquet Hamiltonian as outlined in Sec.\ \ref{sqppr}.

\begin{figure}
\rotatebox{0}{\includegraphics*[width= 0.48 \linewidth]{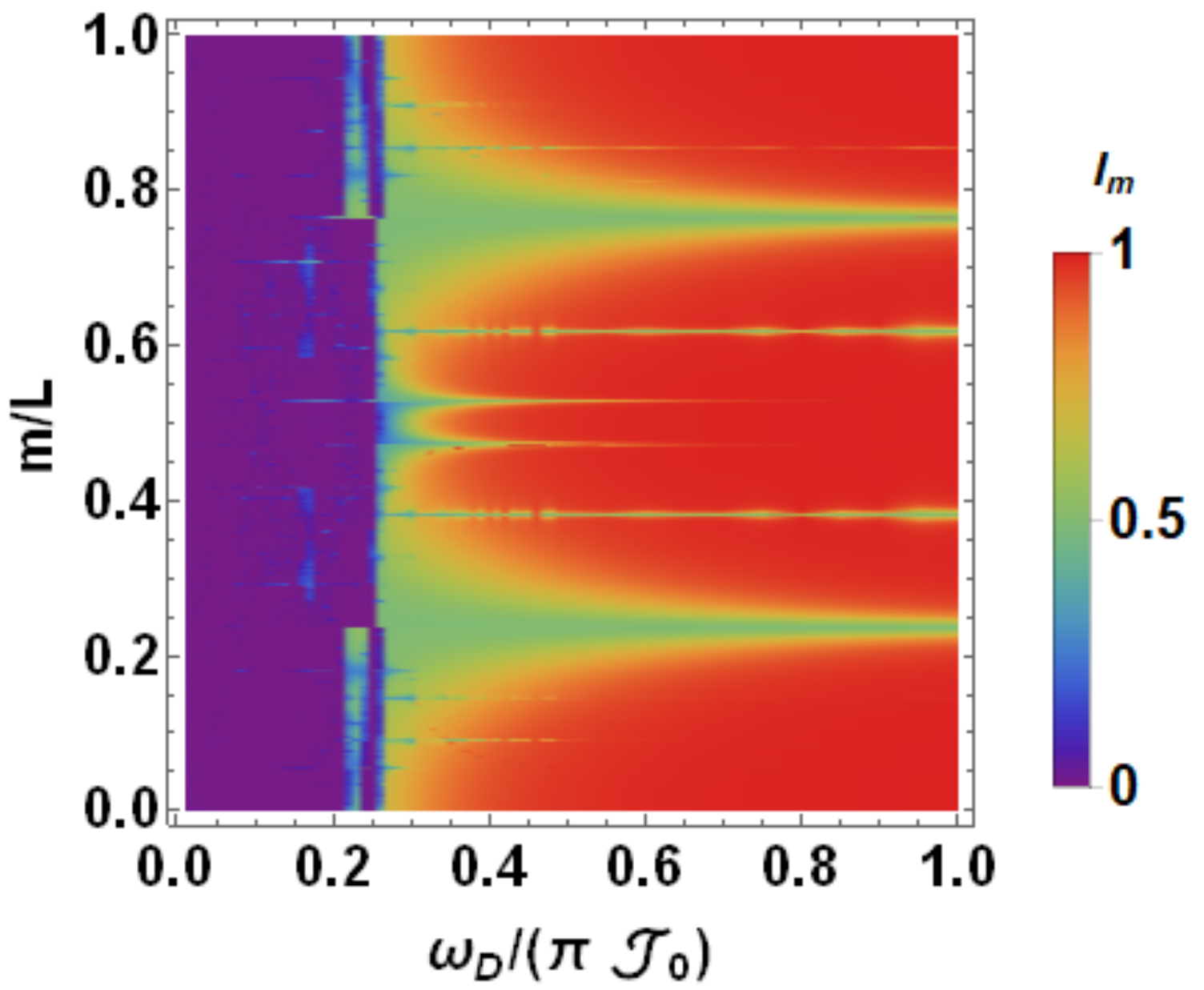}}
\rotatebox{0}{\includegraphics*[width= 0.48 \linewidth]{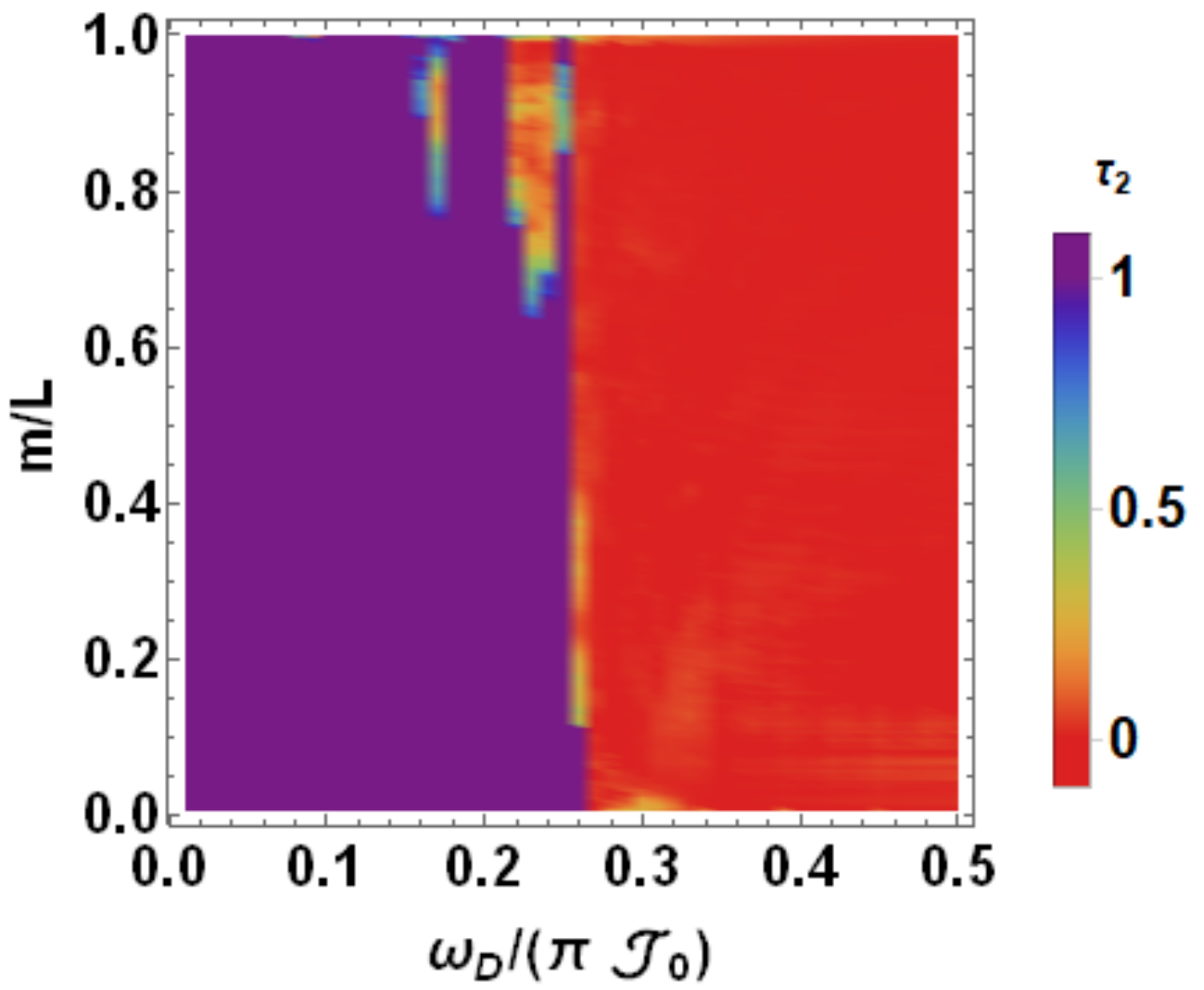}}
\rotatebox{0}{\includegraphics*[width= 0.48 \linewidth]{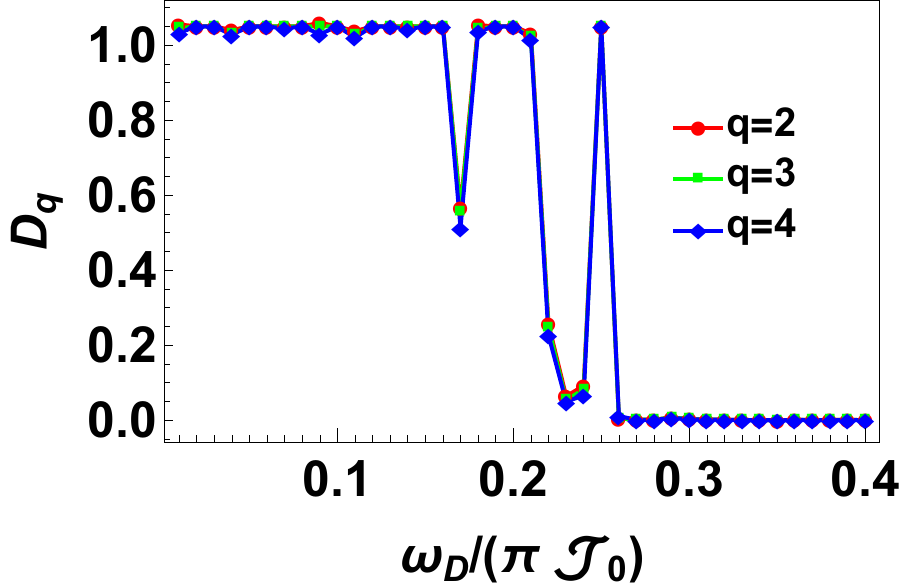}}
\rotatebox{0}{\includegraphics*[width= 0.48 \linewidth]{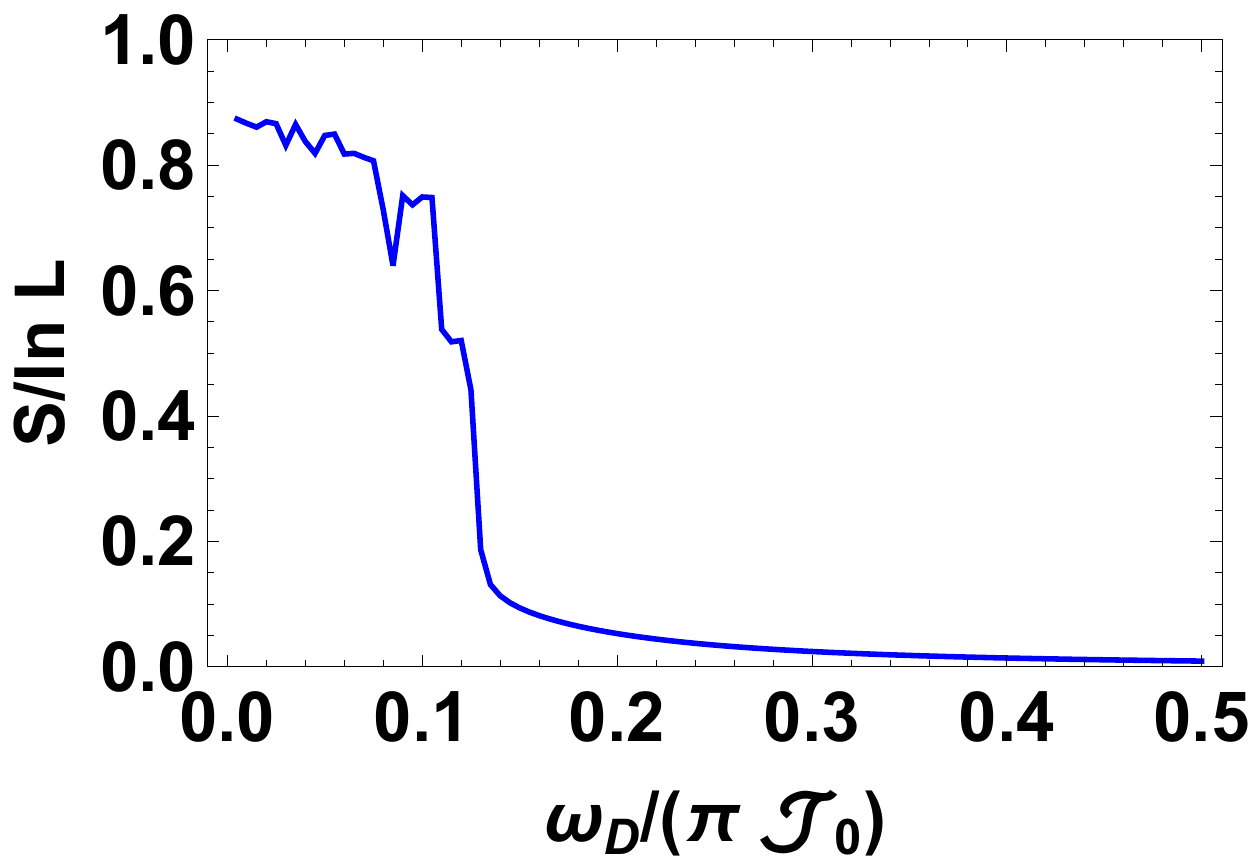}}
\rotatebox{0}{\includegraphics*[width= 0.48 \linewidth]{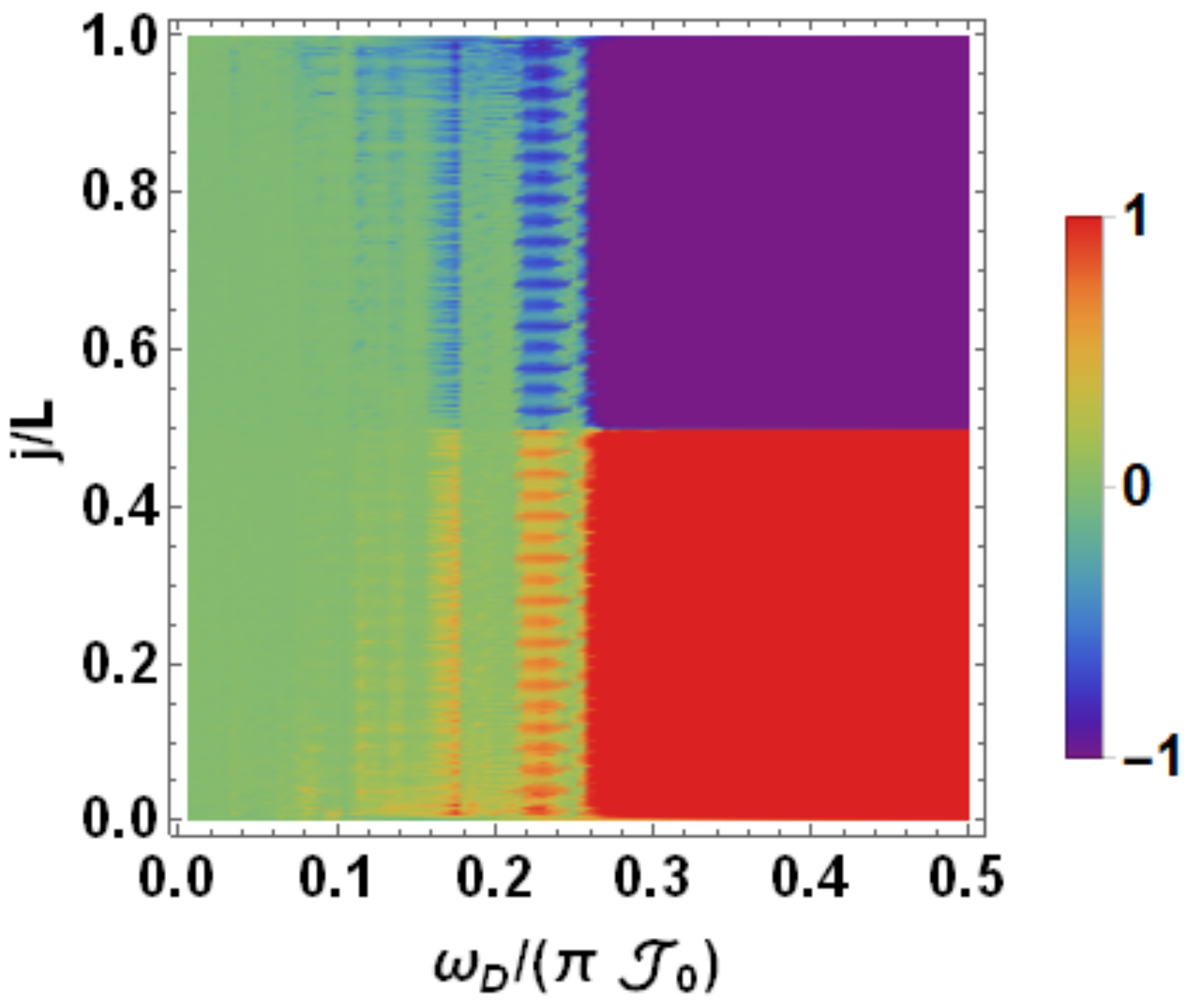}}
\rotatebox{0}{\includegraphics*[width= 0.48 \linewidth]{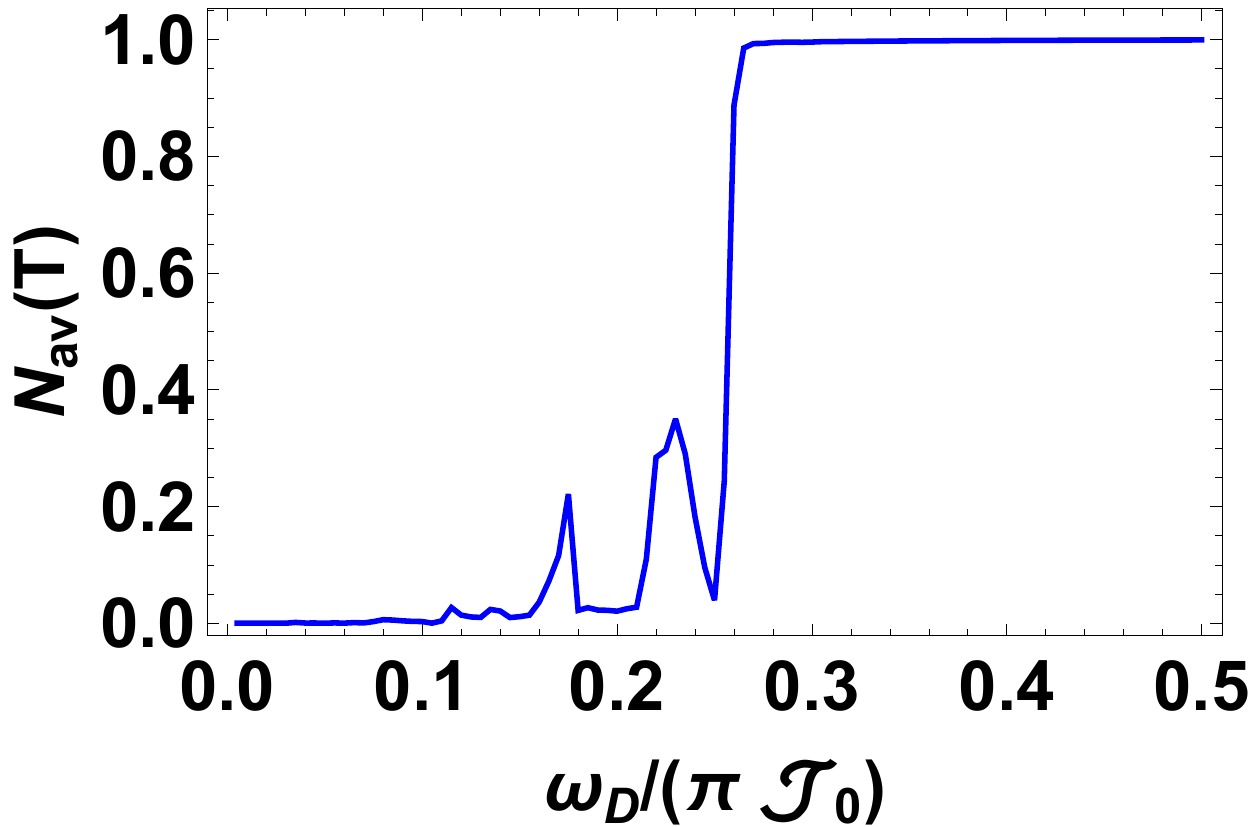}}
\caption{Top Left Panel: Plot of $I_m$ as a function of $m/L$ and
$\omega_D/(\pi \mathcal{J}_0)$. Top Right Panel: Plot of $\tau_2$ as
a function of $m/L$ and $\omega_D/(\pi \mathcal{J}_0)$. Middle Left
Panel: Plot of $D_q$ as a function of $\omega_D/(\pi \mathcal{J}_0)$
for $m/L=0.75$. Middle Right Panel: A plot of $S$ as a function of
$\omega_D/(\pi \mathcal{J}_0)$ showing the signature of
localization-delocalization transition at $\hbar \omega_D/(\pi
\mathcal{J}_0) \simeq 0.25$. Bottom left Panel: Plot of $N_{0j}$ as
a function of $j/L$ and $\omega_D/(\pi \mathcal{J}_0)$ showing
fermion density profile at all sites of the chain in the steady
state as a function of $\omega_D/(\pi \mathcal{J}_0)$. Bottom right
Panel:Plot of $N_{\rm av}$ as a function of $\omega_D/(\pi
\mathcal{J}_0)$. The system sizes used to calculate $\tau_2$ and
$D_q$ is $L=200\sim1600$ in steps of $200$. The other plots are
presented at $L=1024$.Here $V_0/\mathcal{J}_0=0.025$ , all other
parameters are same as in Fig.\ \ref{fig1}. See text for details.}
\label{fig10}
\end{figure}

The results obtained from this procedure is shown in Fig\
\ref{fig10}. We find that the all properties of the driven systems,
such as the presence of a mobility edge in the Floquet spectrum, the
presence of multifractal Floquet eigenstates, their signature in
transport, and the presence of the CAT states remain qualitatively
similar; however, the position of the localization-delocalization
transition shows a significant change. From the plot of $I_m$ as a
function of $\omega_D$ and $m$ in the top left panel of Fig.\
\ref{fig10}, we find that the transition shifts to $\hbar
\omega_D/\mathcal{J}_0 \simeq 0.25 \pi$; the mobility
edge exists over narrower regions (one near $0.16 \pi \le \hbar
\omega_D/\mathcal{J}_0 \le 0.18 \pi$ and another near $0.2 \pi \le
\hbar \omega_D/\mathcal{J}_0 \le 0.25 \pi$) as can be seen from the
plot of $\tau_2$ as a function of $m$ and $\omega_D$ in the top
right panel. A plot of $D_q$ shown in the middle left panel confirms
the presence of multifractal states in these regions. In the middle
right panel, we show the plot of the mean Shannon entropy $S$ as a
function of $\omega_D$. We find that $S$ also bears the signature of
the localization-delocalization transition. The bottom panels show
steady state distribution of particles in this system starting from
the domain wall state. The bottom left panel shows the distribution
of $N_{0j}$ over lattice sites (scaled by system size($L$)) as a
function of the drive frequency. The plot demonstrates the
non-monotonic behavior of $N_{0j}$ as a function of $\omega_D$ just
below the transition. Finally the bottom right panel shows a plot of
$N_{\rm av}$ as a function of $\omega_D$ in the steady state; we
find that it displays signature of the localization-delocalization
transition around $\hbar \omega_D/\mathcal{J}_0=0.25 \pi$ and also
shows peaks at intermediate frequency where the mobility edge
appears in the spectrum. The height of these peaks are less than
unity; this is a consequence of the fact that the entire spectrum is
not localized at these frequencies.

\begin{figure}
\rotatebox{0}
{\includegraphics*[width=0.48\linewidth]{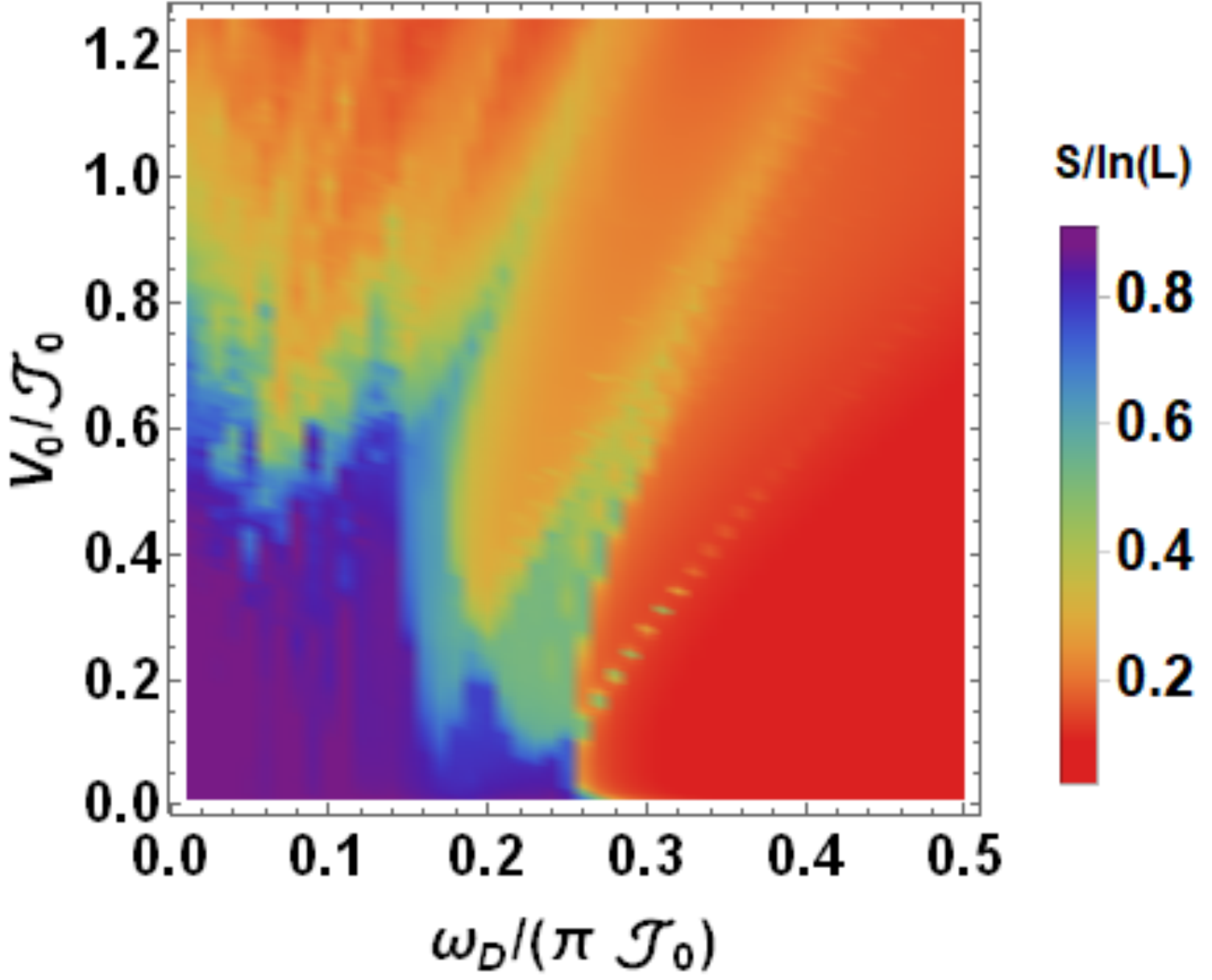}}
\rotatebox{0}{\includegraphics*[width=0.48\linewidth]{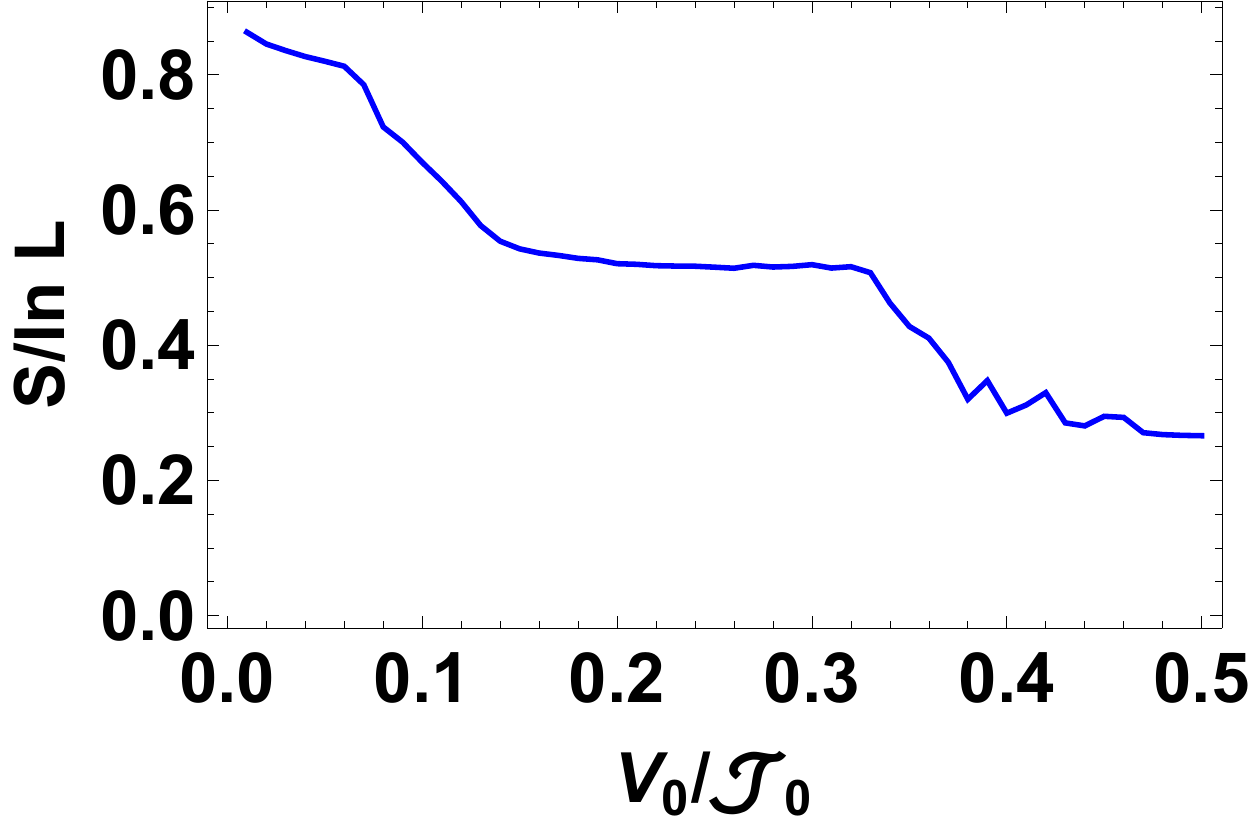}}
\caption{Left: Plot of $S/\ln L$ as a function of $V_0/\mathcal{J}_0$ and
$\omega_D/(\pi \mathcal{J}_0)$ for the sinusoidal protocol with $L=610$. Right: Plot of
$S/\ln L$ as a function of $V_0/\mathcal{J}_0$ for a cut taken at
$\hbar \omega_D/(\pi \mathcal{J}_0) = 0.22$. All other parameters
are same as in Fig.\ \ref{fig1}.} \label{fig10b}
\end{figure}

Finally, we plot the mean Shannon entropy $S/\ln
L$ as a function of $V_0$ and $\omega_D$. We find that for the
sinusoidal protocol the presence of multifractal states occurs in
a reduced area of parameter space compared to that for square pulse
protocol studied earlier. Moreover, we get localized states at a
lower frequency as compared to square pulse protocol for the same
value of $V_0$. From both the graphs we observe that as we increase
$V_0$ the frequency at which all states becomes localized decreases.
This expected since the off-diagonal hopping terms becomes small
compared to the diagonal AA potential term leading to dynamical
localization.

Our results therefore indicate that the localization-delocalization
transition in these systems along with the presence of the CAT and
multifractal state exists for both discreet and continuous
protocols. However, the range of $\omega_D$ for which the
multifractal states exists is significantly reduced in the latter
case.

\section{Floquet Perturbation Theory}
\label{secfpt}

In this section, we aim to obtain an analytic, albeit perturbative,
understanding of several features of the driven AA model found via
exact numerics using FPT which is known to provide accurate results
in the large drive amplitude limit \cite{fl1,fl2,rydref1}. The
square pulse protocol will be treated in Sec.\ \ref{secfptsq} while
the continuous drive protocol will be addressed in Sec.\
\ref{secfptsin}.

\subsection{FPT for square pulse protocol}
\label{secfptsq}

In this section, we shall focus on the square pulse protocol given
by Eq.\ \ref{spulse1} in the large drive amplitude limit $\mathcal{J}_0 \gg
V_0$. In this limit, we consider the contribution from $H_p$ to the
evolution operator $U(T,0)$ as perturbation and develop a systematic
expansion for $U$ following Refs.\ \onlinecite{fl1,fl2,rydref1}. To
this end, we first note that the first term in such an expansion is
given by $U_0$ which can be written as $U_0= \prod_k U_{0 k}$ where
\begin{eqnarray}
U_{0k}(t,0) &=&  e^{i t \mathcal{J}_0 \cos k c_k^{\dagger} c_k}, \quad t
\le T/2 \label{u0eq1} \\
&=& e^{i (T-t) \mathcal{J}_0 \cos k c_k^{\dagger} c_k}, \quad T/2\le t \le T
\nonumber
\end{eqnarray}
Here and in the rest of this section, we have set $\hbar$ to unity.
This leads to $U_0(T,0)= I$ (where $I$ denotes the identity matrix)
and $H_{F0}=0$. The vanishing $H_{F0}$ can be seen to be the
consequence of the symmetric nature of the drive protocol.

The first order perturbative correction of $U_0$, within FPT, is
given by
\begin{eqnarray}
U_1 &=& -i \int_0^T dt U_0^{\dagger}(t,0) H_p U_0(t,0) \label{forfl}
\end{eqnarray}
To evaluate this, we use the number basis in momentum space,
$|k\rangle \equiv |n_k\rangle$, since $U_0$ is diagonal in this
basis. The matrix element of $U_1$ in this basis is then given by
\begin{eqnarray}
\langle k_1|U_1|k_2\rangle &=& \frac{4 V(k_1-k_2)}{\mathcal{J}_0 f(k_1,k_2)}
\sin[\mathcal{J}_0 f(k_1,k_2) T/4] \nonumber\\
&& \times e^{i T \mathcal{J}_0 f(k_1,k_2)/4} \label{u1mat} \\
V(k) &=&  V_0 \sum_j \exp[i k j] \cos(2\pi \eta j) =V(-k), \nonumber\\
f(k_1,k_2) &=&  \cos k_2-\cos k_1 =-f(k_2,k_1)  \nonumber
\end{eqnarray}
where we have set $\phi=0$ without loss of generality. This
indicates that the Floquet Hamiltonian to first order in
perturbation theory is given by \cite{fl1,fl2}
\begin{eqnarray}
H_{F1} &=& i \sum_{ k_1,k_2} \frac{4 V(k_1-k_2)}{\mathcal{J}_0 T f(k_1,k_2)}
\sin[\mathcal{J}_0 f(k_1,k_2) T/4] \nonumber\\
&& \times  e^{i T \mathcal{J}_0 f(k_1,k_2)/4} c_{k_1}^{\dagger} c_{k_2}
\label{f2ham1}
\end{eqnarray}
A similar procedure for $U_2(T,0)$ yields the relation $U_2(T,0)=
U_1(T,0)^2/2$ and thus yield $H_{F2}=0$. The details of this
calculation is similar to that carried out in Ref.\
\onlinecite{rydref1} and is not presented here. In what follows, we
shall analyze $H_{F1}$ (Eq.\ \ref{f2ham1}) with the aim of obtaining
qualitative understanding of the presence of multifractal states in
the Floquet spectrum.

\begin{figure}
\rotatebox{0}{\includegraphics*[width= 0.48 \linewidth]{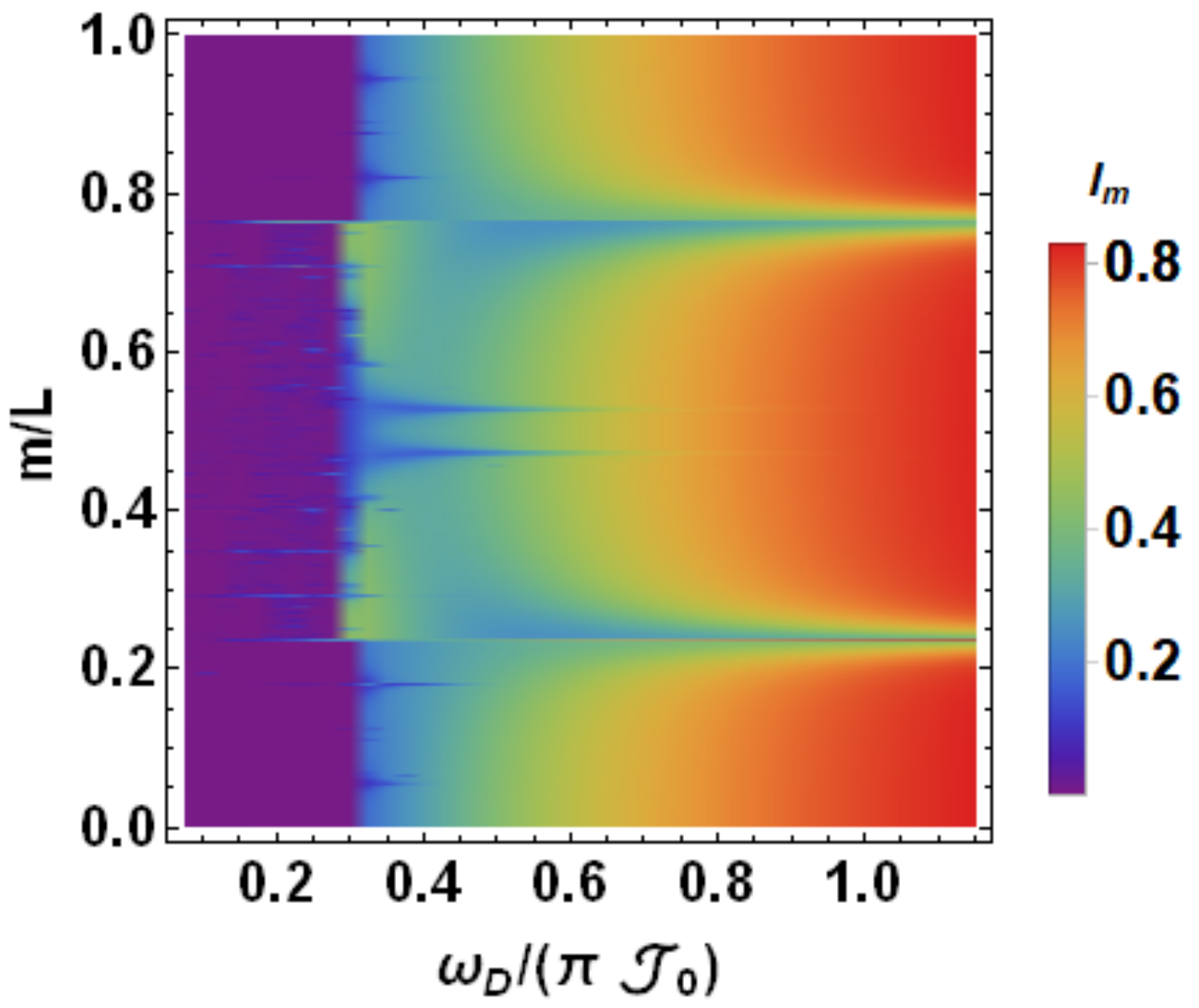}}
\rotatebox{0}{\includegraphics*[width= 0.48 \linewidth]{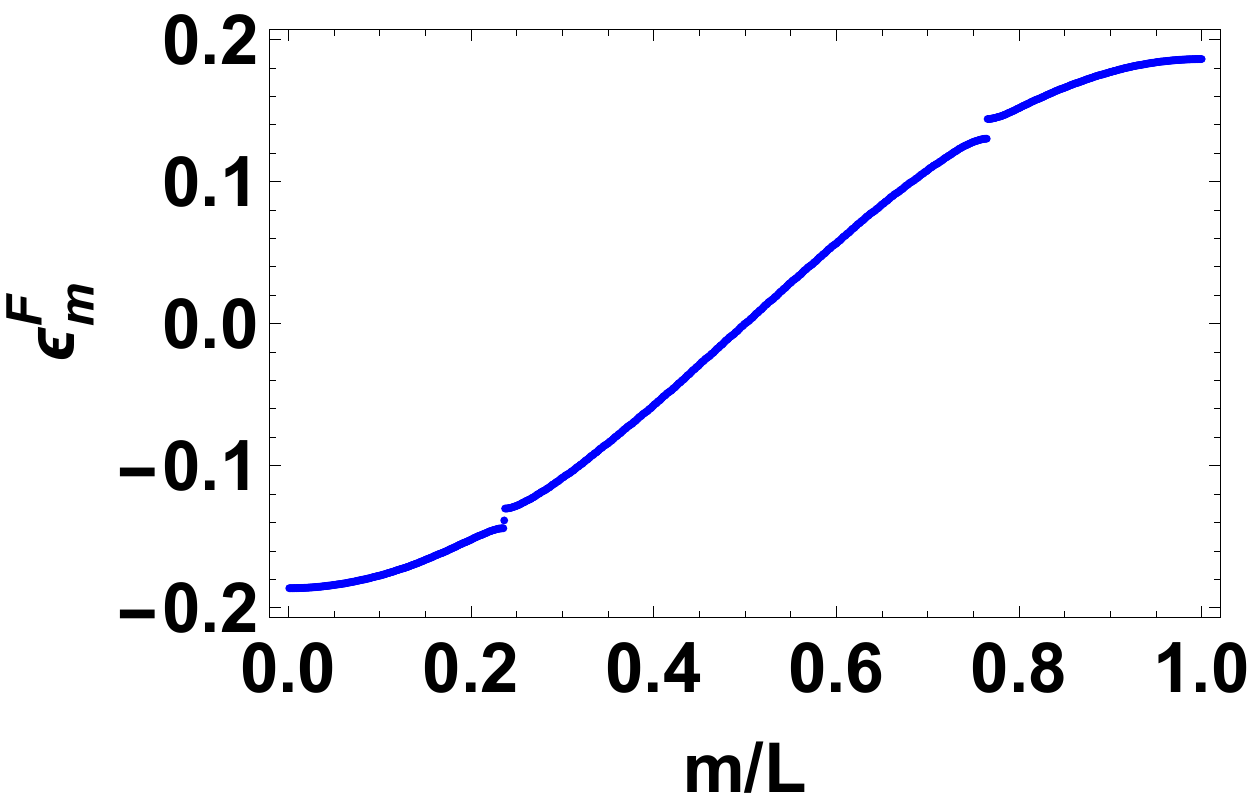}}
\rotatebox{0}{\includegraphics*[width= 0.48 \linewidth]{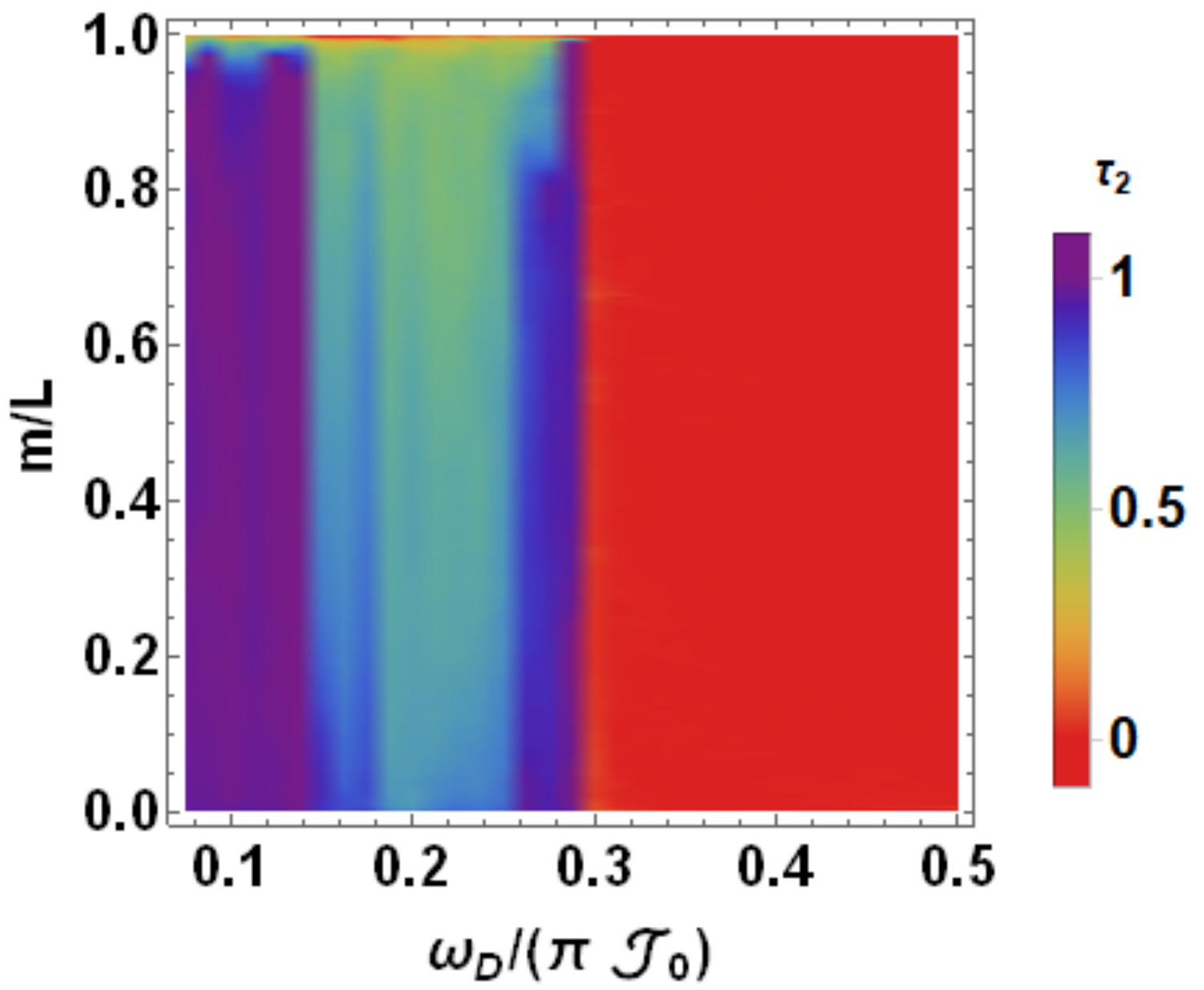}}
\rotatebox{0}{\includegraphics*[width= 0.48 \linewidth]{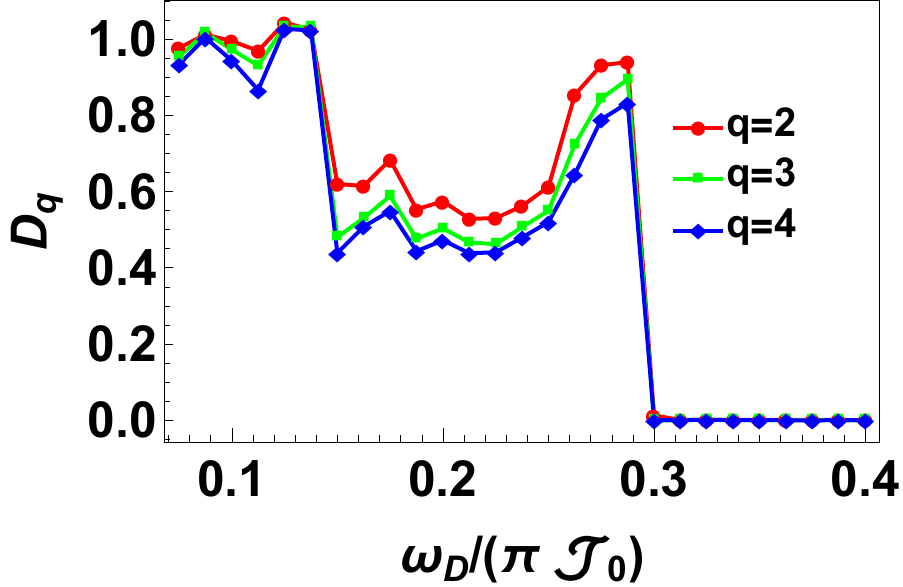}}
\caption{Top Left Panel: Plot of $I_m$ as a function of $m/L$ and
$\omega_D/(\pi\mathcal{J}_0)$. Top Right Panel: Plot of eigenstates $\epsilon_m^F$
obtained by diagonalizing $H_{F1}$ as a function of $m/L$ for $\hbar
\omega_D/(\pi \mathcal{J}_0) = 1$. Bottom Left Panel: Plot of $\tau_2$ as a function
of $m/L$ and $\omega_D/(\pi \mathcal{J}_0)$. Bottom right Panel: Plot of $D_q$ as a
function of $\omega_D/(\pi\mathcal{J}_0)$ for $m/L=0.75$. All other parameters are same
as in Fig.\ \ref{fig1} and Fig.\ \ref{fig5}. See text for details.} \label{fig11}
\end{figure}

A straightforward numerical diagonalization of $H_{F1}$ yields the
Floquet eigenstates and eigenvalues. To study the nature of these
Floquet eigenstates as a function of drive frequency, we plot the
IPR $I_m$ and $\tau_2$ corresponding to these states in Fig.\
\ref{fig11}. The top left panel of this plot shows the plot of $I_m$
as a function of $m/L$ and $\omega_D$.  $I_m$ obtained from the
eigenstates of $H_{F1}$ retains all qualitative feature of the
Floquet eigenstates obtained from exact numerics. In particular, the
plot shows a localization-delocalization transition around $\hbar
\omega_D/\mathcal{J}_0 \simeq 0.3 \pi$ which is close to the exact
value $\hbar \omega_c/\mathcal{J}_0 = 0.3 \pi$. Moreover, the
spectrum indicates the presence of the CAT states in the spectrum;
we have checked that the origin of these state can be tracked back
to the flat regions in the Floquet spectrum as can be seen from the
top right panel of Fig.\ \ref{fig11}. The bottom left panel shows a
plot of $\tau_2$ as a function of $m/L$ and $\omega_D$. We find that
just below the transition, we find a wide range of frequency $0.15
\pi \le \hbar \omega_D/\mathcal{J}_0 \le 0.3 \pi$ where we find
states with $0 < \tau_2 < 1$ signifying the possibility of
existence of multifractal states. However, we note the mobility edge
is now restricted to a very narrow region $0.26 \pi \le \hbar
\omega_D/\mathcal{J}_0 \le 0.3 \pi$; for $0.15 \pi \le \hbar
\omega_D/\mathcal{J}_0 \le 0.25\pi$ we do not find delocalized
states in the spectrum which is contrast with that obtained in exact
numerics. The presence of multifractal states in the spectrum of
$H_{F1}$ is further confirmed by plotting $D_q$ for $q=2,3,4$ as a
function of $\omega_D$ for $m/L=0.75$ in the bottom right panel of
Fig.\ \ref{fig11}; the plot shows clear signature of multifractality
for $0.15 \pi \le \hbar \omega_D/\mathcal{J}_0 \le 0.3 \pi$. Our results
indicate that $H_{F1}$ constitutes semi-analytic expression of a
Floquet Hamiltonian which support multifractal states in its
eigenspectrum.

To understand the origin of these multifractal states, we obtain a
real space representation of $H_{F1}$. A Fourier transform of Eq.\
\ref{f2ham1} yields
\begin{eqnarray}
H_{F1} &=&\sum_{j,j'} H_{jj'} c_j^{\dagger} c_{j'} \nonumber\\
H_{jj'} &=& \int_{-\pi}^{\pi} \frac{dk_1 dk_2}{2\pi} e^{ i(k_1 j -
k_2
j')} \frac{4 i V(k_1-k_2)}{\mathcal{J}_0 T f(k_1,k_2)} \nonumber\\
&& \times \sin[\mathcal{J}_0 f(k_1,k_2) T/4] e^{i T \mathcal{J}_0 f(k_1,k_2)/4} \label{ft1}
\end{eqnarray}
A straightforward calculation outlined in the Appendix leads to an
analytic expression for $H_{jj'}$ given by
\begin{widetext}
\begin{eqnarray}
H_{jj'} &=& \frac{i}{2 \sqrt{2} }\sum_{w=|j-j'|/2}^{\infty}\frac{(
T\mathcal{J}_0)^{2w}}{(2w+1)}\frac{  \sqrt{\pi} }{\kappa^{\prime}(j,j',2w)}
\left[\frac{1}{2^{2w-1}}\begin{pmatrix}
2w \\
w
\end{pmatrix}V(j+j')/2) \right. \nonumber\\
&& \left.
+\frac{(-1)^w}{2^{2w-1}}\sum_{z=0}^{w-1}(-1)^z\begin{pmatrix}
2 w \\
z
\end{pmatrix} (V[(w-z)+(j+j')/2]+V[(j+j')/2-(w-z)])\right] \quad j-j' =2n \nonumber\\
&=& -\frac{1}{ 2\sqrt{2}}\sum_{w=(|j-j'|-1)/2}^{\infty}\frac{( T
\mathcal{J}_0)^{2w+1}}{(2w+2)}\frac{  \sqrt{\pi}}{\kappa^{\prime}(j,j',2w+1)}
 \frac{(-1)^w}{4^w}\sum_{z=0}^w(-1)^w\begin{pmatrix}
 2w+1\\
 z
 \end{pmatrix} \nonumber\\
 && \times
 \left(V[(j+j')/2-(2w+1-2z)/2]-V[(j+j')/2+(2w+1-2z)/2]\right) , \,\,
 j-j'=2n+1
 \label{posFPTsq}
\end{eqnarray}
\end{widetext}
where $n$ is an integer and the function $\kappa'$ is given by
\begin{eqnarray}
&& \kappa^{\prime}(m,n,p)= 1, \quad p=0  \label{kappadef} \\
&& = |m-n|\prod_{s=1,s \neq|m-n|/2}^{p/2}[(m-n)^2-(2s)^2], \quad  p =2k \nonumber \\
&& = \prod_{s=0, s\neq(|m-n|-1)/2}^{(p-1)/2}[(m-n)^2-(2s+1)^2],\,\,
p=2k+1 \nonumber
\end{eqnarray}
for integer $k$. From the expression of $H_{jj'}$ we clearly find
that the Floquet Hamiltonian corresponds to a hopping Hamiltonian
whose range increases with decreasing drive frequency. At very high
frequencies, only the $j=j'$ (on-site) term survives and we get back
the Magnus result. As the frequency is decreased, the amplitude of
terms for which $j \ne j'$ (hopping terms with range $|j-j'|$)
increases. Thus at intermediate frequencies, this corresponds to a
Hamiltonian with on-site quasiperiodic term (corresponding to
$j=j'$) and intermediate range hopping terms (for both odd and even
$|j-j'|$) whose amplitude depends on the drive frequency. It is well
known that similar Hamiltonians, for specific range of hopping
amplitudes, supports multifractal states in their spectrum
\cite{sinharef}. We point out that here the drive frequency may be
used to engineer these amplitudes. Our result thus constitutes an
example of analytic form of a Floquet Hamiltonian which supports
multifractal states.

\subsection{FPT for continuous protocol}
\label{secfptsin}

For the continuous protocol, we choose $\mathcal{J}(t)= \mathcal{J}_0 \cos(\omega_D t)$ so
that for $\mathcal{J}_0 \gg V_0$, $U_0$ is given by
\begin{eqnarray}
U_0(t,0) &=& \exp\left[-\frac{i \mathcal{J}}{\omega_D} \sin (\omega_D t)
\sum_n c_k^{\dagger} c_k \right] \label{contu0}
\end{eqnarray}
where we have set $\hbar =1$. This leads to $U_0(T,0)=I$ and
$H_{F0}=0$. We note that the eigenbasis for $U_0$ is still given by
$|k\rangle \equiv |n_k\rangle$

\begin{figure}
\rotatebox{0}{\includegraphics*[width= 0.48 \linewidth]{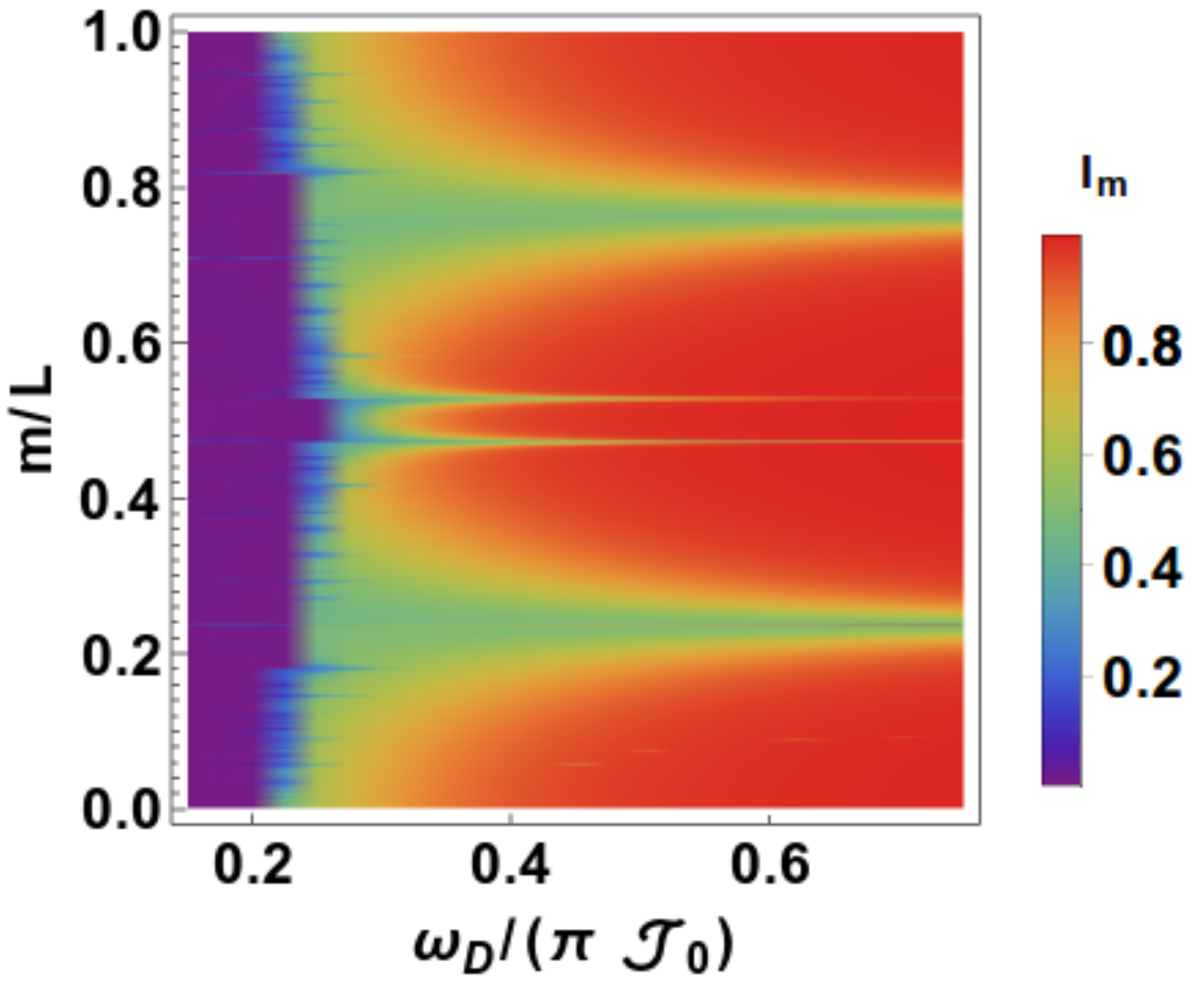}}
\rotatebox{0}{\includegraphics*[width= 0.48 \linewidth]{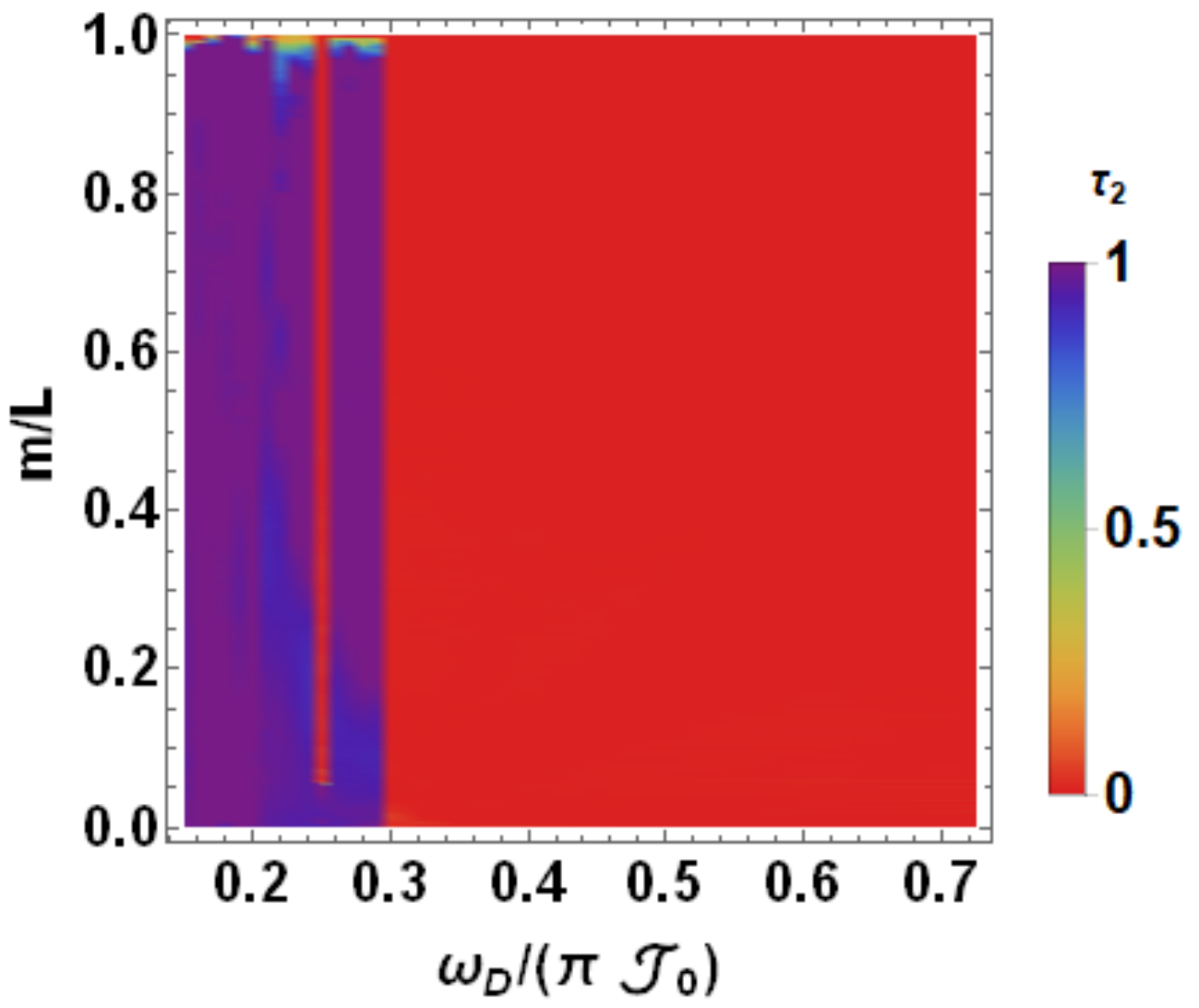}}

\caption{Left Panel: Plot of $I_m$ as a function of $m/L$ and
$\omega_D/(\pi\mathcal{J}_0)$. Right Panel: Plot of $\tau_2$ as a function
of $m/L$ and $\omega_D/(\pi\mathcal{J}_0)$. All other parameters are same
as in Fig.\ \ref{fig1} and Fig.\ \ref{fig5}. See text for details.} \label{fig12}
\end{figure}

The perturbative contribution to the first order term in the Floquet
Hamiltonian is then by Eq.\ \ref{forfl} with $U_0(t,0)$ given by
Eq.\ \ref{contu0}. A straightforward calculation shows that the
matrix elements of $U_1$
\begin{eqnarray}
\langle k_1 |U_1(T,0)|k_2\rangle  &=& -i T V(k_1-k_2) J_0(x_{12})
\label{contu1mat}
\end{eqnarray}
where $J_0$ denotes Bessel functions and $x_{12}= \mathcal{J}_0
f(k_1,k_2)/\omega_D$. Using this, we find that the first order
Floquet Hamiltonian is given by
\begin{eqnarray}
H_{F1} = \sum_{k_1,k_2} V(k_1-k_2) J_0( x_{12}) c_{k_1}^{\dagger}
c_{k_2}
\end{eqnarray}
We note that for $\omega_D \to \infty$, $J_0 \to 1$ and $H_{F1} \to
H_p$ which reproduces the Magnus results. The second order terms can
be computed in an analogous fashion. The computation procedure is
same as charted out in Ref.\ \onlinecite{fl2} and yields
\begin{widetext}
\begin{eqnarray}
H_{2F} &=& \sum_{k_1,k_2,k_3,k_4} \sum_{n=0}^{\infty}
\frac{V(k_1-k_2)V(k_3-k_4)}{(2n+1)\omega_D} \left[J_0(x_{12})
J_{2n+1}(x_{34})- J_0(x_{34}) J_{2n+1}(x_{12}) \right]
c_{k_1}^{\dagger} c_{k_2} c_{k_3}^{\dagger} c_{k_4}.
\end{eqnarray}
\end{widetext}
We note that $H_{2F} \to 0$ as $\omega_D \to \infty$ which is
consistent with the Magnus expansion results which yields a
vanishing second order contribution to $H_F$.

Next, we obtain the Floquet eigenstates and corresponding
quasienergies via numerical diagonalization of $H_F= H_{1F}+H_{2F}$.
The results are shown in Fig.\ \ref{fig12}. Again, qualitative
features like the presence of CAT states, mobility edge and
multifractal states are all captured by the perturbative $H_F$
obtained from FPT. The FPT results also show a significant reduction
in the range of drive frequencies $\omega_D$ that give rise to
multifractal states consistent with the exact numerics for the
sinusoidal protocol.

Finally we obtain a representation of $H_F$ in real space following
an analysis which is identical to that carried out in the previous
section for the square-pulse protocol. For this purpose, we consider
$H_{F1}$ and obtain its analytic form in real space. The details of
the calculation in charted out in the appendix. This yields $H_{1F}=
\sum_{jj'} H_{jj'} c_j^{\dagger} c_{j'}$ where $H_{jj'}=0$ for
$|j-j'|=2m+1$. For $|j-j'|=2m$, it is given by
\begin{widetext}
\begin{eqnarray}
H_{jj'} &=& \sum_{p=(j-j')/2,}^{\infty}\frac{(-1)^p2(2p)! (\mathcal{J}_0 T/(2 \pi))^{2p}
\sqrt{\pi} }{p! 2^{p-1}
4\Gamma(p+1)\kappa(j,j',p)}\left[\begin{pmatrix}
p \\
p/2
\end{pmatrix}V((j+j')/2) \right. \nonumber\\
&& \left. + (-1)^{p/2} \sum_{z=0}^{p/2-1}(-1)^z\begin{pmatrix}
p \\
z
\end{pmatrix} (V[(p/2-z)+(j+j')/2]+V[(j+j')/2-(p/2-z)])\right]
\nonumber\\
\kappa(m,n,p)&=& 1,\quad {\rm for}\, p=0, \quad
\kappa(m,n,p)= (m-n)\prod_{s=1,s
\neq(m-n)/2}^{p}[(m-n)^2-(2s)^2],\quad {\rm otherwise}
\label{sinrs1}
\end{eqnarray}
\end{widetext}
We note that similar to the square pulse protocol, we get a
real-space Floquet whose range increases with decreasing frequency.
The high-frequency limit leads to a completely local Hamiltonian
consistent with the Magnus result. However, for the continuous drive
protocol discussed in this subsection, $H_{1F}$ only induces
next-nearest neighbor couplings. The coupling between odd and higher
neighboring sites which differ by an odd number of lattice sites is
induced by $H_{2F}$. We do not compute this terms here but merely
observe that their contribution would be smaller by at least a
factor of $V_0/\omega_D$. The difference in coupling strength
between sites differing by odd and even number of lattice sites also
explains the reason for the structure of the CAT states. We find
that they are distributed between a site and its next-nearest
neighbor (rather than the expected nearest one). This is clearly a
consequence of having larger $H_{jj'}$ between the next-nearest
neighbor sites compared to the nearest ones.

\section{Discussion}
\label{secdiss}

In this work, we have charted out the phase diagram of the driven AA
model using both square pulse and sinusoidal drive protocols. Our
numerical studies, carried out using exact diagonalization of the
fermionic system, reveals the presence of
localization-delocalization transition in this system occurring at a
critical drive frequency $\omega_c$. Moreover, below $\omega_c$, for
a range of drive frequencies, we find the existence of a mobility
edge which separates delocalized Floquet eigenstates with
quasienergies below the edge from localized or multifractal
eigenstates above it. Our analysis shows the presence of
multifractal states in the Floquet eigenspectrum over a wide range
of drive frequencies. We show that the presence of the mobility edge
leaves its imprint on the transport of the system and on survival
probability and Shannon entropy of the driven fermions. Moreover,
the fermion transport starting from a domain wall state where all
the fermions are localized to the left-half of the chain can discern
the presence of multifractal states in the Floquet eiegnspectrum. We
note that the non-driven AA model does not support mobility edge or
multifractal states in its spectrum; thus our results constitute
dynamical signatures which have no analog in the non-driven model.

The numerical results that we find can be semi-analytically
understood within FPT. Our results regarding this constitutes
derivation of semi-analytic, albeit perturbative, Floquet
Hamiltonians for both square pulse and sinusoidal drive protocols.
We show that these perturbative, semi-analytic Hamiltonians
reproduce the localization-delocalization transition obtained
numerically; moreover, they support CAT and multifractal states in
their eigenspectrum. The reason for the presence of such states can
be understood by obtaining real-space representation of these
Hamiltonians. In real-space, these Floquet Hamiltonians contain
on-site quasiperiodic terms along with hopping terms which connects
between fermions at different sites. We find that the range of
these latter class of terms increase with decreasing drive
frequency. Consequently, these Floquet Hamiltonians belong to a
class of Hamiltonians with Aubrey-Andr\'{e} interactions and
quasi-long range hopping terms. It was shown in Ref.\
\onlinecite{sinharef} that these Hamiltonian support multifractal
states.

Our results indicate that the signature of the
localization-delocalization transition can be obtained by studying
fermionic transport. This allows us to suggest realistic experiment
which can test our theory. We suggest realization of the
Aubrey-Andr\'{e} potential in an optical lattice as done recently in
Ref.\ \onlinecite{exp2}. The drive of the hopping term may be
induced by tuning the laser strength creating the optical lattice
using either of the periodic protocols discussed. In addition, one
can start from a configuration where the fermions in the lattice are
confined to the left-half of the chain. Our prediction is that there
will be critical drive frequency $\omega_c$ below which the system
will eventually delocalize. This will be reflected in a sharp drop
in the value of $N_{\rm av}^2$ as sketched in Fig.\ \ref{fig7}.
Moreover, for a range of frequencies below $\omega_c$, $N_{\rm
av}^2$ will remain between its values for localized ($N_{\rm
av}^2=1$) and delocalized ($N_{\rm av}^2=0)$ states signifying the
presence of the mobility edge.

In conclusion, we have studied the driven Aubrey-Andr\'{e} model and
showed the presence of a drive-induced localization-delocalization
transition. Our results indicate the presence of mobility edge and
multifractal states in the Floquet eigenstates; their existence can
be seen from analytic, perturbative form of $H_F$ which we derive
using a Floquet perturbation theory which represents a resummation
of an infinite class of terms in the Magnus expansion. We show that
the presence of this mobility edge is reflected in fermionic
transport and suggest experiments which can test our theory.

\begin{acknowledgments}

The authors acknowledges related discussions at ICTS, Bengaluru
during the program Thermalization, Many body localization and
Hydrodynamics (Code: ICTS/hydrodynamics2019/11). The work of A.S.
is partly supported through the Max Planck Partner Group program
between the Indian Association for the Cultivation of Science
(Kolkata) and the Max Planck Institute for the Physics of Complex
Systems (Dresden).

\end{acknowledgments}
\appendix

\section{Real space representation of $H_F$}
\label{appa}

\subsection{Square pulse}
\label{appsqpulse}

We start from Eq. \ref{ft1} from the main text,
\begin{eqnarray}
H_{F1} &=&\sum_{j,j'} H_{jj'} c_j^{\dagger} c_{j'} \nonumber\\
H_{jj'} &=& \int_{-\pi}^{\pi} \frac{dk_1 dk_2}{\pi} e^{ i(k_1 j -
k_2
j')} \frac{4 i V(k_1-k_2)}{\mathcal{J}_0 T f(k_1,k_2)} \nonumber\\
&& \times \sin[\mathcal{J}_0 f(k_1,k_2) T/4] e^{i T \mathcal{J}_0 f(k_1,k_2)/4} \label{ft11}
\end{eqnarray}

We shift to the center of momentum coordinates,
$q=\frac{k_1-k_2}{2}$ and $r=\frac{k_1+k_2}{2}$ where $dq dr
=\frac{1}{2} dk_1 dk_2$ and write
\begin{eqnarray}
H_{jj'}&=&\frac{1}{2 \pi}\int_{-\pi}^{\pi}\int_{-\pi}^{\pi} dq dr e^{i[q(j+j')+r(j-j')]} \frac{4 i V(2q)}{\mathcal{J}_0 T g(q,r)}\nonumber \\
&& \times\sin[\mathcal{J}_0 g(q,r) T/4] e^{i T \mathcal{J}_0 g (q,r)/4}
\label{integral}
\end{eqnarray}
Now expanding the oscillatory part, we write
\begin{eqnarray}
\sin[\mathcal{J}_0 g(q,r) T/4] e^{i T \mathcal{J}_0 g (q,r)/4}
=\frac{T}{4}\sum_{p=0}^{\infty}\frac{(i T \mathcal{J}_0
g(q,r)/2)^{p}}{(p+1)!}  \nonumber\\
\end{eqnarray}
Hence Eq. \ref{integral} can be written as,
\begin{eqnarray}
H_{jj'}&=&\frac{1}{2 \pi}\sum _{p=0}^{\infty}\frac{(i T
\mathcal{J}_0/2)^{p}}{(p+1)!}\int_{-\pi}^{\pi}\int_{-\pi}^{\pi} dq
dr \nonumber \\
&&\times e^{i[q(j+j')+r(j-j')]} i V(2q) (2 \sin q \sin r)^{p}
\end{eqnarray}
The next task is to perform the integrals. Performing the integral
over $r$ first, we get
\begin{eqnarray}
H_{jj'}&=&\frac{1}{2\pi}\sum _{p=0}^{\infty}\frac{(i T \mathcal{J}_0/2)^{p}}{(p+1)!}\int_{-\pi}^{\pi} dq  e^{i[q(j+j')]} \\
&&\times i V(2 q) (\sin q )^{p}\frac{2^{p+1}p!(-i)^p \sin[(j-j')\pi]
}{\kappa(j,j',p)} \nonumber
\end{eqnarray}
where
\begin{eqnarray}
\kappa(j,j',p)&=&(j-j')\prod_{s=1}^{p/2}[(j-j')^2-(2s)^2],\hspace{0.2in} p=2k \nonumber \\
&=&\prod_{s=0}^{(p-1)/2}[(j-j')^2-(2s+1)^2],\hspace{0.2in} p=2k+1
\nonumber\\
\end{eqnarray}
for any integer $k$. It is to be noted that only when $s=|j-j'|/2$
for $p$ even and $s=(|j-j'|-1)/2$ for $p$ odd, the integrals give a
finite contribution. Hence the summation over $p$ must start from
$p=|j-j'|$. This gives,
\begin{eqnarray}
H_{jj'}&=&\frac{i}{2\pi}\sum _{p=|j-j'|}^{\infty}\frac{(i T \mathcal{J}_0/2)^{p}}{(p+1)!}\int_{-\pi}^{\pi} dq  e^{i[q(j+j')]}\nonumber \\
&&\times V(2 q) (\sin q )^{p}\frac{2^{p+1}p!(-i)^p \pi }{\kappa^{\prime}(j,j',p)}
\label{qintegral}
\end{eqnarray}
where ,\begin{eqnarray}
&&\kappa^{\prime}(j,j',p)=1,\hspace{0.2in} \textit{p=0}  \\
&=&|j-j'|\prod_{s=1,s \neq|j-j'|/2}^{p/2}[(j-j')^2-(2s)^2],\hspace{0.1in} p=2k \nonumber \\
&=&\prod_{s=0,
s\neq(|j-j'|-1)/2}^{(p-1)/2}[(j-j')^2-(2s+1)^2],\hspace{0.1in}
p=2k+1\nonumber
\end{eqnarray}
Next, we perform the integral over $q$ using standard trigonometric
identities. First one should separate out the even and odd parts of
the integral, and note that when $j-j'$ is even, $p$ necessarily is
always even as the rest of the terms integrate to $0$ and similarly
for $j-j'$ odd. Hence for $j-j'$ even, assuming $p=2 w$, we get
\begin{widetext}
\begin{equation}
H_{jj'}=\frac{i}{2\pi}\sum_{w=|j-j'|/2}^{\infty}\frac{( T \mathcal{J}_0)^{2w}}{(2w+1)}\frac{ \pi }{\kappa^{\prime}(j,j',2w)}\int_{-\pi}^{\pi} dq  \cos[q(j+j')]V(2 q)
 [\frac{1}{2^{2w}}\begin{pmatrix}
2w \\
w
\end{pmatrix}+\frac{(-1)^w}{2^{2w-1}}\sum_{z=0}^{w-1}(-1)^z\begin{pmatrix}
2 w \\
z
\end{pmatrix} \cos[2(w-z)q]]
\end{equation}
\end{widetext}
For odd $j-j'$, we consider $p=2 w+1$  and obtain
\begin{widetext}
\begin{eqnarray}
H_{jj'} &=& -\frac{ 1 }{2\pi}\sum_{w=(|j-j'|-1)/2}^{\infty}\frac{( T
\mathcal{J}_0)^{2w+1}}{(2w+2)}\frac{ \pi
}{\kappa^{\prime}(j,j',2w+1)}\int_{-\pi}^{\pi} dq  \sin[q(j+j')]V(2
q) \nonumber\\
&& \times \frac{(-1)^w}{4^w}\sum_{z=0}^w(-1)^w\begin{pmatrix}
 2w+1\\
 z
 \end{pmatrix}\sin[(2w+1-2z)q]
\end{eqnarray}
\end{widetext}
Using the inverse Fourier transform $\frac{1}{\sqrt{2
\pi}}\int_{-\pi}^{\pi} V(2 q) e^{i 2 q x}=V(x)$ and integrating over
$q$, we get  Eq. \ref{posFPTsq} of the main text.

\subsection{Sinusoidal pulse}

For this drive protocol we start from
\begin{eqnarray}
H_{jj'} &=& \frac{1}{2\pi}\int_{-\pi}^{\pi} \int_{-\pi}^{\pi}d k_1 d
k_2e^{i[k_1 j -k_2 j']}V(k_1-k_2)\nonumber\\
&& \times J_0(\mathcal{J}_0 f(k_1,k_2)/\omega_D)
\end{eqnarray}
As in the case of square protocol, we switch to relative and center
of mass momenta and obtain
\begin{eqnarray}
H_{jj'} &=& \frac{1}{4\pi}\int_{-\pi}^{\pi} \int_{-\pi}^{\pi}d q d r
e^{i[q(j+j')+r(j-j')]}V(2 q)\nonumber\\
&& \times J_0(\mathcal{J}_0 g(q,r)/\omega_D) \label{position}
\end{eqnarray}
where $g(q,r)=-2 (\sin q \sin r)$. Next, we use the expansion of
$J_0(x)$,
\begin{equation}
J_0(x)=\sum_{p=0}^{\infty}\frac{(-1)^p}{p! \Gamma(p+1)}(x/2)^{2p}
\label{bessel}
\end{equation}
Substituting Eq. \ref{bessel} in Eq.\ \ref{position} we find,
\begin{eqnarray}
H_{jj'}&=&\sum_{p=0}^{\infty}\frac{1}{4\pi}\int_{-\pi}^{\pi} \int_{-\pi}^{\pi}d q d r e^{i[q(j+j')+r(j-j')]} V(2 q)\nonumber \\
&&\times \frac{(-1)^p}{p! \Gamma(p+1)}(\mathcal{J}_0 \sin q \sin
r/\omega_D)^{2p}
\end{eqnarray}
Integrating over $r$ we get,
\begin{eqnarray}
H_{mn}&=&-\sum_{p=0}^{\infty}\frac{1}{4\pi}\int_{-\pi}^{\pi} dq  e^{i[q(j+j')]}V(2 q) \\
&&\times (\mathcal{J}_0\sin q/\omega_D )^{2 p}\frac{(-1)^p 2(2p)!
\sin[(j-j')\pi] }{p! \Gamma(p+1)\kappa(j,j',p)} \nonumber
\end{eqnarray}
where,
\begin{eqnarray}
\kappa(j,j',p)&=&|j-j'|\prod_{s=1}^{p}[(j-j')^2-(2s)^2],\hspace{0.2in}  \nonumber
\end{eqnarray}
And noting that the summation can only start from $p=(m-n)/2$ we write,
\begin{eqnarray}
H_{jj'}&=&-\sum_{p=|j-j'|/2,|j-j'| even}^{\infty}\frac{1}{4\pi}\int_{-\pi}^{\pi} dq  e^{i[q(j+j')]}V(2 q)\nonumber \\
&& \times (\mathcal{J}_0 T\sin q/(2 \pi) )^{2 p}\frac{(-1)^p2(2p)! \pi }{p! \Gamma(p+1)\kappa^{\prime}(j,j',p)}
\end{eqnarray}
where ,\begin{eqnarray}
\kappa^{\prime}(j,j',p)&=&1,\hspace{0.2in} \textit{p=0}  \\
&=&|j-j'|\prod_{s=1,s \neq(j-j')/2}^{p}[(j-j')^2-(2s)^2]\nonumber \\
 &&{\rm otherwise} \nonumber
\end{eqnarray}
and we have replaced $\omega_D$ by $2 \pi/T$. One can immediately
see if $|j-j'|$ is odd then no term contributes and $H_{jj'}=0$.
Then one can integrate over $q$ as well to get Eq. \ref{sinrs1} of
the main text.  The expressions of the second order term in $H_F$ is
quite complicated and we have not analyzed their form in position
space. However, we note that these terms are of the form $\sim
\sum_{n=0}^{\infty} [J_{2n+1}(x) J_0(y)-J_0(x)J_{2n+1}(y)]/(2n+1)$.
Thus from the expansion of $J_n(x)$, it can be seen that these terms
would actually give rise to odd powers of $\sin q$. This means that
here the terms of $H_{jj'}$ where $j-j'$ odd will be non-zero. The
consequence of this is discussed in the main text.

\section{Approach to Steady state}
\label{appa1}

\begin{figure}
\rotatebox{0}{\includegraphics*[width= 0.48 \linewidth]{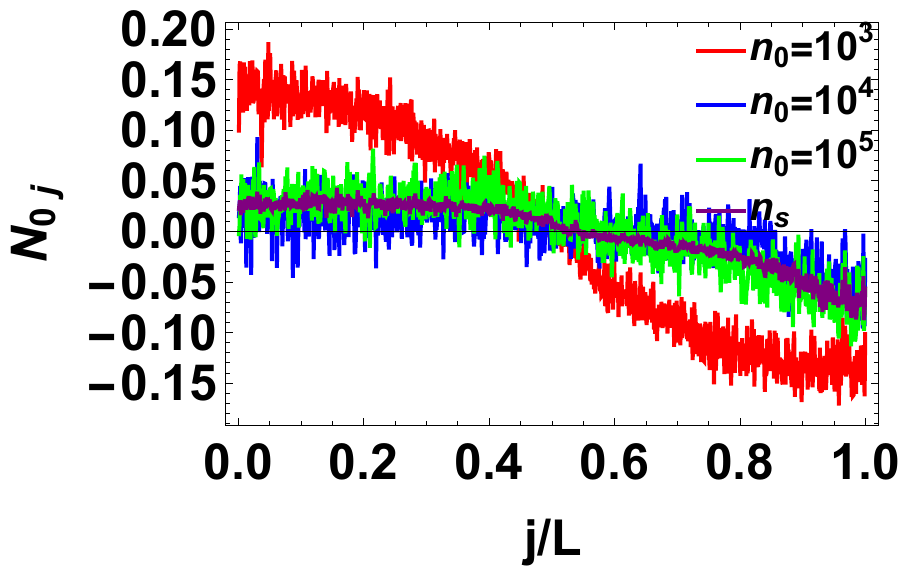}}
\rotatebox{0}{\includegraphics*[width= 0.48 \linewidth]{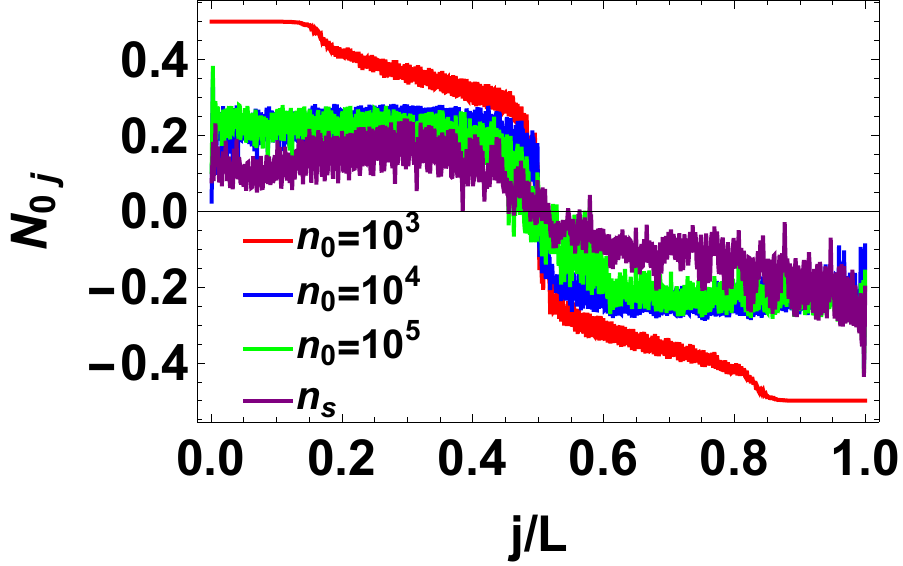}}
\rotatebox{0}{\includegraphics*[width= 0.48 \linewidth]{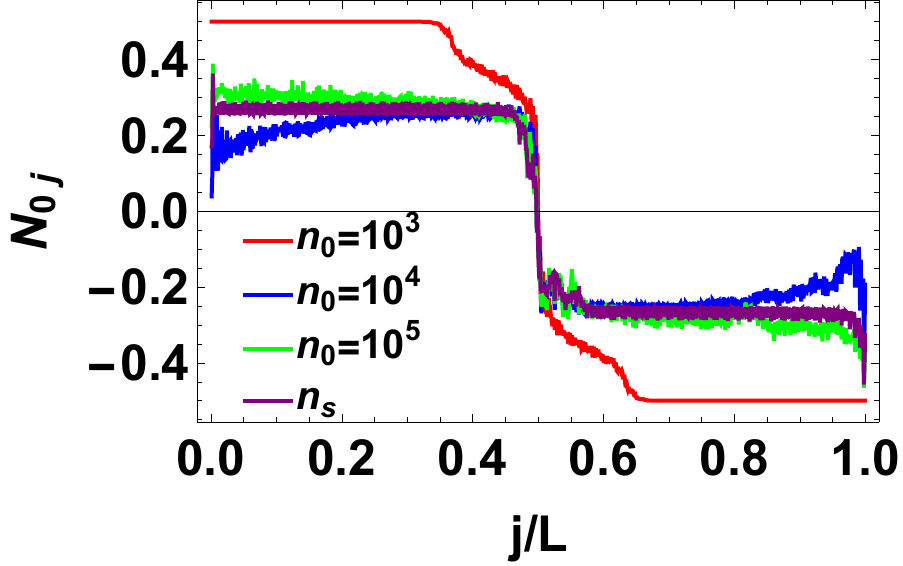}}
\rotatebox{0}{\includegraphics*[width= 0.48 \linewidth]{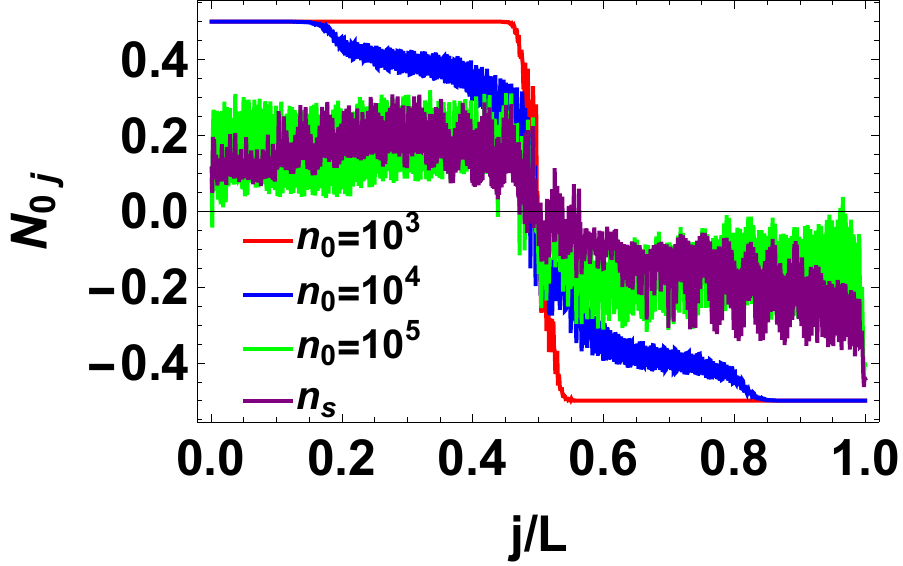}}
\rotatebox{0}{\includegraphics*[width= 0.48 \linewidth]{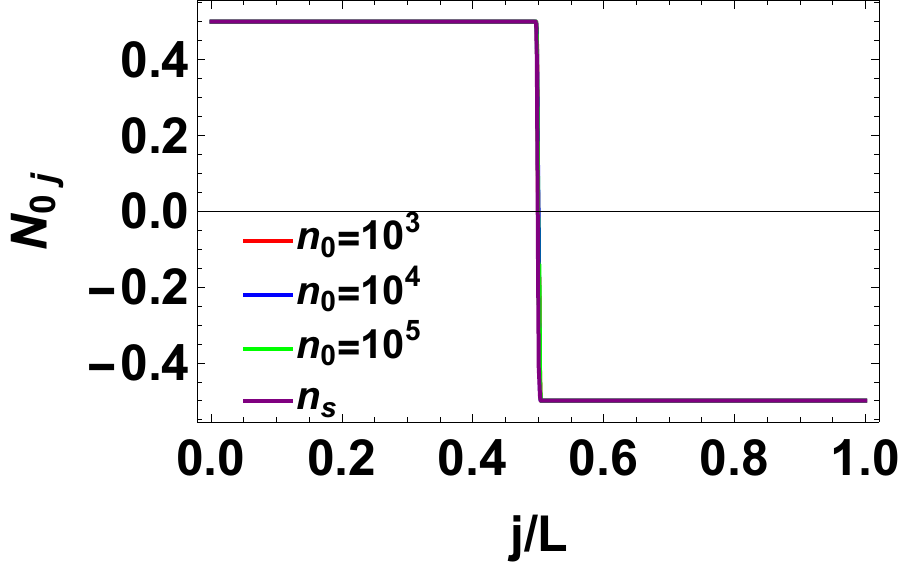}}
\rotatebox{0}{\includegraphics*[width= 0.48 \linewidth]{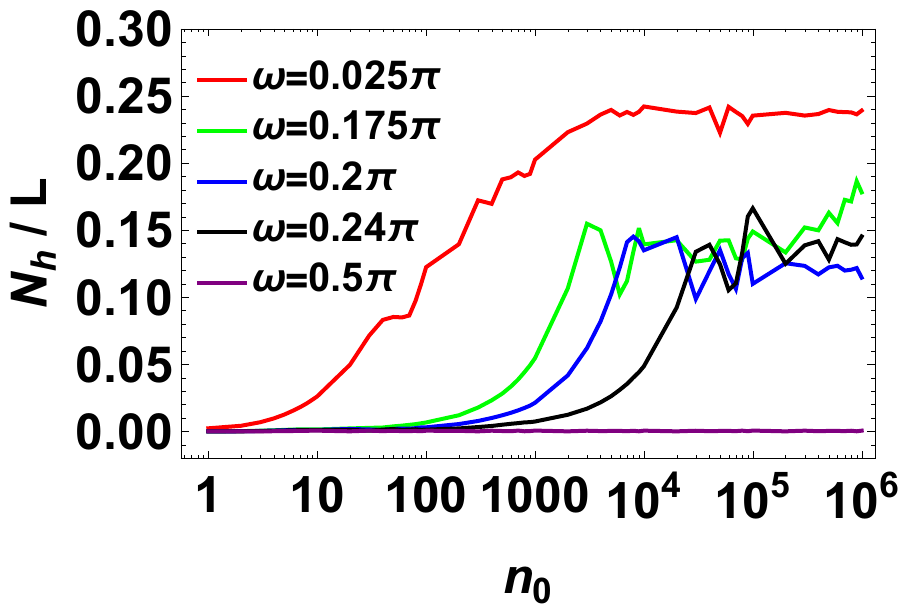}}
\caption{Top Left Panel: Plot of $N_{0j}$ as a function of $j/L$ for
$\omega_D/(\pi \mathcal{J}_0)$=0.025 where the full spectrum is
delocalized. Top Right Panel: Plot of $N_{0j}$ as a function of
$j/L$ for $\omega_D/(\pi \mathcal{J}_0)$=0.175 at which there is a
mobility edge between delocalized and multifractal states. Middle
Left Panel: Plot of $N_{0j}$ as a function of $j/L$ for
$\omega_D/(\pi \mathcal{J}_0)$=0.20 at which there is a mobility
edge between delocalized and localized states. Middle Right Panel:
Plot of $N_{0j}$ as a function of $j/L$ for $\omega_D/(\pi
\mathcal{J}_0)$=0.24 at which there is a mobility edge between
delocalized and multifractal states. This drive frequency is near
the critical frequency $\omega_c$. Bottom left Panel: Plot of
$N_{0j}$ as a function of $j/L$ for $\omega_D/(\pi
\mathcal{J}_0)$=0.5 where the full spectrum is localized. Bottom
right Panel:Plot of the number of particles present in the right
half (beginning from domain wall initial state) as a function of
number of drive cycles $n_0$ for various representative drive
frequencies. All other parameters are same as in Fig.\ \ref{fig1}.
See text for details.} \label{figapp1}
\end{figure}

In this appendix we discuss, in brief, the approach of our model subjected to
square pulse drive, to the steady states shown in the main text, starting from
the domain wall initial state given by,
\begin{eqnarray*}
|\psi_{\rm init}\rangle = |n_{1}=1, ... n_{L/2}=1, n_{L/2+1}=0, ...
n_{L}=0\rangle  \label{initstate}
\end{eqnarray*}
To this effect we study the distribution of fermion number density
$N_{j}=\langle \psi(n_0 T)|\hat{n}_j-1/2|\psi(n_0 T)\rangle$, where
$\psi(n_0 T)=U(n_0 T,0) |\psi_{\rm init}\rangle$ and
$\hat{n}_j=c_j^{\dagger}c_j$ at different number of cycles $n_0$.
Fig.\ \ref{figapp1} shows the distributions studied for different
drive frequencies. For low drive frequencies ($\omega_D/(\pi
\mathcal{J}_0)=0.025$) where the entire spectrum was shown to be
delocalized, it is seen that this quantity attains its steady state
value for a smaller number of cycles compared to other cases. As the
drive frequency is increased and we reach the region with mobility
edge, transport becomes slower as can be seen from the top right and
middle left panels which show results for $\omega_D/(\pi
\mathcal{J}_0)=0.175$ and $\omega_D/(\pi \mathcal{J}_0)=0.20$
respectively. However, it is seen that while $10^3$ drive cycles is
not enough to reach close to the steady states, $10^4$ cycles is
enough even for the drive frequency ($\omega_D/(\pi
\mathcal{J}_0)=0.20$) which supports a mobility edge between
localized and delocalized states. However for a drive frequency
higher than that which also supports multifractal and delocalized
states ($\omega_D/(\pi \mathcal{J}_0)=0.24$), $n_0=10^4$ gives the
impression the system is localized from the distribution. Only at
extremely large number of cycles $n_0=10^5$ does the system give the
expected behavior of the steady state. This is possibly due to the
proximity of this drive frequency to the critical frequency
$\omega_D/(\pi \mathcal{J}_0)=0.30$. The bottom right panel of Fig.
\ref{figapp1} shows the evolution of $N_h=\sum_{j=L/2+1}^L
\hat{n_j}$ with time, i.e., the transport of particles from the
left-half of the system to the right-half. It also shows that only
after a sufficiently long time scale does $N_h$ for $\omega_D/(\pi
\mathcal{J}_0)=0.24$ overtake $\omega_D/(\pi \mathcal{J}_0)=0.20$
which is expected in the steady state as the latter frequency
supports delocalized and localized eigenfunctions and hence,
particle transport should show a suppression compared to the former
drive frequency which supports delocalized and multifractal
eigenfunctions.

\begin{figure}
\rotatebox{0} {\includegraphics*[width=0.49\linewidth]{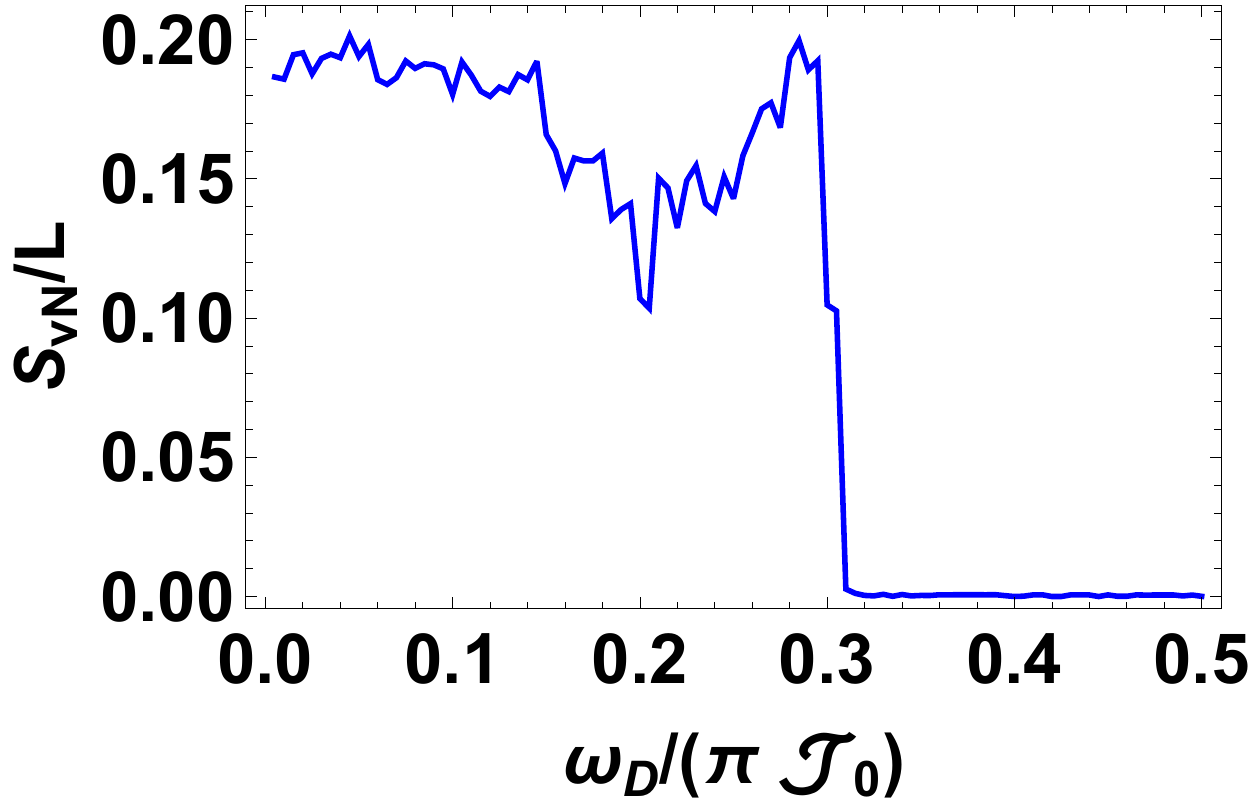}}
{\includegraphics*[width= 0.49 \linewidth]{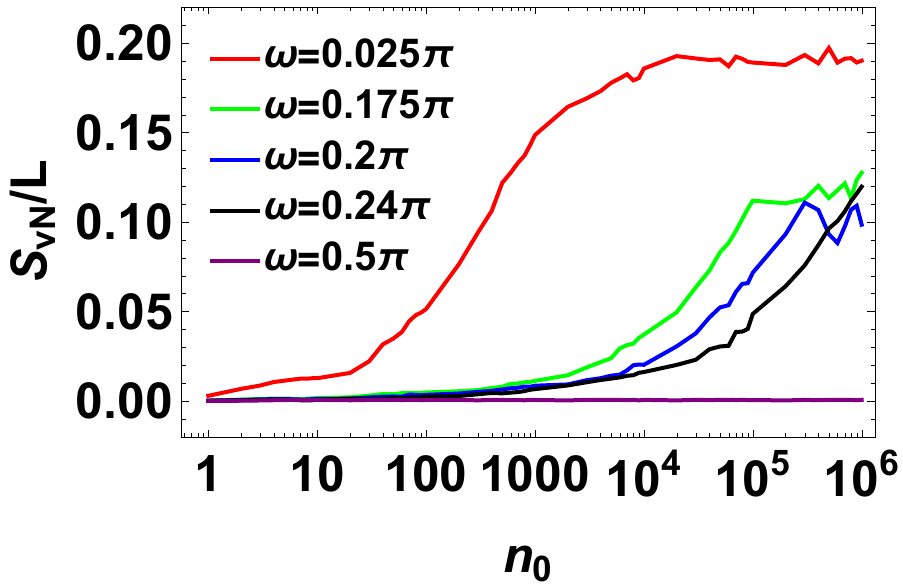}}
\caption{Left panel: Plot of half chain von-Neumann entanglement
entropy($S_{vN}$) in the steady state starting from a domain wall
initial state scaled by the system length $L$ as a function of drive
frequency $\omega_D/(\pi \mathcal{J}_0)$ showing the signature of
delocalized, localized and mixture of delocalized and multifractal
and delocalized and localized states. Right Panel: Plot of half
chain entanglement as a function of number of drive cycles $n_0$ for
various representative drive frequencies showing how it reaches the
steady state value. All other parameters are same as in Fig.\
\ref{fig1}.} \label{figapp2}
\end{figure}

In Fig. \ref{figapp2} we study whether similar feature show up in
the half chain entanglement entropy of the system starting from the
domain wall state. To calculate the stroboscopic time evolution of
von-Neumann entropy we first calculate the time-evolution of the
two-point correlation function ($\langle c_i^{\dagger}c_j\rangle
(t)$) in the Heisenberg picture, and then use the technique outlined
in Ref. \onlinecite{henley} to extract the von-Neumann entropy
between the left half of the system (between sites $1$ and $L/2$)
and the right half ($L/2+1$ and $L$). To calculate the steady state
entanglenment entropy, we utilize the steady state correlators
calculated using the procedure outlined in the main text and then
use the method of Ref. \onlinecite{henley}. The steady state
entanglement shows the expected features of a dip when the system's
eigenstates change from being fully delocalized to delocalized and
multifractal and then, to delocalized and localized as the drive
frequency is increased from $\omega_D \approx 0$. Finally, the
steady state entanglement becomes almost zero when the system
becomes fully localized beyond $\omega_c$. However, as with the
number density in the right-half of the system, the entanglement
also requires larger times to approach its steady state value as we
tune the drive frequency to be closer to the critical frequency,
showing a behavior similar to the transport of particles to the
right half of the system.

\end{document}